%% file: B2G-19-005_temp.tex
\pdfoutput=1

\documentclass[11pt,twoside,a4paper,cmspaper,final,collab]{cms-tdr}
\def\svnVersion{57a7ed0-D}\def\svnDate{2020/12/13}\def\cmsCernNoTag{CERN-EP-2020-154}\def\cmsCernDate{\today}\def\cmsMessage{"Published in Physical Review D as \href{http://dx.doi.org/10.1103/PhysRevD.102.112004}{\doi{10.1103/PhysRevD.102.112004}.}"}

\begin{document}\cmsNoteHeader{B2G-19-005}

\hyphenation{had-ron-i-za-tion}
\hyphenation{cal-or-i-me-ter}
\hyphenation{de-vices}
\newlength\cmsTabSkip\setlength{\cmsTabSkip}{1ex}
\newlength\masslimitswidth\ifthenelse{\boolean{cms@external}}{\setlength{\masslimitswidth}{0.39\textwidth}}{\setlength{\masslimitswidth}{0.45\textwidth}}
\ifthenelse{\boolean{cms@external}}{\providecommand{\masslimitsone}{upper\xspace}}{\providecommand{\masslimitsone}{upper left\xspace}} 
\ifthenelse{\boolean{cms@external}}{\providecommand{\masslimitstwo}{middle\xspace}}{\providecommand{\masslimitstwo}{upper right\xspace}} 
\ifthenelse{\boolean{cms@external}}{\providecommand{\masslimitsthree}{lower\xspace}}{\providecommand{\masslimitsthree}{lower\xspace}} 
\newcommand{\BbH}{\ensuremath{\PB \to \PQb\PH}\xspace}
\newcommand{\BbZ}{\ensuremath{\PB \to \PQb\PZ}\xspace}
\newcommand{\BtW}{\ensuremath{\PB \to \PQt\PW}\xspace}
\newcommand{\BrBbH}{\ensuremath{\mathcal{B}(\PB \to \PQb\PH)}\xspace}
\newcommand{\BrBbZ}{\ensuremath{\mathcal{B}(\PB \to \PQb\PZ)}\xspace}
\newcommand{\bhbh}{\ensuremath{\PQb\PH\PQb\PH}\xspace}
\newcommand{\bhbz}{\ensuremath{\PQb\PH\PQb\PZ}\xspace}
\newcommand{\bzbz}{\ensuremath{\PQb\PZ\PQb\PZ}\xspace}
\newcommand{\mvlq}{\ensuremath{m_\text{VLQ}}\xspace}
\newcommand{\chisq}{\ensuremath{\chi^2}\xspace}
\newcommand{\chimod}{\ensuremath{\chi^2_\text{mod}}\xspace}
\newcommand{\chimodndf}{\ensuremath{\chi^2_\text{mod}/\text{ndf}}\xspace}
\DeclareRobustCommand{\VQY}{{\HepParticle{Y}{}{}}\Xspace}
\DeclareRobustCommand{\VQT}{{\HepParticle{T}{}{}}\Xspace}
\renewcommand{\floatpagefraction}{0.7}

\cmsNoteHeader{B2G-19-005}

\title{A search for bottom-type, vector-like quark pair production in a fully hadronic final state in proton-proton collisions at \texorpdfstring{$\sqrt{s} = 13\TeV$}{sqrt(s) = 13 TeV}}

\date{\today}

\abstract{A search is described for the production of a pair of bottom-type vector-like quarks (VLQs), each decaying into a \PQb or \PAQb quark and either a Higgs or a \PZ boson, with a mass greater than 1000\GeV. The analysis is based on data from proton-proton collisions at a 13\TeV center-of-mass energy recorded at the CERN LHC, corresponding to a total integrated luminosity of 137\fbinv. As the predominant decay modes of the Higgs and \PZ bosons are to a pair of quarks, the analysis focuses on final states consisting of jets resulting from the six quarks produced in the events. Since the two jets produced in the decay of a highly Lorentz-boosted Higgs or \PZ boson can merge to form a single jet, nine independent analyses are performed, categorized by the number of observed jets and the reconstructed event mode. No signal in excess of the expected background is observed. Lower limits are set on the VLQ mass at 95\% confidence level equal to 1570\GeV in the case where the VLQ decays exclusively to a \PQb quark and a Higgs boson, 1390\GeV for when it decays exclusively to a \PQb quark and a \PZ boson, and 1450\GeV for when it decays equally in these two modes. These limits represent significant improvements over the previously published VLQ limits.}

\hypersetup{pdfauthor={CMS Collaboration},pdftitle={A search for bottom-type, vector-like quark pair production in a fully hadronic final state in proton-proton collisions at sqrt(s)= 13 TeV},pdfsubject={CMS},pdfkeywords={CMS, physics, vector-like quark, VLQ}}

\maketitle

\section{Introduction}

One of the biggest puzzles in elementary particle physics concerns the large difference between the electroweak scale and the Planck scale, and the related problem of the unexpectedly low value of the Higgs boson mass~\cite{tHooft:1979rat}. In the standard model (SM), the Higgs boson \PH is assumed to be a fundamental scalar (spin-0) particle. Unlike the fundamental fermions (leptons and quarks) and the vector gauge bosons, the corrections to the Higgs boson mass due to vacuum energy fluctuations are quadratic, driving the Higgs boson mass to the cutoff value of the vacuum energy fluctuations. In the absence of any new physics below the Planck scale, this cutoff is about $10^{19}\GeV$. In that case, the Higgs boson mass would naturally be expected to be seventeen orders of magnitude greater than its measured mass of 125\GeV.

Although supersymmetry provides an elegant solution to this problem~\cite{Wess:1973kz,Fayet:1976cr}, the lack of evidence for the production of supersymmetric particles at the CERN LHC indicates that, if supersymmetry is realized in nature, it is broken at an energy scale greater than a few \TeVns and, therefore, does not solve the fine tuning of the 125\GeV Higgs boson mass. Several alternative theories have been proposed for solving this fine tuning problem. These theories include composite Higgs models~\cite{Georgi:1974yw,Kaplan:1983sm,Agashe:2004rs}, in which the Higgs boson is not a fundamental particle, but rather contains constituents bound by a new type of gauge interaction, and little Higgs models~\cite{ArkaniHamed:2001nc,ArkaniHamed:2002qy}, in which the Higgs boson is a pseudo-Nambu--Goldstone boson that arises from spontaneous breaking of a global symmetry at the \TeVns energy scale. Both of these types of models predict a new class of vector-like fermions~\cite{delAguila:1982fs} with the same charges as the SM fermions, but with purely vector current couplings to the weak gauge bosons. In composite Higgs models, the vector-like quarks (VLQs) are excited bound-state resonances, while in little Higgs models they are fundamental particles that cancel loop divergences. 

Since the VLQs are nonchiral, Lagrangian mass terms not arising from Yukawa couplings to the Higgs field are allowed, thereby avoiding the constraints on heavy, sequential fourth-generation quarks set by the measured cross section for Higgs boson production at the LHC~\cite{Aad:2019mbh,Sirunyan:2018sgc}. Requiring VLQs to have renormalizable couplings to the SM quarks permits only four types of VLQs, defined by their charge $q$: $q = -1/3$ (\PB),  $q= +2/3$ (\VQT), $q = -4/3$ (\PX), and $q = +5/3$ (\VQY)~\cite{Aguilar-Saavedra:2013qpa}. These are arranged into seven multiplets: two singlets (\VQT and \PB), three doublets ($\VQT\PB$, $\PX\VQT$, and $\VQY\PB$), and two triplets ($\PX\VQT\PB$ and $\VQT\PB\VQY$)~\cite{delAguila:2000rc}. This analysis focuses on the $q = -1/3$ (\PB) type of VLQ. 

The branching fractions $\mathcal{B}$ of the \VQT and \PB are model specific and depend upon the VLQ multiplet configuration, the mass of the VLQ, and the coupling of the VLQ to chiral quarks~\cite{Aguilar-Saavedra:2013wba}. In general, up-type quark mass eigenstates will be mixtures of the chiral up-type quarks with the \VQT VLQ, while down-type quark mass eigenstates will be mixtures of the chiral down-type quarks and the \PB VLQ. Precision measurements of the couplings of the first and second generation SM quarks constrain their mixings with VLQs and indicate that the only sizable couplings of the \VQT and \PB VLQs allowed are to SM quarks of the third generation, although couplings to other quarks are not excluded~\cite{Atre:2008iu,Atre:2011ae,delAguila:2000rc}. In this analysis, we assume the \PB VLQ has three decay modes: \BbZ, \BbH, and \BtW. In most models, for a \PB VLQ mass greater than the current limit of approximately 1000\GeV, there is a small difference between \BrBbZ and \BrBbH, depending on the VLQ mass, but the difference is essentially zero for masses greater than 2000\GeV. The expected values of \BrBbZ and \BrBbH also depend upon the multiplet configuration. They are 50\% for both the $\PX\VQT\PB$ triplet and the $\PB\VQY$ doublet, and 25\% for the $\VQT\PB\VQY$ triplet and the \PB singlet.
The branching fractions for the $\VQT\PB$ doublet depend upon the mixing of the \VQT and \PB VLQs with chiral quarks. If the $\VQT\PQt$ mixing is zero, the \BbH and \BbZ branching fractions are 50\%. If the $\VQT\PQt$ and $\PB\PQb$ mixing are equal, these branching fractions are 25\%. If the $\PB\PQb$ mixing is zero, these branching fractions are zero~\cite{Aguilar-Saavedra:2013qpa}.

Results from both the ATLAS and CMS Collaborations using events with fully hadronic final states, based on data from proton-proton ($\Pp\Pp$) collisions at $\sqrt{s} = 13\TeV$ with an integrated luminosity of 36\fbinv, have excluded a pair-produced \PB with mass up to approximately 1100\GeV. The ATLAS analysis~\cite{Aaboud:2018wxv} was based on a classification of event signatures using neural networks, while the CMS analysis~\cite{Sirunyan:2019sza} used event shapes to identify Lorentz-boosted objects. The ATLAS (CMS) results exclude masses, at 95\% confidence level (\CL), up to 1010 (980), 710 (1070), and 950 (1025)\GeV for the 100\% \BbH, 100\% \BbZ, and the $\PB\VQY$ doublet cases, respectively. In addition, an ATLAS analysis~\cite{Aaboud:2018pii} combining both fully hadronic and leptonic channels excludes values of the \PB mass up to 1140\GeV for the $\PB\VQY$ doublet case. The analysis presented here improves on these results by using the full 137\fbinv data set collected by CMS in 2016--2018, and by fully reconstructing the event kinematics, thereby allowing the mass of the \PB to be reconstructed.

\section{Analysis overview}

\begin{figure*}[hbtp]
\centering
\includegraphics[width=0.8\textwidth]{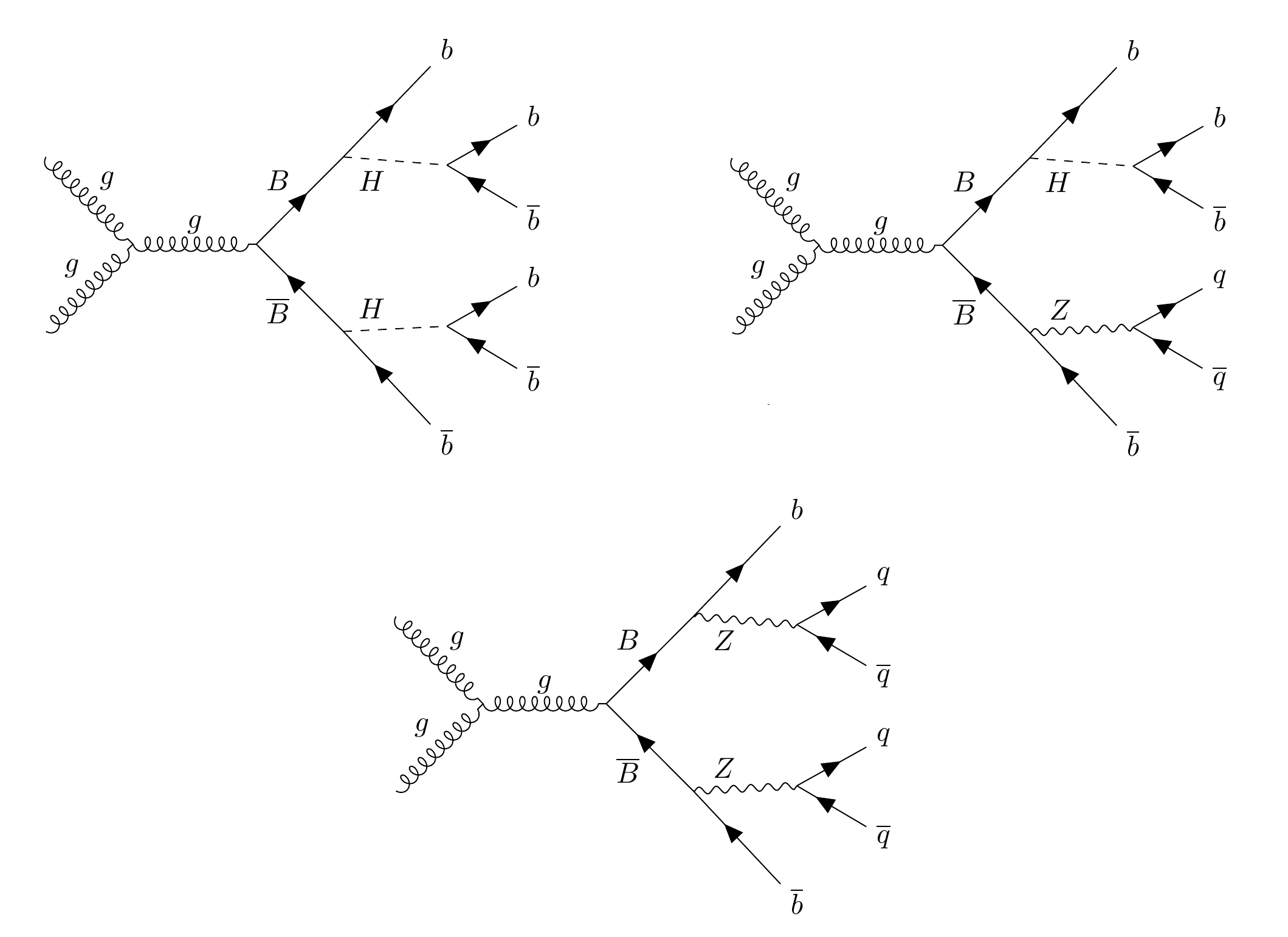}
\caption{Dominant diagrams of the pair production of bottom-type VLQs (\PB) that subsequently decay to a \PQb or \PAQb quark and either a Higgs or \PZ boson. In events targeted by this analysis, the \PZ boson then decays to a pair of quarks, where \PQq denotes any quark other than a top quark, while the Higgs boson decays to \PQb quarks. Upper left: \bhbh mode, upper right: \bhbz mode, lower: \bzbz mode.}
\label{fig:feynman_graph}
\end{figure*}

This analysis involves a search for the production of a pair of bottom-type VLQs with mass greater than 1000\GeV, using data from $\Pp\Pp$ collisions at $\sqrt{s} = 13\TeV$ at the LHC collected by the CMS detector during 2016--2018, corresponding to an integrated luminosity of 137\fbinv. The analysis is focused on events in which each of the VLQs decays to a \PQb or \PAQb quark and to either a Higgs or \PZ boson. Since the dominant decay modes of the Higgs and \PZ bosons are to a quark and antiquark pair, we select final states consisting of jets resulting from the quarks and antiquarks produced in the decays of the VLQs and subsequent decays of the two bosons. The events are categorized into three modes, depending on the daughter bosons: \bhbh, \bhbz, and \bzbz. Figure~\ref{fig:feynman_graph} shows the dominant Feynman diagrams for these three modes.

Background from SM processes (predominantly ``multijet'' events, consisting solely of jets produced through the strong interaction) is reduced by requiring that the jets are consistent with the production of a pair of bosons (either Higgs or \PZ), that the reconstructed VLQs have equal masses, and that some of the jets are tagged as originating from \PQb quarks. For a highly boosted Higgs or \PZ boson, the two jets resulting from its daughter quarks might merge into a single reconstructed jet. In order to include these events, three orthogonal, fully independent analyses are carried out using exclusive sets of events categorized by the observed jet multiplicity: 4, 5, or 6 jets. The final result is obtained by combining these three independent analyses. In this paper, we use ``jet tagging requirements'' to refer to both single jets tagged as being from a \PQb quark, and merged jets tagged as containing a \bbbar pair.

To select the correct assignment of reconstructed jets to parent particles, a modified \chisq metric, \chimod, is used. The \chimod value is determined by the differences between the masses of the two reconstructed bosons and the mass of the Higgs or \PZ boson, normalized by their resolutions, and by the reconstructed fractional mass difference of the two VLQs. The event mode is assigned as \bhbh, \bhbz, or \bzbz, depending on which gives the smallest value of \chimod. An upper cutoff on the value of \chimod is applied to remove background.

The expected background is first determined by fitting the distribution of the number of events as a function of the reconstructed VLQ mass, before jet tagging requirements are applied, so this sample is overwhelmingly background dominated. The fraction of background expected to remain after jet tagging requirements are applied, called the background jet-tagged fraction, is measured using events with VLQ candidate masses in the range 500--800\GeV, in which a VLQ signal has already been excluded, and then corrected for a possible dependence on the VLQ mass by using a control region with a higher \chimod value. Both the \chimod selection and jet tagging requirements are simultaneously optimized for maximal sensitivity to a potential signal. This optimization is done separately for each event mode and jet multiplicity. For the final result, all event mode and jet multiplicity analyses are combined using the procedure in Ref.~\cite{CMS-NOTE-2011-005} to obtain VLQ mass limits as a function of \BrBbH and \BrBbZ, as described further in Section~\ref{sec:results}.

\section{The CMS detector}

The central feature of the CMS apparatus is a superconducting solenoid of 6\unit{m} internal diameter, providing a magnetic field of 3.8\unit{T}. Within the solenoid volume are a silicon pixel and strip tracker, a lead tungstate crystal electromagnetic calorimeter (ECAL), and a brass and scintillator hadron calorimeter (HCAL), each composed of a barrel and two endcap sections. Forward calorimeters extend the pseudorapidity ($\eta$) coverage provided by the barrel and endcap detectors. Muons are detected in gas-ionization chambers embedded in the steel flux-return yoke outside the solenoid.

The electromagnetic calorimeter consists of 75\,848 lead tungstate crystals, which provide coverage in $\abs{\eta} < 1.48$ in a barrel region and $1.48 < \abs{\eta} < 3.0$ in two endcap regions. In the region $\abs{\eta} < 1.74$, the HCAL cells have widths of 0.087 in pseudorapidity and 0.087 in azimuth ($\phi$). In the $\eta$-$\phi$ plane, and for $\abs{\eta} < 1.48$, the HCAL cells map on to $5{\times}5$ arrays of ECAL crystals to form calorimeter towers projecting radially outwards from close to the nominal interaction point. For $\abs{\eta} > 1.74$, the coverage of the towers increases progressively to a maximum of 0.174 in $\Delta \eta$ and $\Delta \phi$. Within each tower, the energy deposits in ECAL and HCAL cells are summed to define the calorimeter tower energies, which are subsequently used to provide the energies and directions of hadronic jets. When combining information from the entire detector, the jet energy resolution amounts typically to 15--20\% at 30\GeV, 10\% at 100\GeV, and 5\% at 1\TeV~\cite{Khachatryan:2016kdb}.

Events of interest are selected using a two-tiered trigger system~\cite{Khachatryan:2016bia}. The first level, composed of custom hardware processors, uses information from the calorimeters and muon detectors to select events at a rate of around 100\unit{kHz} within a fixed time interval of about 4\mus. The second level, known as the high-level trigger (HLT), consists of a farm of processors running a version of the full event reconstruction software optimized for fast processing, and reduces the event rate to around 1\unit{kHz} before data storage.

A more detailed description of the CMS detector, together with a definition of the coordinate system used and the relevant kinematic variables, can be found in Ref.~\cite{Chatrchyan:2008zzk}. 

\section{Data and simulated events}

The data used in this analysis were collected during the 2016--2018 LHC running periods and correspond to an integrated luminosity of 137\fbinv~\cite{CMS-PAS-LUM-17-001,CMS-PAS-LUM-17-004,CMS-PAS-LUM-18-002}.

Signal events with pair production of VLQs were simulated using the Monte Carlo generator \MGvATNLO~\cite{Alwall:2014hca}, version v2.3.3 (v2.4.2) for samples corresponding to 2016 (2017--2018) data, at leading order with the NNPDF3.0 parton distribution functions (PDFs)~\cite{Ball:2014uwa}. The generated VLQ masses $m_{\PB}$ cover the range 1000--1800\GeV in steps of 100\GeV. Hadronization of the underlying partons was simulated using \PYTHIA v8.212~\cite{Sjostrand:2014zea} with the CUETP8M1 tune~\cite{Khachatryan:2015pea} for samples corresponding to 2016 data, and with the CP5 tune~\cite{Sirunyan:2019dfx} for samples corresponding to 2017 and 2018 data. Corrections of the cross sections to next-to-next-to-leading order and next-to-next-to-leading logarithmic soft-gluon resummation were obtained using \textsc{Top++} 2.0~\cite{Czakon:2013goa} with the MSTW2008NNLO68CL parton distribution set from the \textsc{lhapdf} 5.9.0 library~\cite{Whalley:2005nh,Sirunyan:2018omb}. To simulate the effect of additional $\Pp\Pp$ interactions within the same or nearby bunch crossings (``pileup''), \PYTHIA v8.226 with a total inelastic $\Pp\Pp$ cross section of 69.2\unit{mb}~\cite{Aaij:2018okq} was used. Following event generation, the \GEANTfour package~\cite{Agostinelli:2002hh,Allison:2006ve} was used to simulate the CMS detector response. Scale factors corresponding to jet energy corrections, jet energy resolutions~\cite{Khachatryan:2016kdb}, pileup, and jet tagging~\cite{CMS-DP-2018-058,Sirunyan:2017ezt} are applied to the simulated signal events so that the corresponding distributions agree with those in data.

\section{Jet reconstruction and tagging}
\label{sec:jets}

The global event reconstruction, also called the particle-flow event reconstruction~\cite{Sirunyan:2017ulk}, aims to reconstruct and identify each individual particle in an event, with an optimized combination of all subdetector information. In this process, the identification of the particle type (photon, electron, muon, charged hadron, or neutral hadron) plays an important role in the determination of the particle direction and energy. First, photons, electrons, and muons are identified using ECAL energy clusters, tracks in the tracker, and hits in the muon system. Then, charged hadrons are identified as charged particle tracks neither identified as electrons, nor as muons. Finally, neutral hadrons are identified as HCAL energy clusters not linked to any charged hadron trajectory, or as a combined ECAL and HCAL energy excess with respect to the expected charged hadron energy deposit. The energy of charged hadrons is determined from a combination of the track momentum and the corresponding ECAL and HCAL energies, corrected for the response function of the calorimeters to hadronic showers, and the energy of neutral hadrons is obtained from the corresponding corrected ECAL and HCAL energies.

For this analysis, two types of hadronic jets are clustered from these reconstructed particles, using the infrared and collinear safe anti-\kt algorithm~\cite{Cacciari:2008gp, Cacciari:2011ma}. The first type, ``AK4 jets'', uses a distance parameter $\Delta R = \sqrt{\smash[b]{(\Delta \eta)^2 + (\Delta \phi)^2}}$ of 0.4. However, since merged jets from a boosted Higgs or \PZ boson decay may be wider, a second set, using a distance parameter of 0.8 (``AK8 jets'') is also reconstructed from the same set of input particles. Jet momentum is determined as the vectorial sum of all particle momenta in the jet, and is found from simulation to be, on average, within 5 to 10\% of the true momentum over the whole transverse momentum (\pt) spectrum and detector acceptance. Pileup can contribute additional tracks and calorimetric energy depositions to the jet momentum. The pileup per particle identification algorithm (PUPPI)~\cite{Bertolini:2014bba} is used to mitigate the effect of pileup at the reconstructed particle level, making use of local shape information, event pileup properties, and tracking information. A local shape variable is defined, which distinguishes between collinear and soft diffuse distributions of other particles surrounding the particle under consideration. The former is attributed to particles originating from the hard scatter and the latter to particles originating from pileup interactions. Charged particles identified as originating from pileup vertices are discarded. For each neutral particle, a local shape variable is computed using the surrounding charged particles compatible with the primary vertex within the tracker acceptance ($\abs{\eta} < 2.5$), and using both charged and neutral particles in the region outside of the tracker coverage. The momenta of the neutral particles are then rescaled according to their probability to originate from the primary interaction vertex deduced from the local shape variable, superseding the need for jet-based pileup corrections~\cite{CMS-PAS-JME-16-003}. Jet energy corrections are derived from simulation to bring the measured response of jets to that of particle level jets on average. In situ measurements of the momentum balance in dijet, $\text{photon} + \text{jet}$, $\PZ + \text{jet}$, and multijet events are used to account for any residual differences in the jet energy scale between data and simulation~\cite{Khachatryan:2016kdb}. Additional selection criteria are applied to each jet to remove jets potentially dominated by anomalous contributions from various subdetector components or reconstruction failures. 

This analysis only uses AK4 jets with $\pt > 50\GeV$ and AK8 jets with $\pt > 200\GeV$, both within $\abs{\eta} < 2.4$. For AK8 jets, the constituents are reclustered using the Cambridge--Aachen algorithm~\cite{Dokshitzer:1997in,Wobisch:1998wt}. The ``modified mass drop tagger'' algorithm~\cite{Dasgupta:2013ihk,Butterworth:2008iy}, also known as the ``soft drop'' algorithm, with angular exponent $\beta = 0$, soft cutoff threshold $z_{\text{cut}} < 0.1$, and characteristic radius $R_{0} = 0.8$~\cite{Larkoski:2014wba}, is applied to remove soft, wide-angle radiation from the jet. This results in a jet mass that, in the case of large mass, more accurately corresponds to the mass of the mother particle from which the jet originated. For events with boosted Higgs and \PZ bosons, the AK8 soft-drop mass is used to obtain the mass of the merged jet.

The event jet multiplicity is determined by the number of AK4 jets passing the requirements above. In signal events, decays of the VLQ pair and subsequent decays of the Higgs and \PZ boson daughters yield a total of six quarks. In all three event modes, at least two of these are \PQb quarks. Considering the predominant $\PH \to \bbbar$ decay, for the \bhbh event mode, all six are \PQb quarks; for the \bhbz event mode, four or six are \PQb quarks; while for the \bzbz event mode, two, four or six are \PQb quarks. We note that of the hadronic \PZ boson decays, 15\% are to a $\bbbar$ pair; however, $\PZ \to \ccbar$ decays also have a significant chance of passing the tagging requirements, as discussed in Section~\ref{sec:eventsel}, and are also included (along with other possible \PZ boson decay modes) in the signal efficiency. If all six quark jets are individually reconstructed, a 6-jet event is produced; if two jets merge into a single reconstructed jet, this produces a 5-jet event; and if two merged jets are produced, then a 4-jet event results. Note that the VLQ reconstruction does not consider the possibility of additional jets produced by initial state or final state radiation.

Because of the large number of \PQb jets in signal events, \PQb tagging is a powerful tool to significantly reduce the background from SM processes. Individual jets are tagged using the DeepJet \PQb discriminant~\cite{CMS-DP-2018-058} applied to AK4 jets, while merged jets from $\bbbar$ pairs are double \PQb tagged using the algorithm in Ref.~\cite{Sirunyan:2017ezt}, developed in the context of $\PH \to \bbbar$ searches, applied to AK8 jets.

\section{Event selection}
\label{sec:eventsel}

The events used in the analysis are first selected online by the CMS trigger system. The HLT trigger used requires the total \pt measured in the calorimeters to be at least 900 (1050)\GeV for the 2016 (2017--2018) data set. Offline, events with $\HT > 1350\GeV$ are selected, where \HT is defined as the scalar sum of the jet \pt for all AK4 jets with $\pt > 50\GeV$ and $\abs{\eta} < 2.4$. The requirement is set higher than the trigger threshold to avoid effects due to trigger turn on. In order to minimize bias when measuring the efficiency of the HLT triggers, the efficiencies are measured in a data set collected by an orthogonal trigger, which requires the event to have a single muon. For all years, the measured trigger efficiency for events with $\HT > 1350\GeV$ is at least 99.6\%. Table~\ref{table:HT_eff} shows the efficiency for simulated VLQ signal events after the \HT requirement for each of the three jet multiplicity channels and for each of three VLQ masses (1000, 1200, and 1400\GeV).

\begin{table}[hbtp]
\centering
\topcaption{Signal efficiencies of the offline \HT selection, in \%, for each of the jet multiplicity channels, for three VLQ masses (1000, 1200, and 1400\GeV). The efficiency is the fraction of events in each jet multiplicity category satisfying the $\HT > 1350\GeV$ selection. Statistical uncertainties are negligible and therefore omitted.}
\label{table:HT_eff} 
\begin{scotch}{lccc}
VLQ mass [{\GeVns}] & 4 jets & 5 jets & 6 jets\\
\hline
1000 & 95.1 & 89.4 & 81.4 \\
1200 & 98.5 & 96.2 & 91.3 \\
1400 & 99.5 & 98.4 & 95.0 \\
\end{scotch}
\end{table}

The number of tagged jets required to select an event, as well as the working points for the taggers used, are optimized separately for each of the three jet multiplicities and event modes, in order to maximize the expected signal sensitivity. For the working points selected, the single \PQb tagger has an efficiency of 82\% for \PQb jets in simulated \ttbar events with $\pt > 30\GeV$ and a mistag rate of 1\% for light quarks (\PQu, \PQd, or \PQs) and approximately 17\% for charm quarks~\cite{CMS-DP-2018-058}. The double \PQb tagger has an efficiency of 75\% in simulated $\PH \to \bbbar$ events, and a mistag rate of 10\% in simulated inclusive multijet events and 33\% in simulated $\PH \to \ccbar$ events~\cite{Sirunyan:2017ezt}, where a mistag in the double \PQb tag case means that at least one non-\PQb quark subjet is present in the tagged jet. The number of tags required depends on the jet multiplicity as follows: in the 6-jet case, four AK4 jets are required to have a \PQb tag, except in the \bzbz event mode, for which three tags are required. In the 5-jet case, three of the AK4 jets not associated with the merged decay products are required to be \PQb tagged; no double \PQb tag requirement is applied to the AK8 jet associated with the merged decay. In the 4-jet case, two of the nonmerged AK4 jets are required to have a \PQb tag, and one of the merged jets is required to have a double \PQb tag, except in the \bzbz event mode, for which no double \PQb tag is required. These requirements are summarized in Table~\ref{table:tagreqs}.

\begin{table}[hbtp]
\centering
\topcaption{Summary of the minimum number of single and double \PQb tags required for each jet multiplicity and event mode.}
\label{table:tagreqs}
\begin{scotch}{clccc}
Jet multiplicity & Tag & \bhbh & \bhbz & \bzbz \\
\hline
\multirow{2}{*}{4 jets} & Single \PQb & 2 & 2 & 2 \\
 & Double \PQb & 1 & 1 & 0 \\[\cmsTabSkip]

\multirow{2}{*}{5 jets} & Single \PQb & 3 & 3 & 3 \\
 & Double \PQb & 0 & 0 & 0 \\[\cmsTabSkip]

6 jets & Single \PQb & 4 & 4 & 3 \\
\end{scotch}
\end{table}

\section{Event reconstruction}
\label{sec:eventreco}

In the case when the two jets produced from a \PH/\PZ boson decay are individually resolved, the mass of the parent boson can be estimated from the invariant mass of the two jets. In the case where the two jets are merged, the parent boson mass is instead estimated using the soft-drop mass of the AK8 jet. Only those AK8 jets that are within $\Delta R < 0.3$ of an AK4 jet are used. However, if a second AK4 jet is within $\Delta R < 0.6$ of the AK8 jet, this overlap could cause the AK8 jet mass to be misreconstructed, so in this case the AK8 jet is discarded and the two AK4 jets are treated as a resolved dijet boson candidate. 

A central feature in the analysis is the selection of the correct way of combining jets in order to reconstruct the parent particles; this is a difficult task because of the large number of jets in the event. In 6-jet events, there are two pairs of jets originating from \PH/\PZ boson decay; and three jets (including two from the \PH/\PZ decay) associated with each VLQ decay. In 5-jet events, there is a pair of jets associated with one \PH/\PZ boson and a merged jet associated with the other \PH/\PZ boson; each of these is associated with one of the remaining two jets to form a VLQ candidate.  In 4-jet events, there is a merged jet associated with each \PH/\PZ boson, each of which is paired to one of the remaining two jets to form a VLQ candidate. The final reconstructed VLQ mass, \mvlq, is defined as the average mass of the two individual reconstructed VLQs in the event.

The number of possible ways to combine the jets to reconstruct the two \PH/\PZ bosons and then to combine these with the two remaining jets to form the VLQ candidates is 720, 120, and 24 for the 6-, 5-, and 4-jet multiplicities, respectively. However, many of these different combinations simply involve different permutations among the jets that constitute the individual VLQs, and these permutations do not affect the reconstructed VLQ mass. The numbers of combinations that give distinct VLQ masses are only 10, 10, and 3 for the 6-, 5-, and 4-jet multiplicities, respectively. For the 5- and 4-jet multiplicities, however, the jet associated with the merged \PH/\PZ boson decay products distinguishes different combinations, since this jet is treated differently from the others by having a double \PQb tag, rather than a single \PQb tagging requirement. This doubles the number of distinct 5-jet combinations and quadruples the number of distinct 4-jet combinations. The final number of distinct combinations is then 10, 20, and 12 for the 6-, 5-, and 4-jet multiplicities, respectively.

\begin{figure*}[hbtp]
\centering
\includegraphics[width=0.32\textwidth]{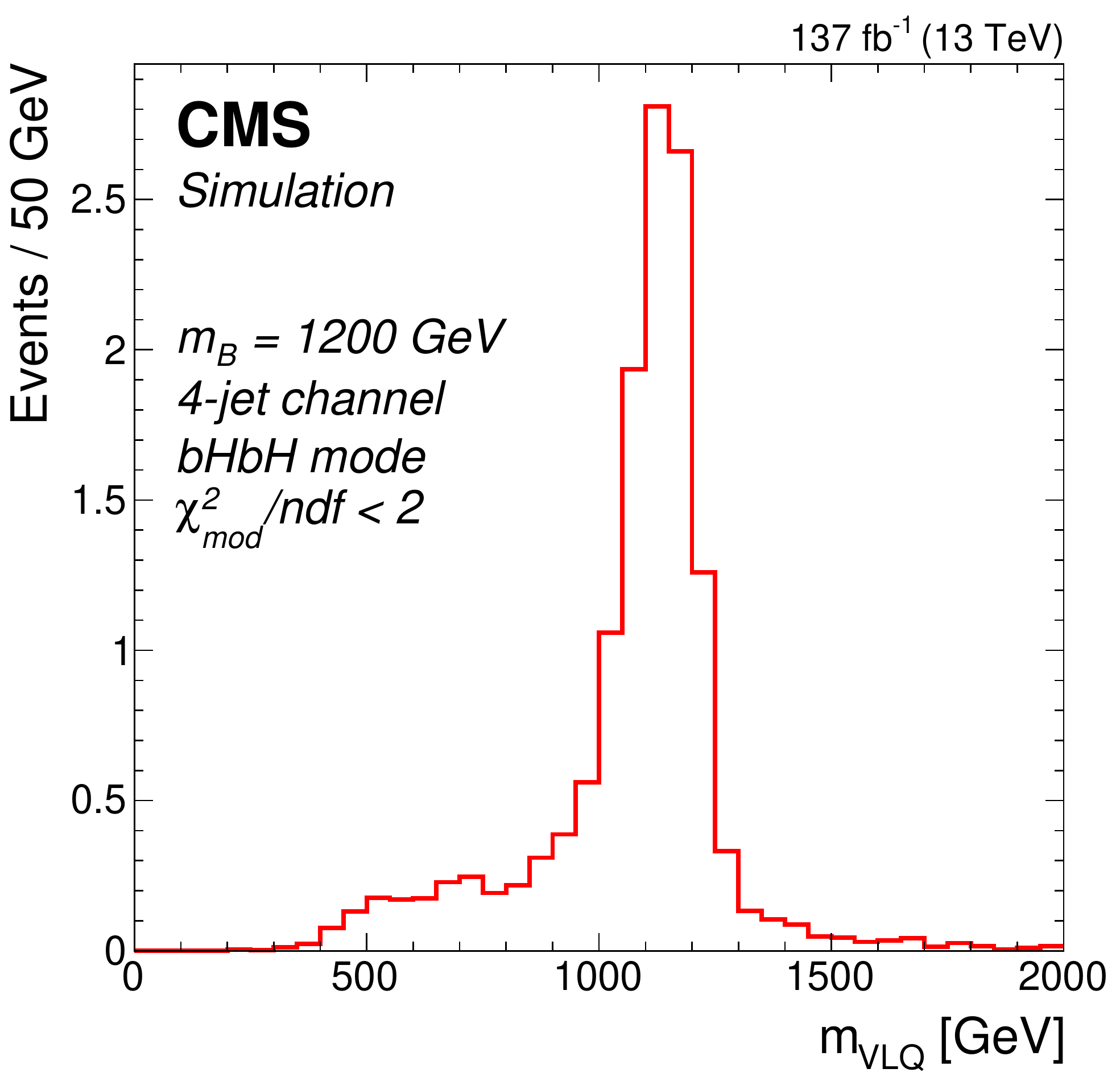}
\includegraphics[width=0.32\textwidth]{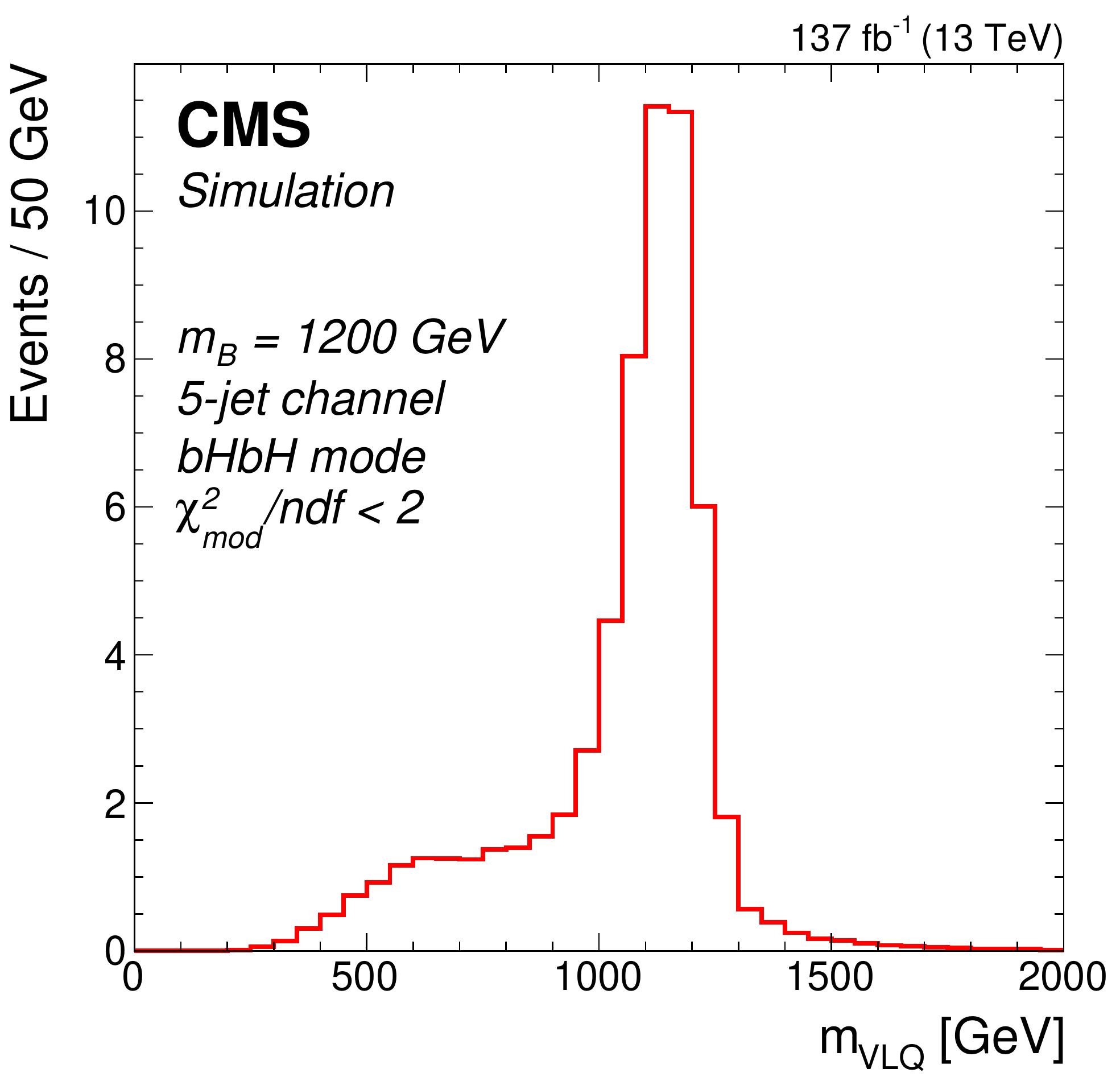}
\includegraphics[width=0.32\textwidth]{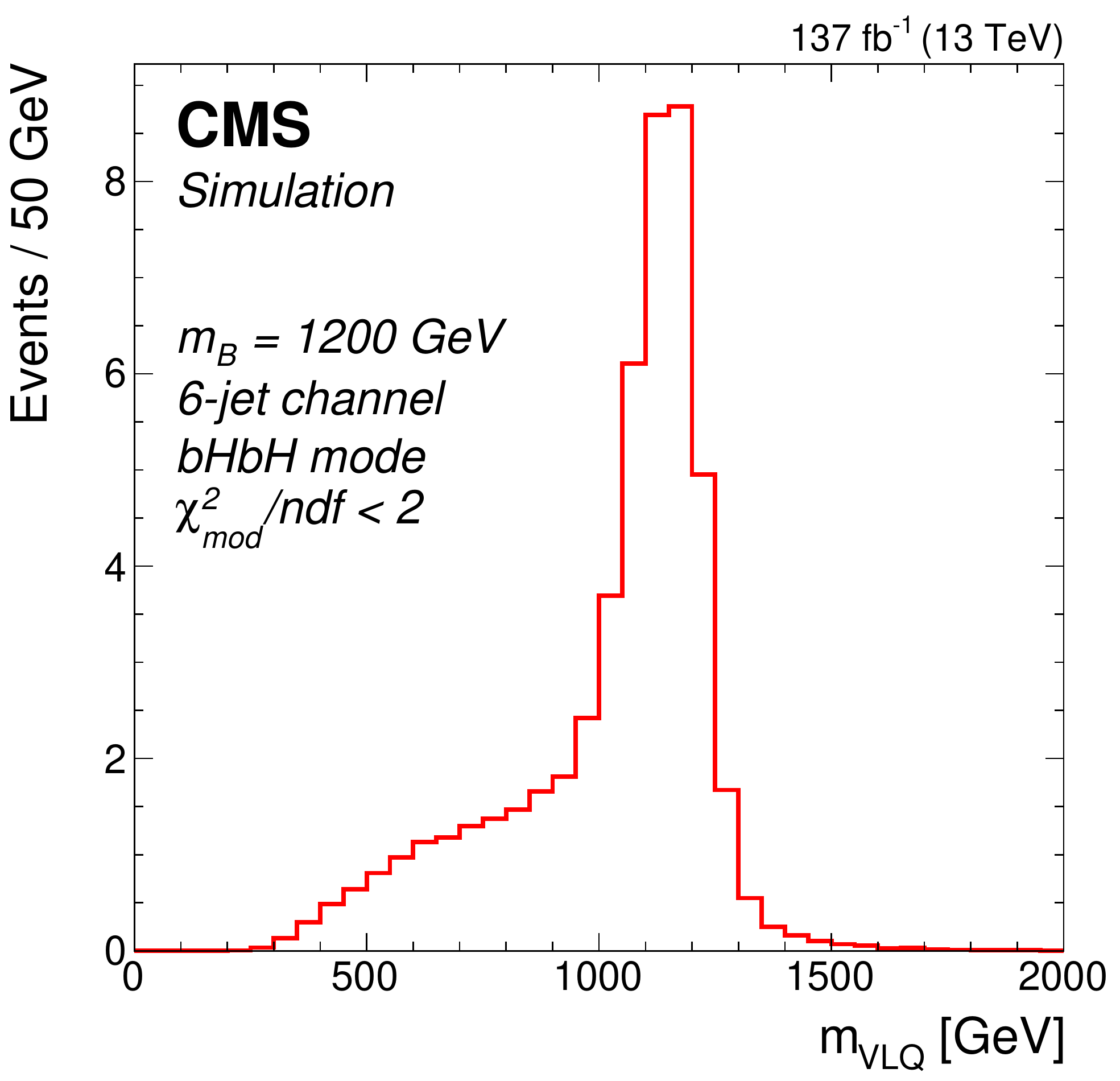}
\includegraphics[width=0.32\textwidth]{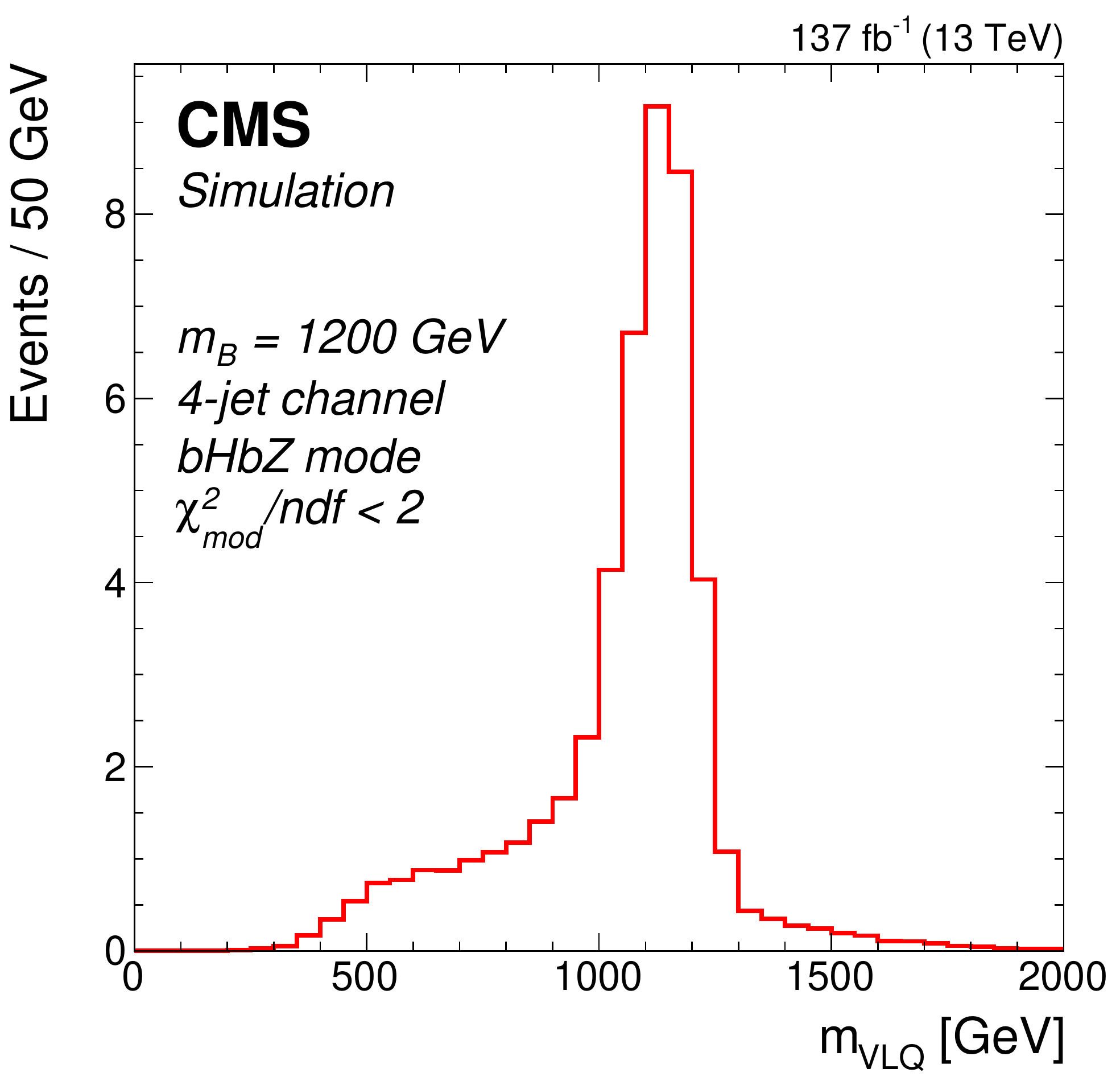}
\includegraphics[width=0.32\textwidth]{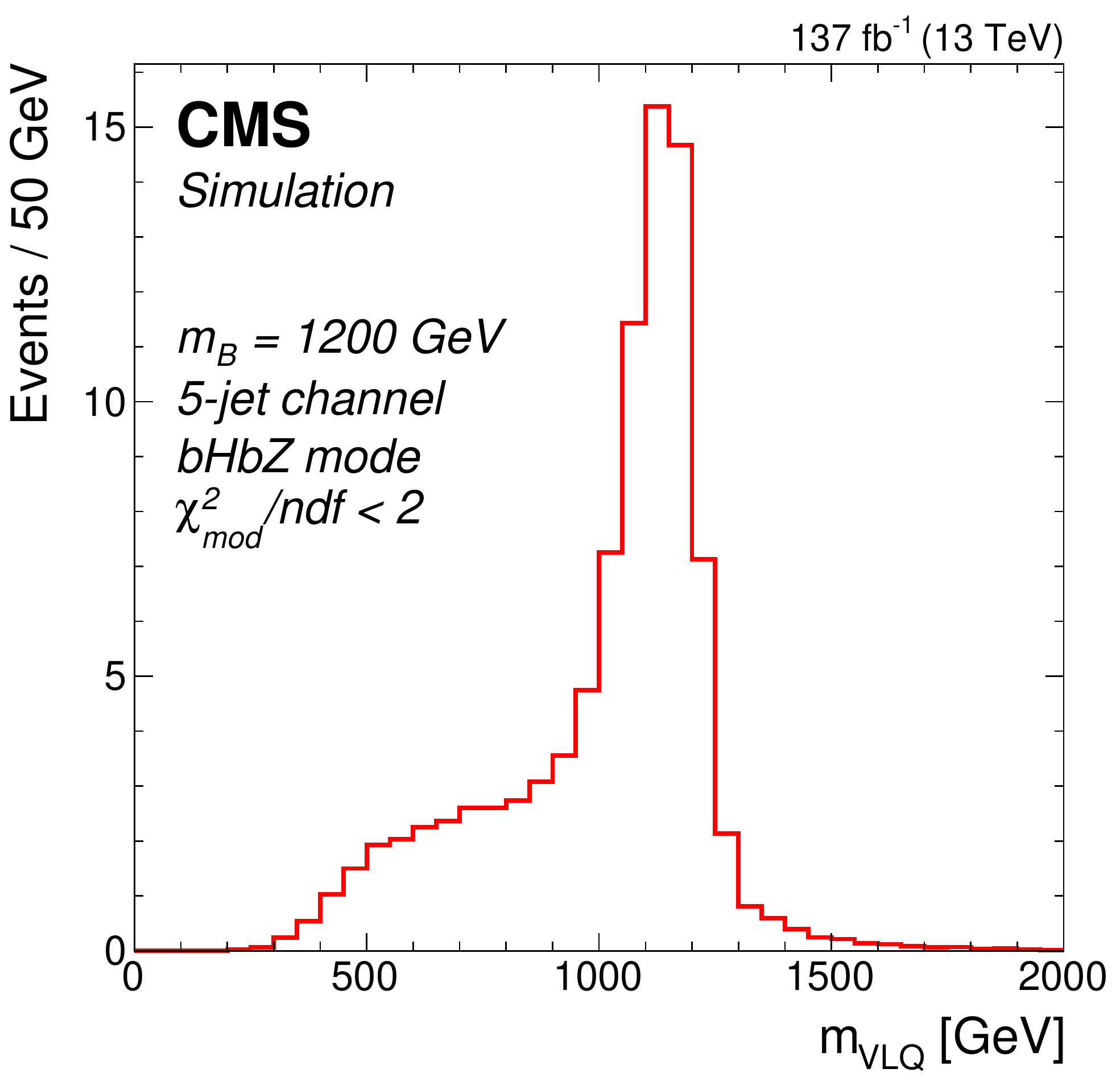}
\includegraphics[width=0.32\textwidth]{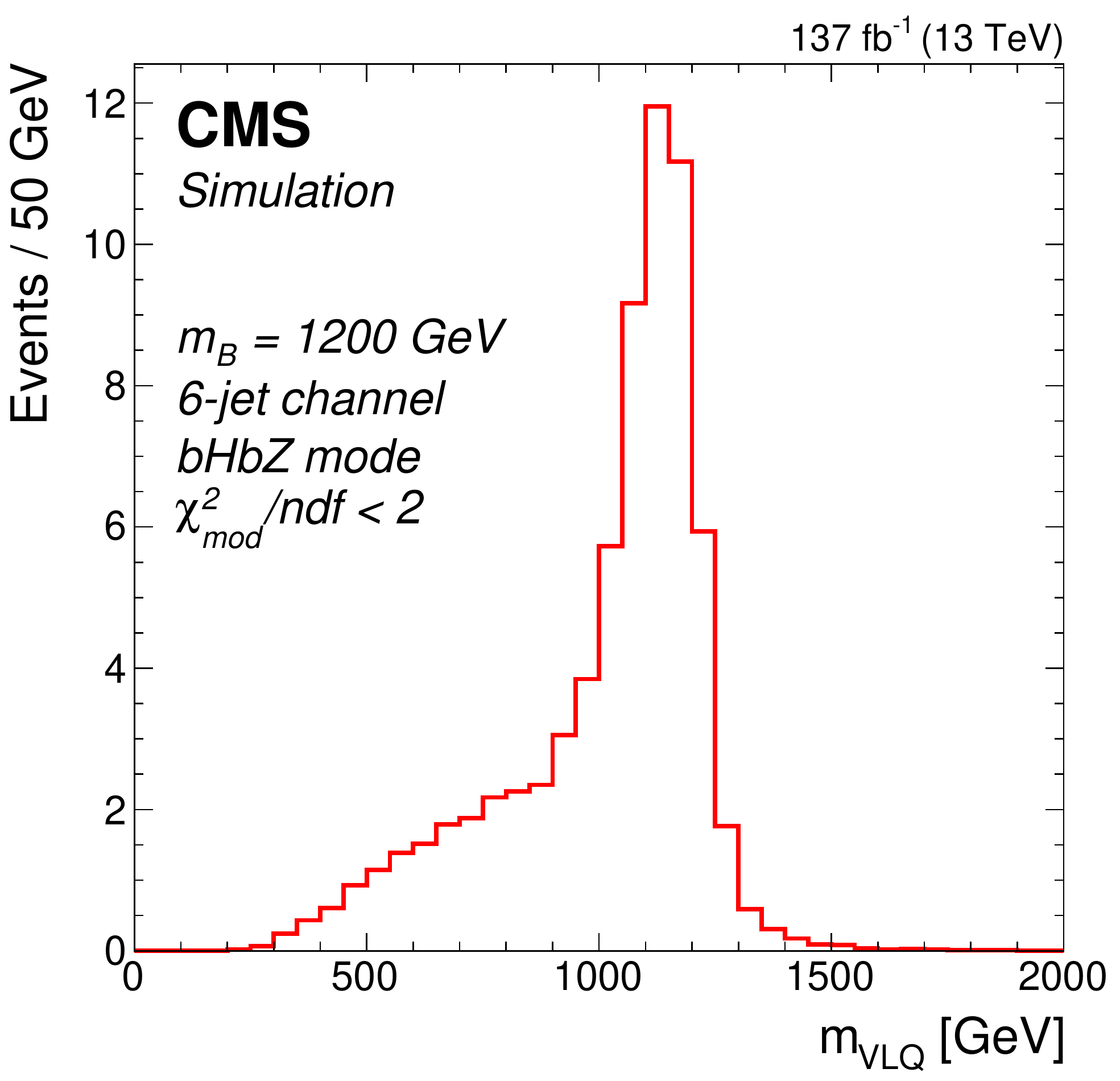}
\includegraphics[width=0.32\textwidth]{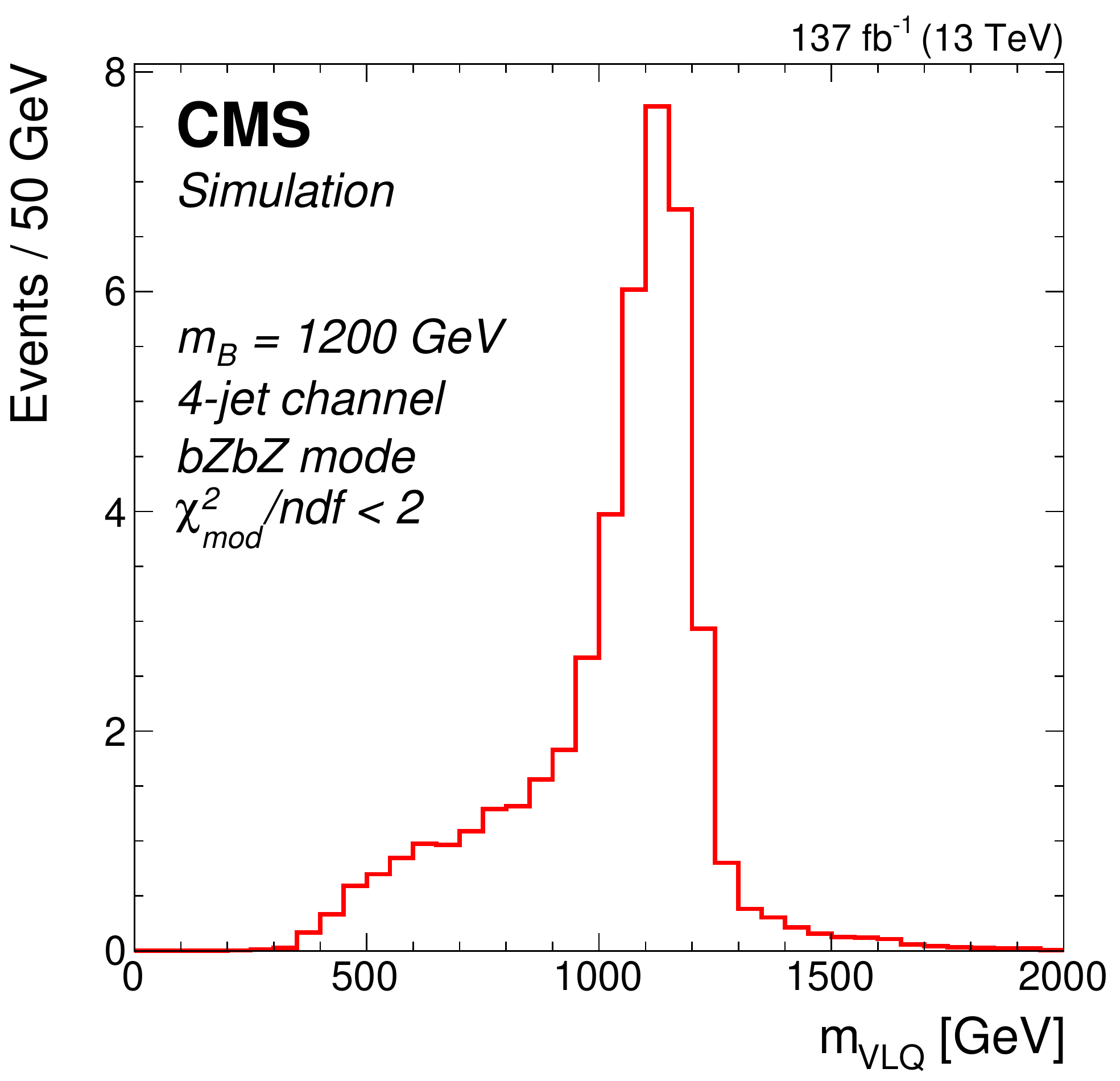}
\includegraphics[width=0.32\textwidth]{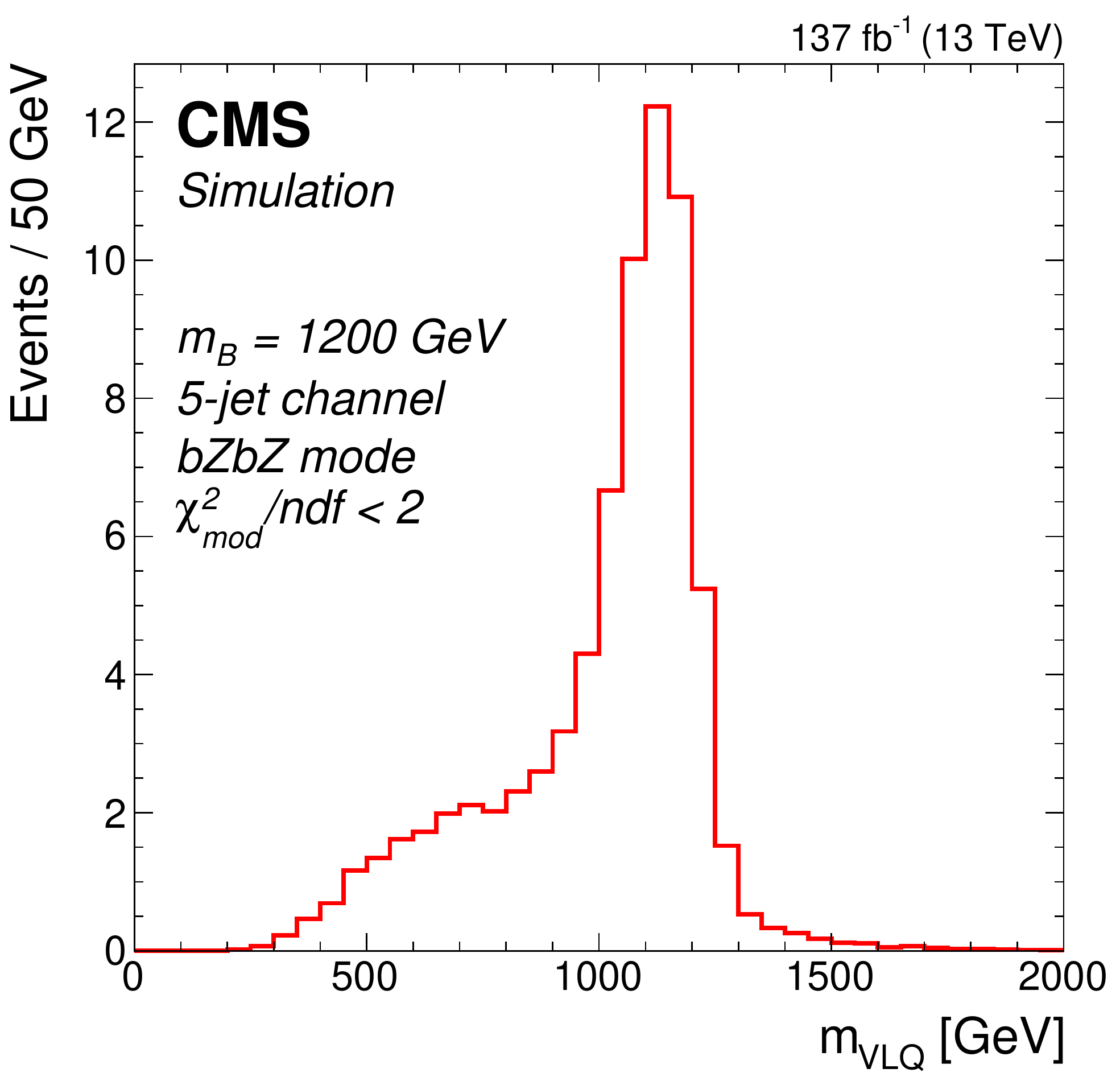}
\includegraphics[width=0.32\textwidth]{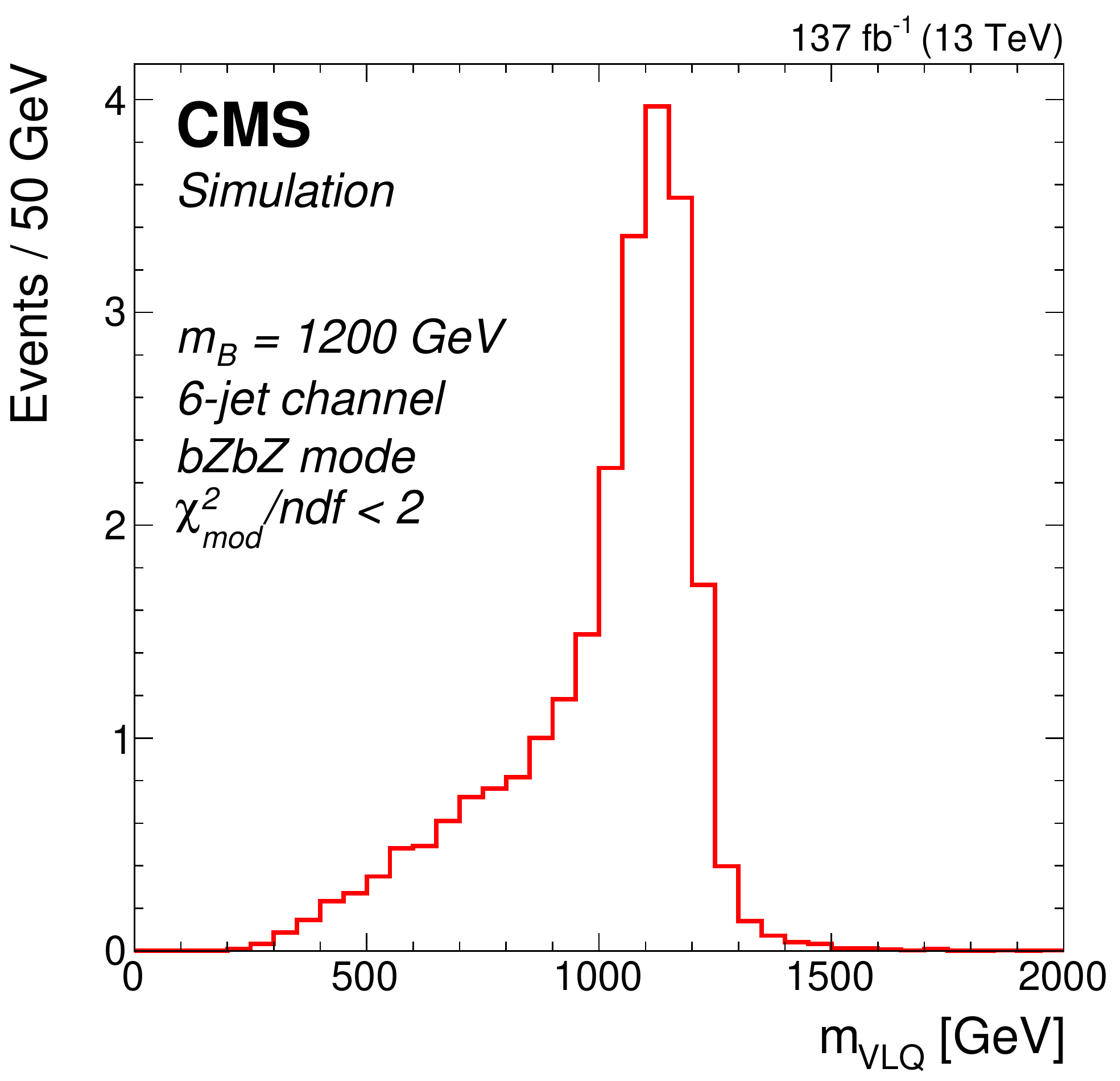}
\caption{Distributions of \mvlq for simulated signal events with a generated VLQ mass $m_{\PB} = 1200\GeV$. A requirement of $\chimodndf < 2$ is applied to the events. Mass distributions for 4-jet (left), 5-jet (center), and 6-jet (right) events are shown for the three event modes: \bhbh (upper row), \bhbz (middle row), and \bzbz (lower row).}
\label{fig:chi2_mass} 
\end{figure*}

For each jet combination, \chimod is determined using Eqs.~\ref{eq:chimodsix}--\ref{eq:chimodfour} below, which depend on the measured mass of a dijet Higgs or \PZ boson candidate ($m_{\text{dijet}}$), the measured soft-drop mass of a merged-jet Higgs or \PZ boson candidate  ($m_{\text{merged}}$), and the fractional mass difference of the two VLQ candidates ($\Delta m_{\text{VLQ}}$), where $\Delta m_{\text{VLQ}}$ is the difference of the masses of the two VLQ candidates divided by the average mass of the two.  The only use of AK8 jets in this calculation is to determine the mass of the merged \PH/\PZ candidates using the soft-drop mass of the matched AK8 jet. All other quantities are determined using AK4 jet kinematics.
\begin{widetext}\begin{linenomath}\begin{align}
  {\text{6-jet events: }} \chimod &= \frac{(m_{\text{dijet}_1} - \overline{m}_{\text{dijet}})^2}{\sigma^2_{m_{\text{dijet}}}} + \frac{(m_{\text{dijet}_2} - \overline{m}_{\text{dijet}})^2}{\sigma^2_{m_{\text{dijet}}}} + \frac{(\Delta m_{\text{VLQ}} - \overline{\Delta m}_{\text{VLQ}})^2}{\sigma^2_{\Delta m_{\text{VLQ}}}},\label{eq:chimodsix}\\
  {\text{5-jet events: }} \chimod &= \frac{(m_{\text{dijet}} - \overline{m}_{\text{dijet}})^2}{\sigma^2_{m_{\text{dijet}}}} + \frac{(m_{\text{merged}} - \overline{m}_{\text{merged}})^2}{\sigma^2_{m_{\text{merged}}}} + \frac{(\Delta m_{\text{VLQ}} - \overline{\Delta m}_{\text{VLQ}})^2}{\sigma^2_{\Delta m_{\text{VLQ}}}},\label{eq:chimodfive}\\
  {\text{4-jet events: }} \chimod &= \frac{(m_{\text{merged}_1} - \overline{m}_{\text{merged}})^2}{\sigma^2_{m_{\text{merged}}}} + \frac{(m_{\text{merged}_2} - \overline{m}_{\text{merged}})^2}{\sigma^2_{m_{\text{merged}}}} + \frac{(\Delta m_{\text{VLQ}} - \overline{\Delta m}_{\text{VLQ}})^2}{\sigma^2_{\Delta m_{\text{VLQ}}}}.\label{eq:chimodfour}
\end{align}\end{linenomath}\end{widetext}
The means ($\overline{m}$ and $\overline{\Delta m}_{\text{VLQ}}$) and standard deviations ($\sigma_m$ and $\sigma_{\Delta m_{\text{VLQ}}}$) of the parameters used in these expressions are determined from simulated signal events in which the jets are matched to the generator-level quarks and \PH/\PZ bosons. These quantities are derived separately for each jet multiplicity, but do not depend on the simulated signal mass. For each parameter, the central core of the distribution is fit with a Gaussian function, whose mean and standard deviation are then used as the parameters in the expressions for \chimod. As the distribution of the merged Higgs boson mass is asymmetrical, two Gaussian functions are separately fit above and below the peak of the distribution. Since the underlying distributions used in these expressions have non-Gaussian tails and are in some cases asymmetric, the values of \chimod are not exactly distributed as a \chisq variable. However, the difference is small, and \chimod is only used to select events, so these deviations do not affect the analysis. Choosing the jet combination that has the lowest value of \chimod gives a high probability of identifying the correct jet combination, and allows \mvlq to be reconstructed. In simulation, this can then be compared with the generated \PB mass $m_{\PB}$. This is indicated in Fig.~\ref{fig:chi2_mass}, which shows the average value of the reconstructed VLQ mass for the jet combination with the lowest \chimod for simulated signal events with $m_{\PB} = 1200\GeV$. In most cases, the VLQ mass is correctly reconstructed. We observe that the reconstructed mass distribution is consistently peaked at a value about 5\% lower than the generated mass, for all generated signal mass values. The low-side tail is due to the presence of incorrectly reconstructed events. 

\begin{figure*}[hbtp]
\centering 
\includegraphics[width=0.32\textwidth]{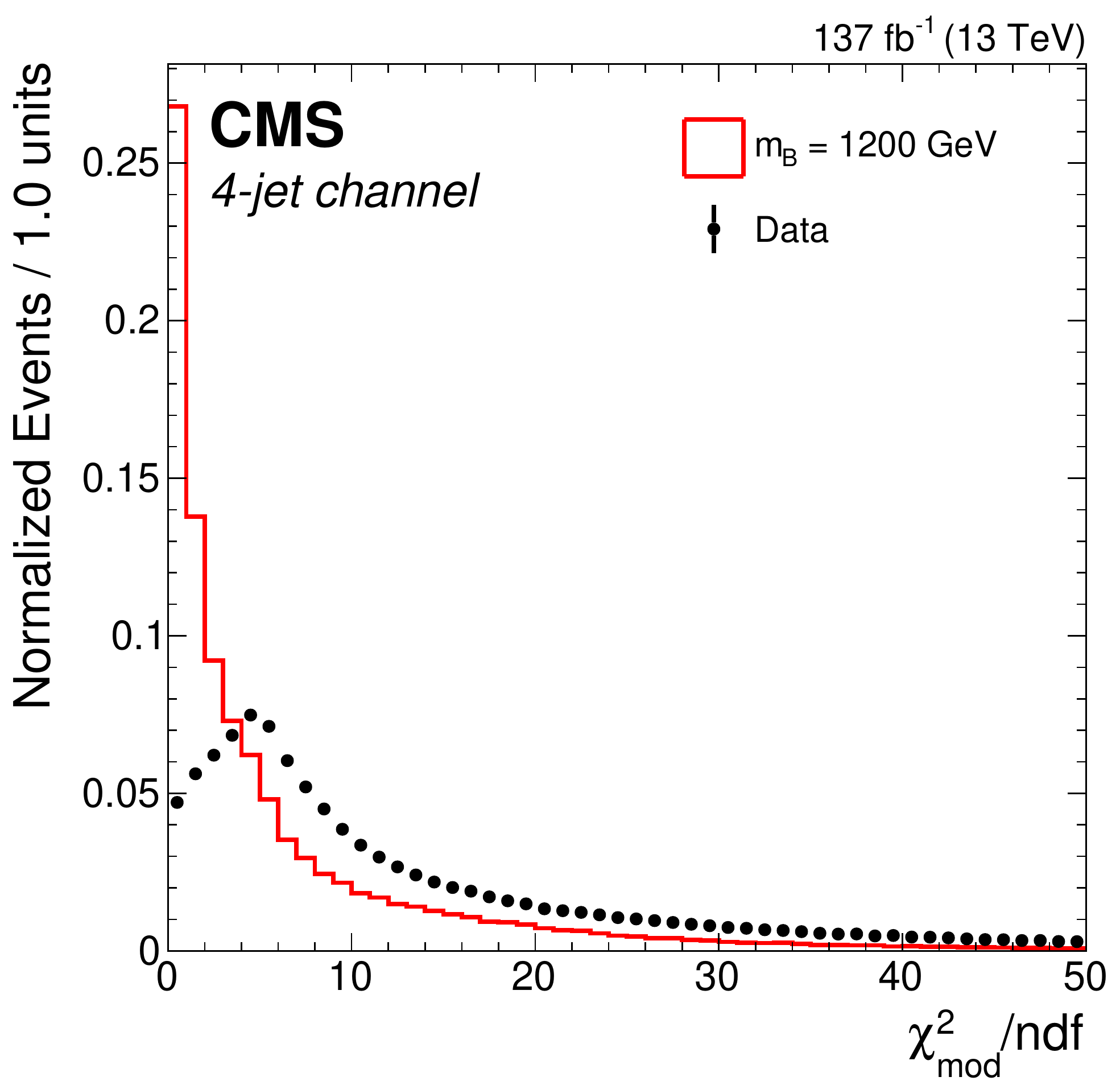}
\includegraphics[width=0.32\textwidth]{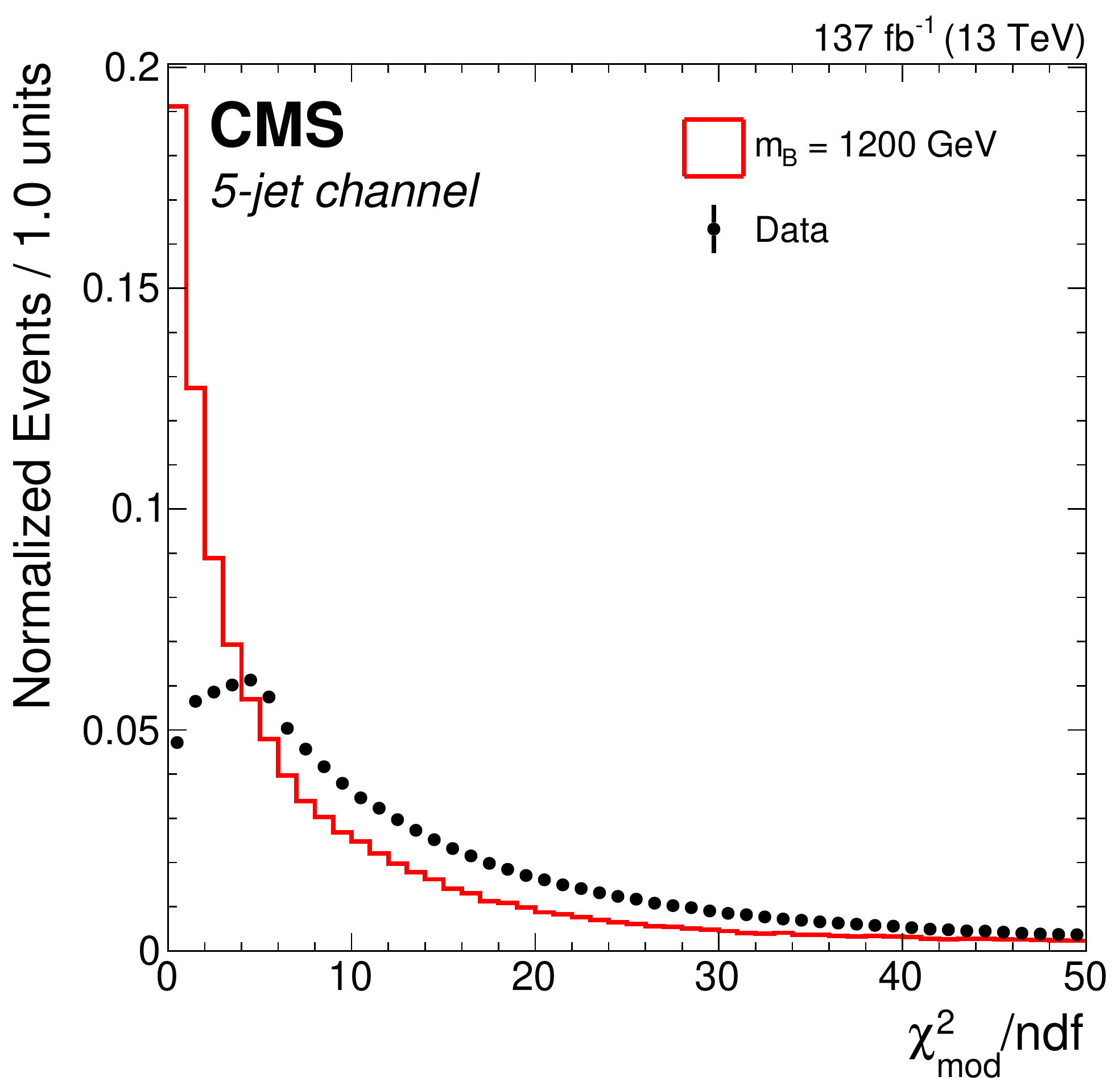}
\includegraphics[width=0.32\textwidth]{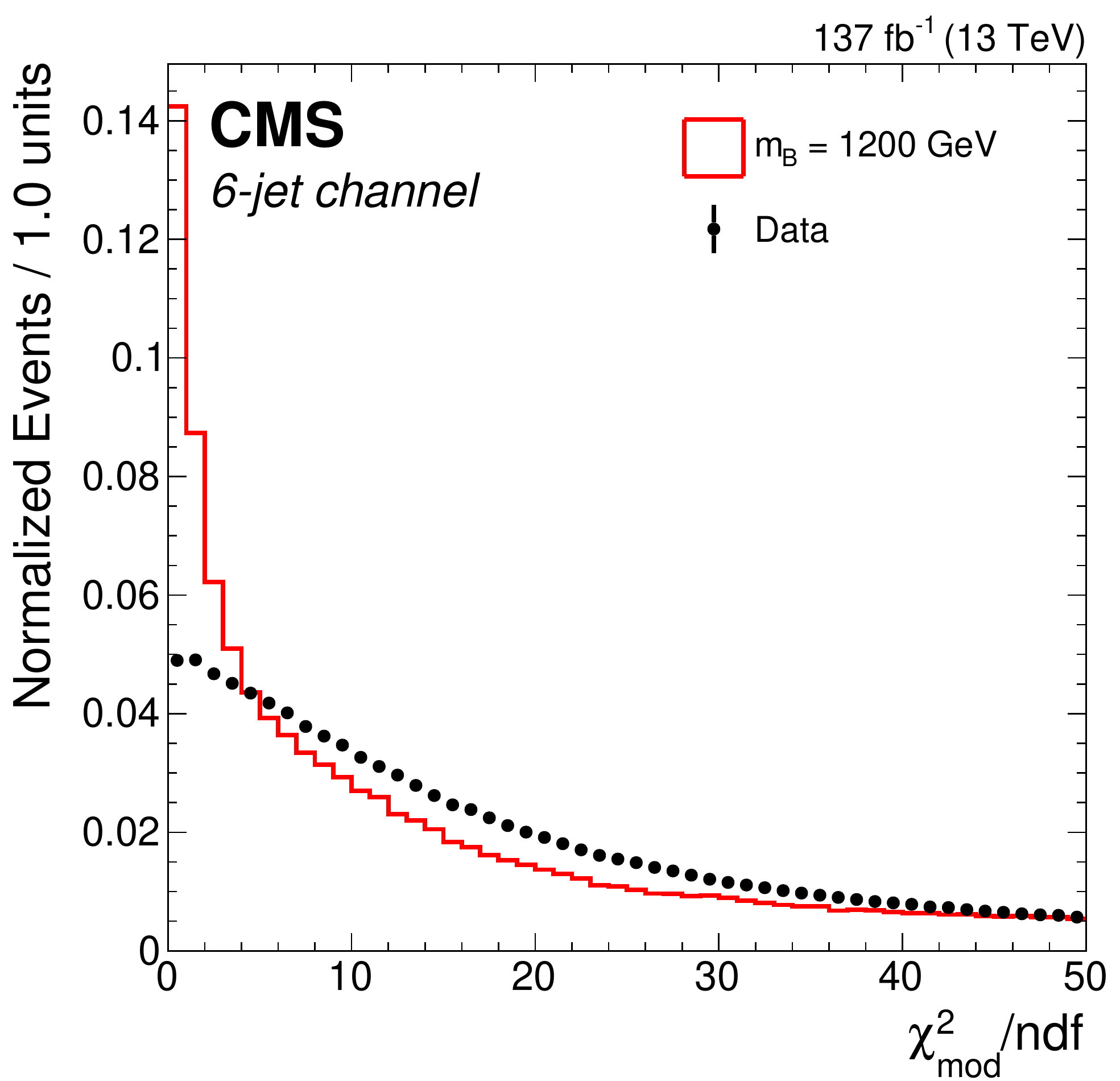}
\caption{Distribution of \chimodndf for the best jet combination for simulated 1200\GeV VLQ events (red histogram) and data (black points), for 4-jet (left), 5-jet (center), and 6-jet events (right). The simulated signal events and data events are normalized to the same integral value within the displayed \chimod range.}
\label{fig:chi2}
\end{figure*}

Figure~\ref{fig:chi2} shows the distributions of \chimodndf, where ndf is the number of degrees of freedom, for the best jet combination (\ie, the combination with the lowest \chimod), from simulated 1200\GeV VLQ signal events, at each jet multiplicity. Each \chimod expression has three degrees of freedom, one for each term. The distributions for the data are also shown for comparison. In these plots, the simulated signal and data distributions are normalized to the same integral value within the displayed \chimod range. This figure demonstrates that requiring a small \chimod value for the best jet combination provides an effective method for removing background.

The \chimod value is also used to select the event mode. There are three possible decays of the bottom-type VLQ: \BbH, \BbZ, and \BtW.  This results in six possible modes for the $\PB\PAB$ pair production events: \bhbh, \bhbz, \bzbz, $\PQt\PW\PQb\PH$, $\PQt\PW\PQb\PZ$, and $\PQt\PW\PQt\PW$. The latter three modes involve one or two decays of a VLQ to a \PQt quark and a \PW boson. These events either have a jet multiplicity greater than six, or contain leptons and missing transverse energy from the \PW decays. Although this analysis is not optimized for sensitivity to these events, events with \BtW, if present, have some probability to be selected as one of the other three event modes and can affect the sensitivity of the analysis. These events are included in the signal simulation and are added according to their reconstruction efficiency. For events that satisfy the \HT requirement and that are categorized as either 4-, 5-, or 6-jet multiplicity events, the \chimod described above is calculated for each of the three event modes: \bhbh, \bhbz, or \bzbz, and the mode of the event is selected as the one that has the best \chimod value. Events are categorized by their jet multiplicity and their reconstructed mode, regardless of the underlying decay mode for simulated signal events.

Before examining the data in the potential signal region, the event selection parameters (the jet tagging parameters and \chimod) are optimized. This optimization is performed by varying the parameters and selecting the values that maximize the sensitivity to a 1600\GeV VLQ signal. The mass of 1600\GeV is chosen because it is the point with maximum sensitivity of the analysis; however, the optimized parameters are largely independent of the point chosen. The optimized jet tagging parameters are described in Section~\ref{sec:eventsel}, and the optimized \chimodndf values are shown below in Table~\ref{table:cut_optim}. With the optimized selection, the overall signal efficiency measured in simulation for a generated VLQ mass of 1600\GeV is approximately 5\% in the $\BrBbZ = 100\%$ scenario, increasing to 10\% for $\BrBbH = 100\%$.

\begin{table}[hbtp!]
\centering
\topcaption{Optimized values of the \chimodndf selection as a function of jet multiplicity and event mode.}
\label{table:cut_optim} 	
\begin{scotch}{ccccccc}
      & \multicolumn{3}{c}{Jet multiplicity} \\
Event mode  & 4    & 5   & 6 \\
\hline
\bhbh       & 5.5  & 2   & 2.75 \\
\bhbz       & 2.75 & 2.5 & 2.5 \\
\bzbz       & 2    & 2   & 2.25 \\
\end{scotch}
\end{table}

\begin{figure*}[hbtp]
\centering
\includegraphics[width=0.32\textwidth]{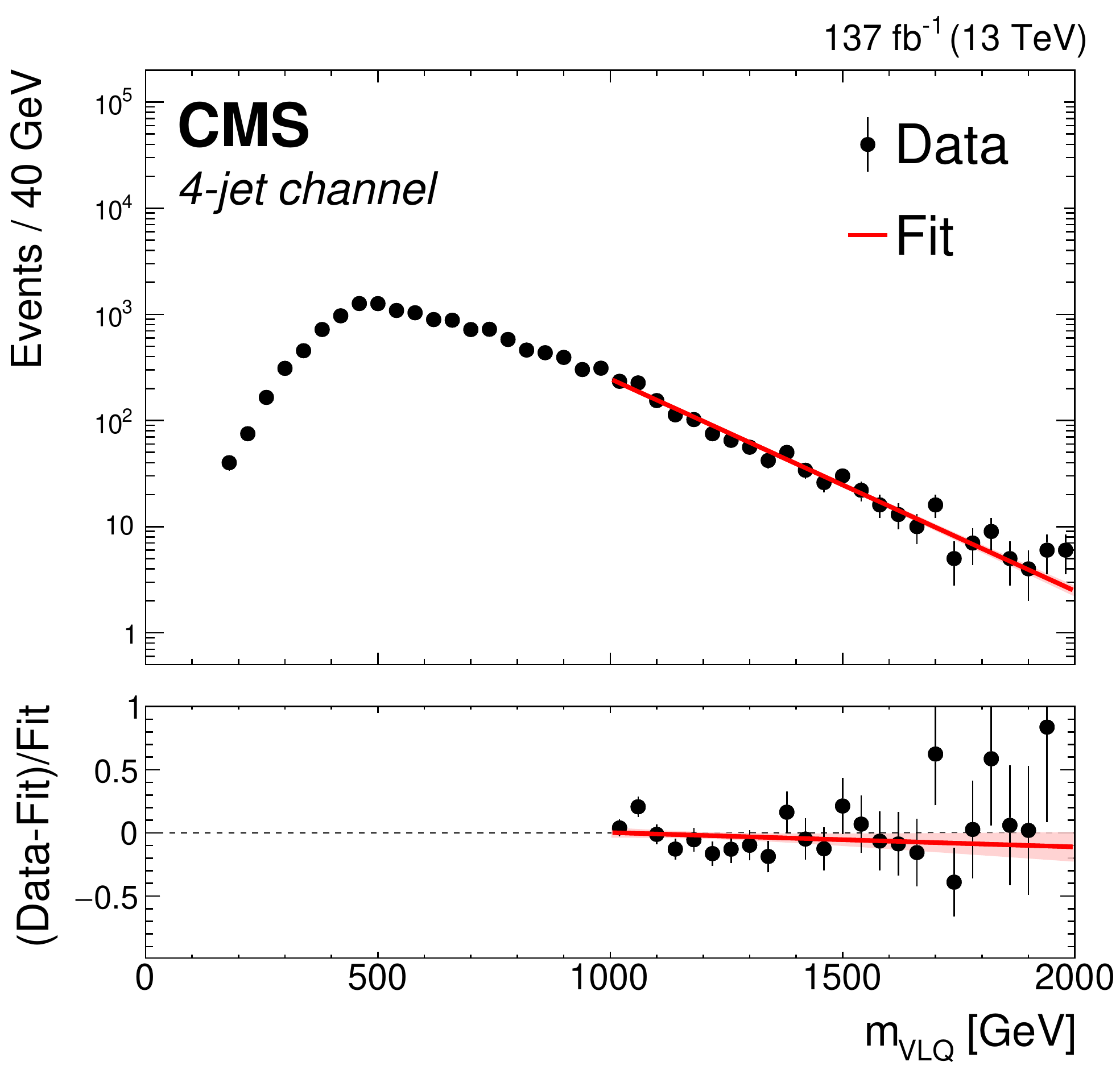}
\includegraphics[width=0.32\textwidth]{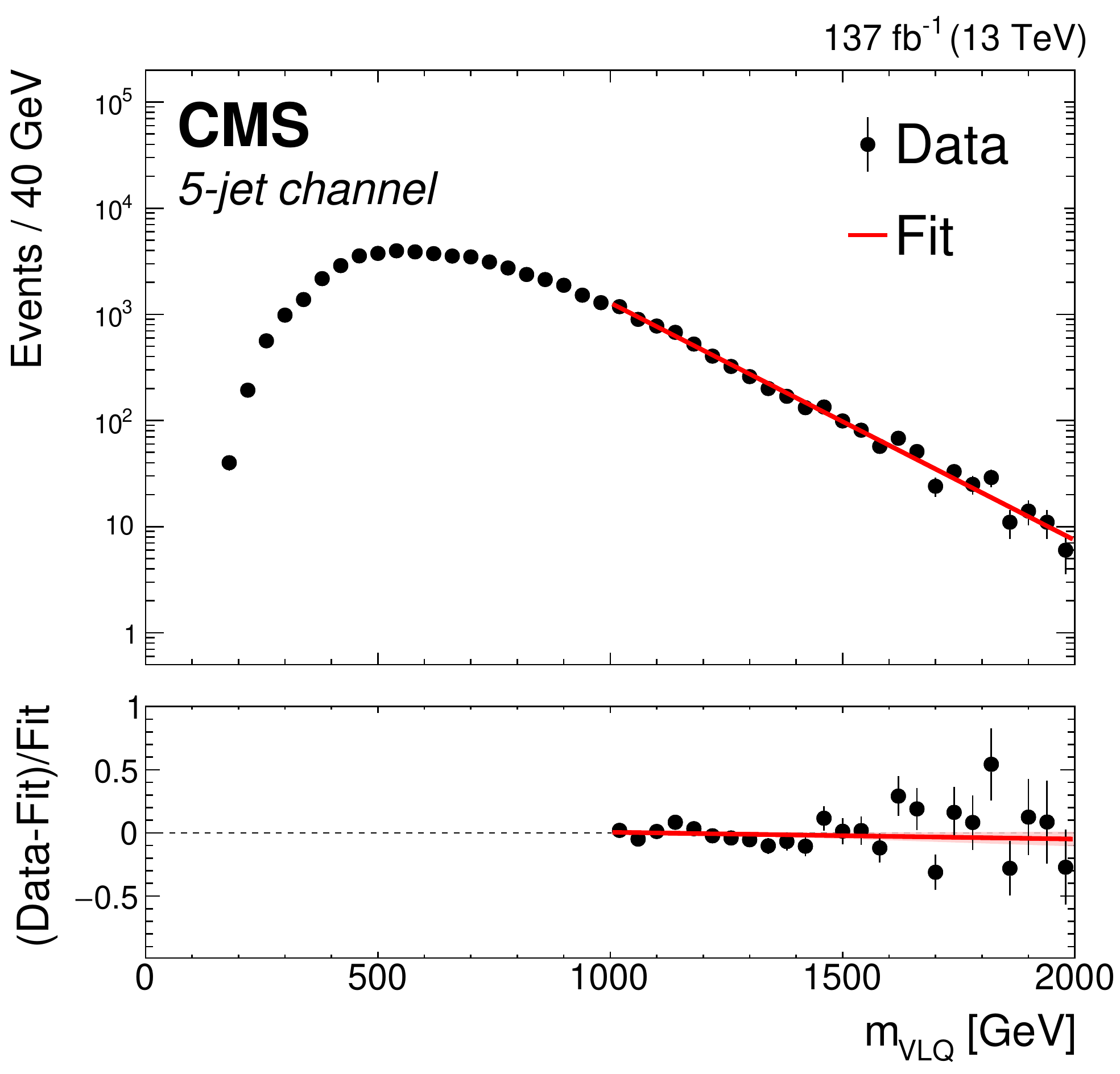}
\includegraphics[width=0.32\textwidth]{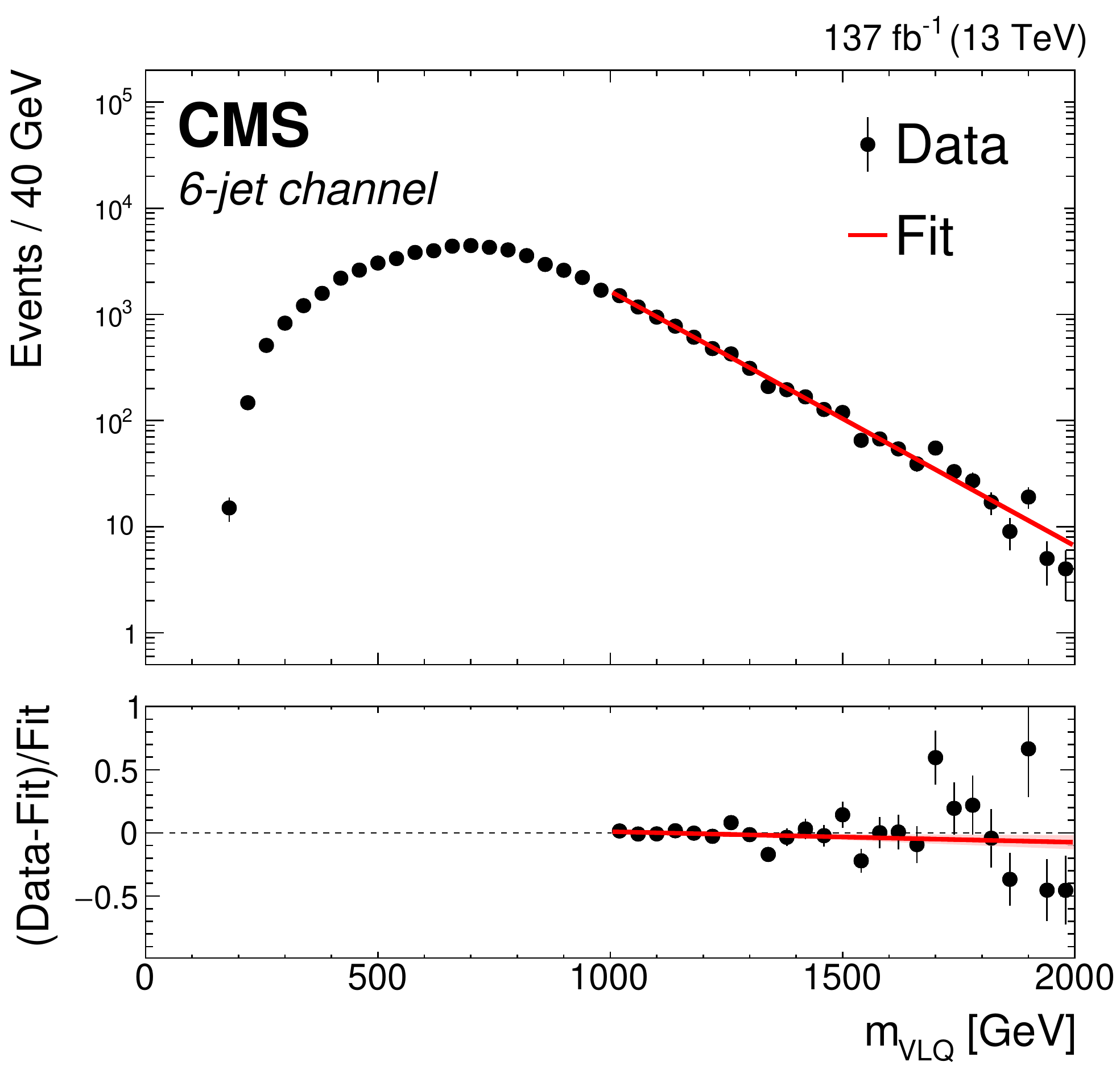}
\caption{Distributions of \mvlq for the jet combination with the lowest \chimod in 4-jet (left), 5-jet (center), and 6-jet (right) multiplicity events. The red lines show the exponential fit in the range 1000--2000\GeV. The lower panels show the fractional difference, $(\text{data}-\text{fit})/\text{fit}$.}
\label{fig:bckg}
\end{figure*}

\section{Background estimation}\label{section:background}

The expected background is estimated in a low-mass sideband region in data, using the ratio of the number of events passing the tag requirements to the number before these requirements are applied. The estimation is done separately for each of the three event modes, jet multiplicities, and three data-taking years, for a total of 27 cases.

Figure~\ref{fig:bckg} shows the distribution of \mvlq for the jet combination with the lowest \chimod for each of the three jet multiplicities. All events shown in this plot are required to pass a selection of $\chimodndf < 4$.   The falloff in the distribution at lower masses is due to the $\HT > 1350\GeV$ requirement. The distributions are then fit with an exponential function for VLQ candidate masses greater than 1000\GeV; in all three cases, the function (shown by the red line) agrees with the data. An $F$-test~\cite{10.2307/1913018} shows that a more complex model, namely an exponential plus constant background, offers no significant improvement over the exponential distribution. The lower plots show the fractional difference between the data and the fit. At this stage, since there is no requirement made on jet tagging, the ratio of background to signal event acceptance is more than two orders of magnitude larger than after jet tagging, so the fits are insensitive to any possible signal events in the data.

The background jet-tagged fraction (BJTF) is the fraction of background events that remain after the jet tagging requirements, as described in Section~\ref{sec:jets}, are applied. Since the BJTF for events with $\mvlq > 1000\GeV$ could be biased due to signal events that might be in the data, the BJTF is initially determined only for events in which \mvlq is between 500 and 800\GeV, which is below the current lower exclusion limit on the VLQ mass~\cite{Sirunyan:2019sza,Aaboud:2018wxv}. Table~\ref{table:redfac} shows the BJTF for data events with \mvlq in the range 500--800\GeV for each of the three event modes and three jet multiplicities.

\begin{table}[hbtp]
\centering
\topcaption{Values of the BJTF for data events with \mvlq in the range 500--800\GeV for each of the three event modes and three jet multiplicities.}
\label{table:redfac} 
\begin{scotch}{cccc}
&\bhbh & \bhbz & \bzbz \\ 
\hline
4 jets & $0.0042 \pm 0.0014$ & $0.0019 \pm 0.0004$ & $0.0025 \pm 0.0004$ \\
5 jets & $0.0041 \pm 0.0003$ & $0.0036 \pm 0.0002$ & $0.0048 \pm 0.0009$ \\
6 jets & $0.0019 \pm 0.0002$ & $0.0019 \pm 0.0002$ & $0.0020 \pm 0.0005$ \\
\end{scotch}
\end{table}

\begin{figure*}[p]
\centering
\includegraphics[width=0.32\textwidth]{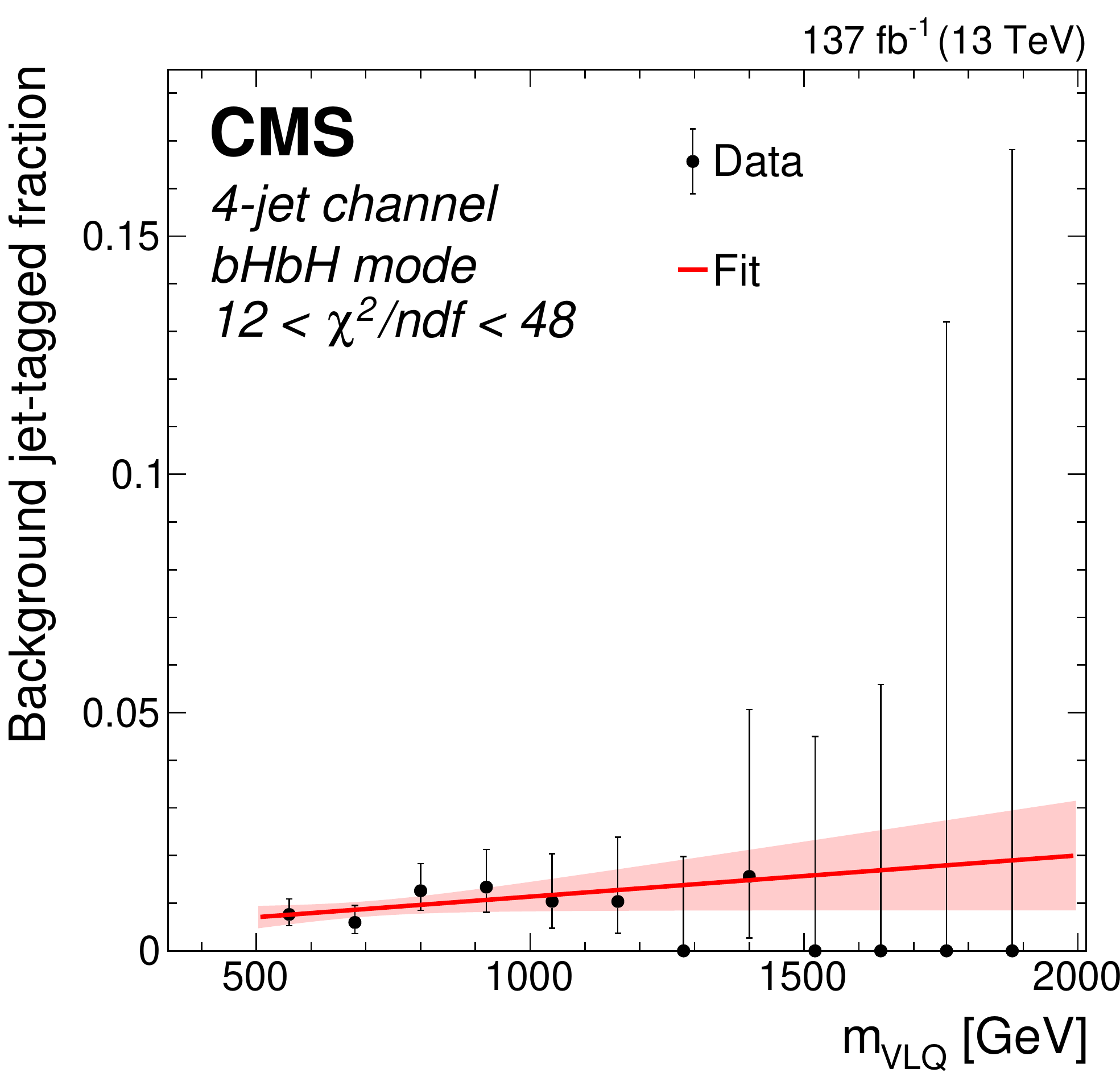}
\includegraphics[width=0.32\textwidth]{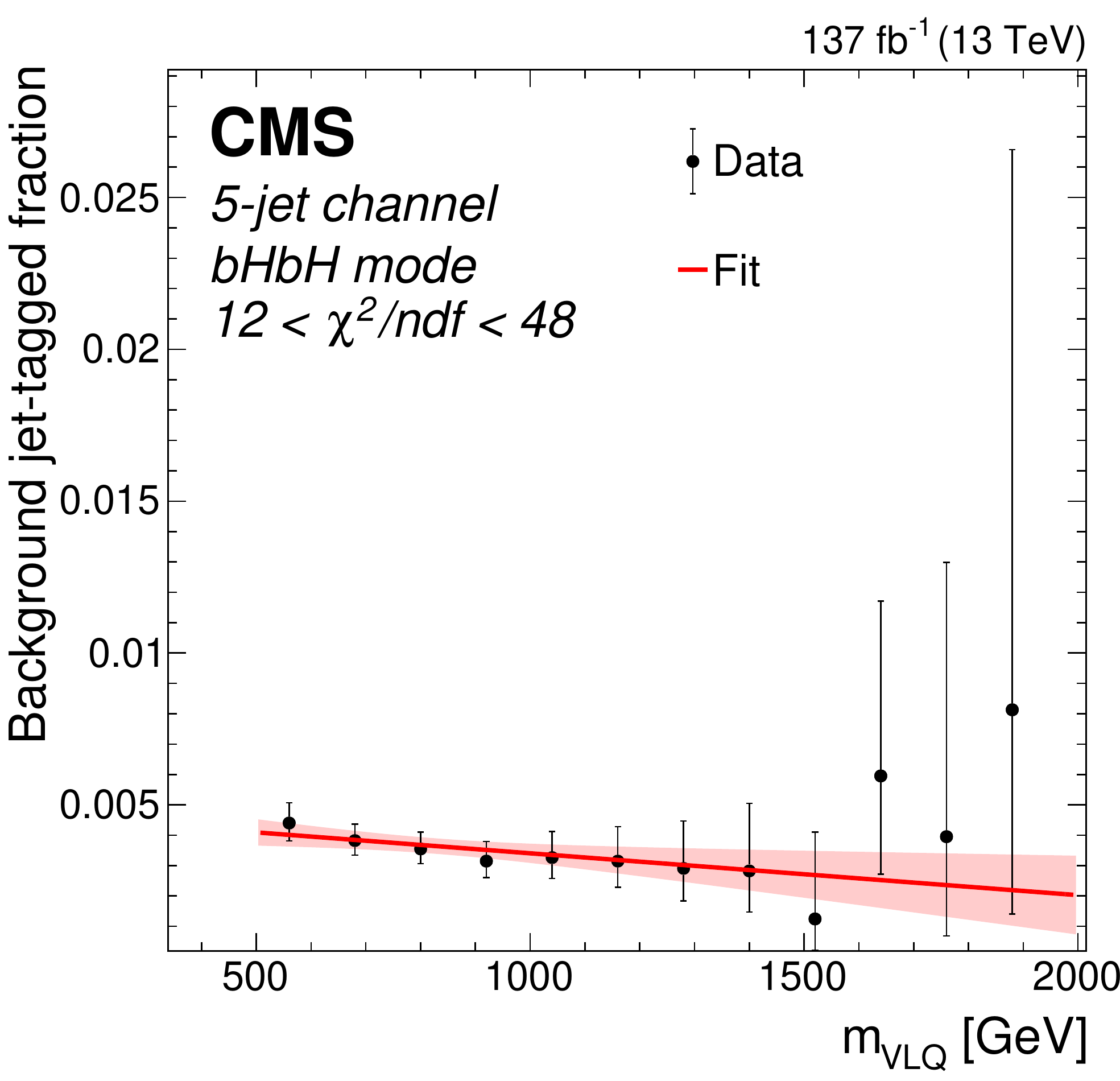}
\includegraphics[width=0.32\textwidth]{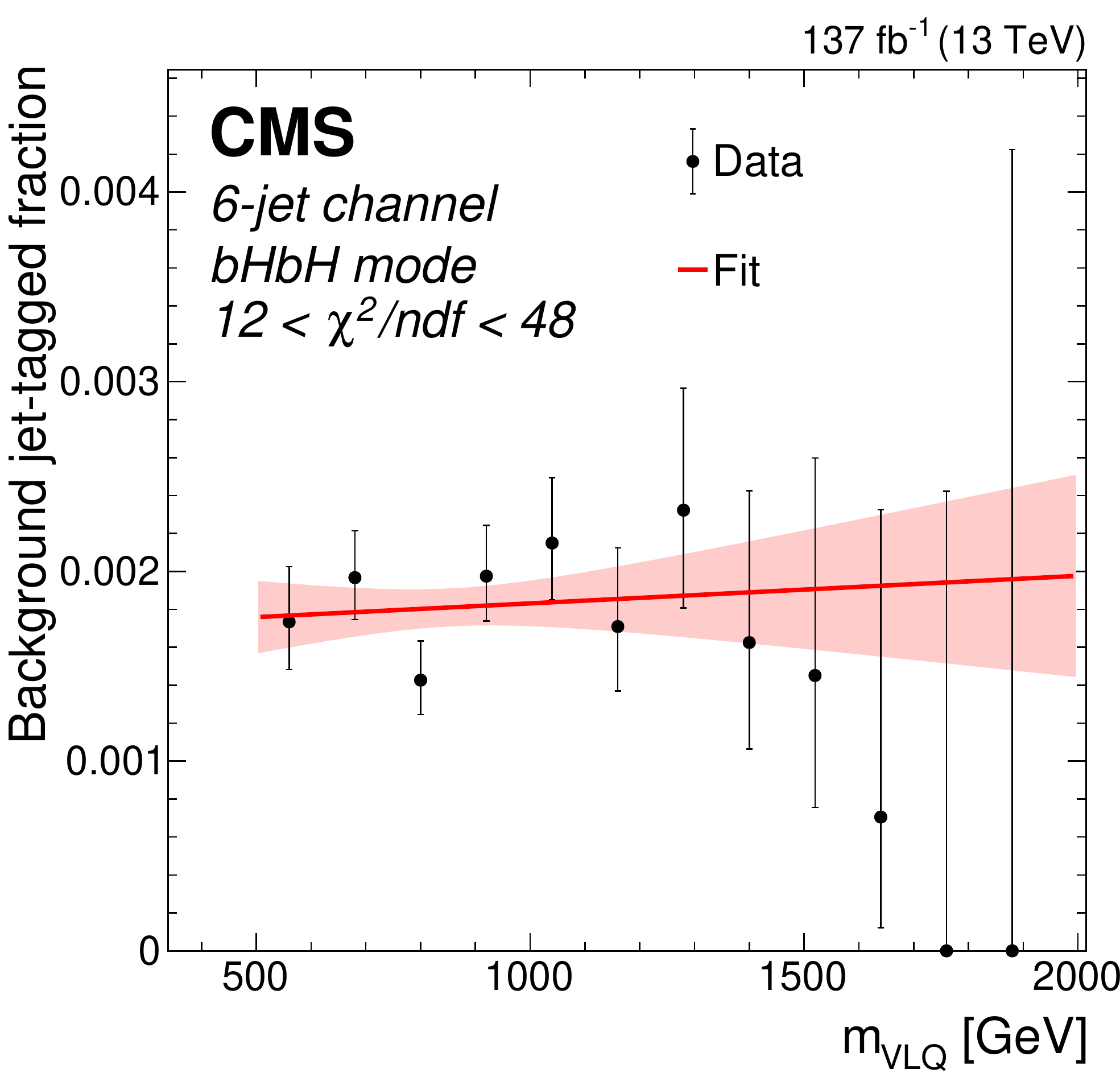}
\includegraphics[width=0.32\textwidth]{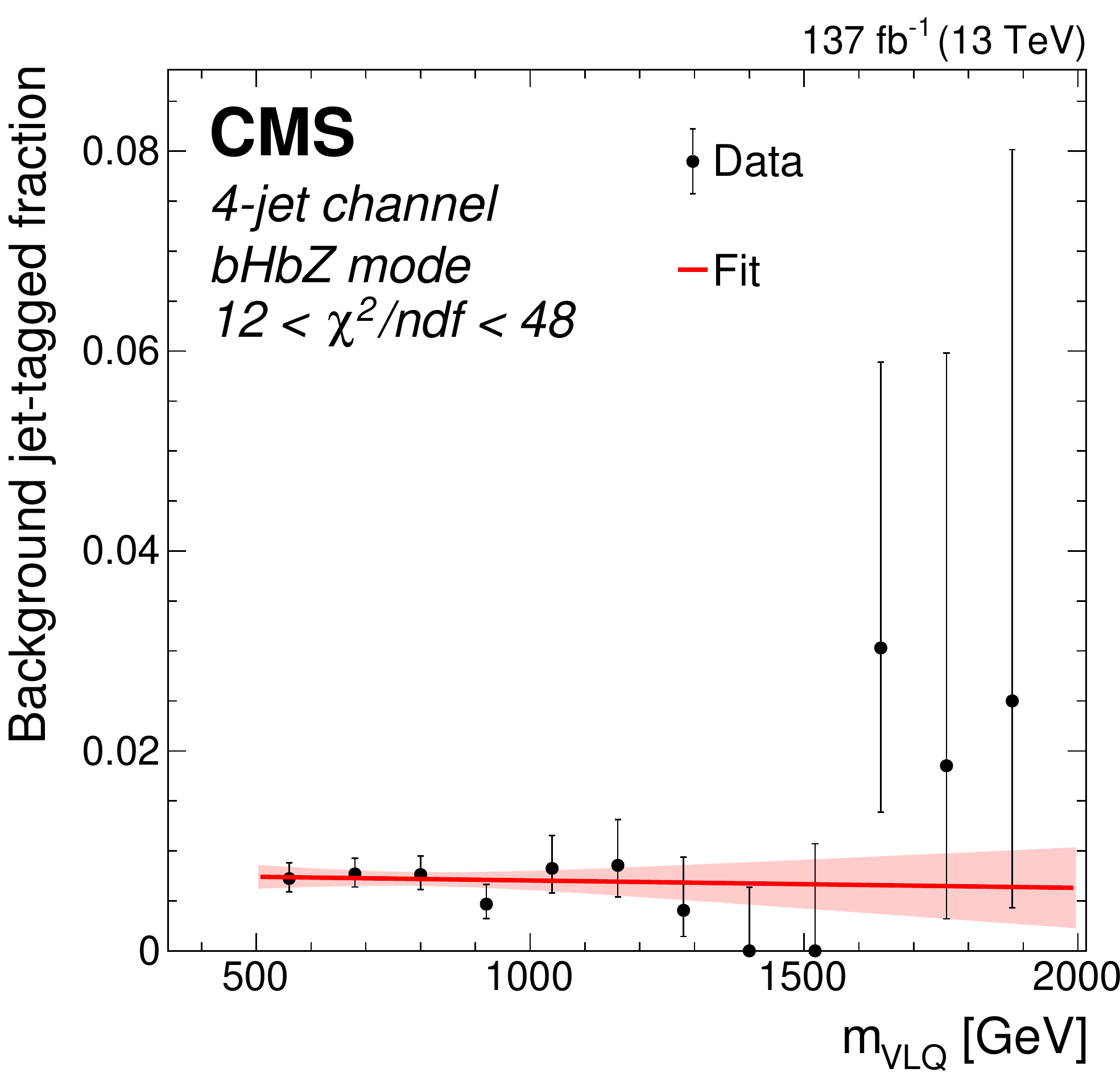}
\includegraphics[width=0.32\textwidth]{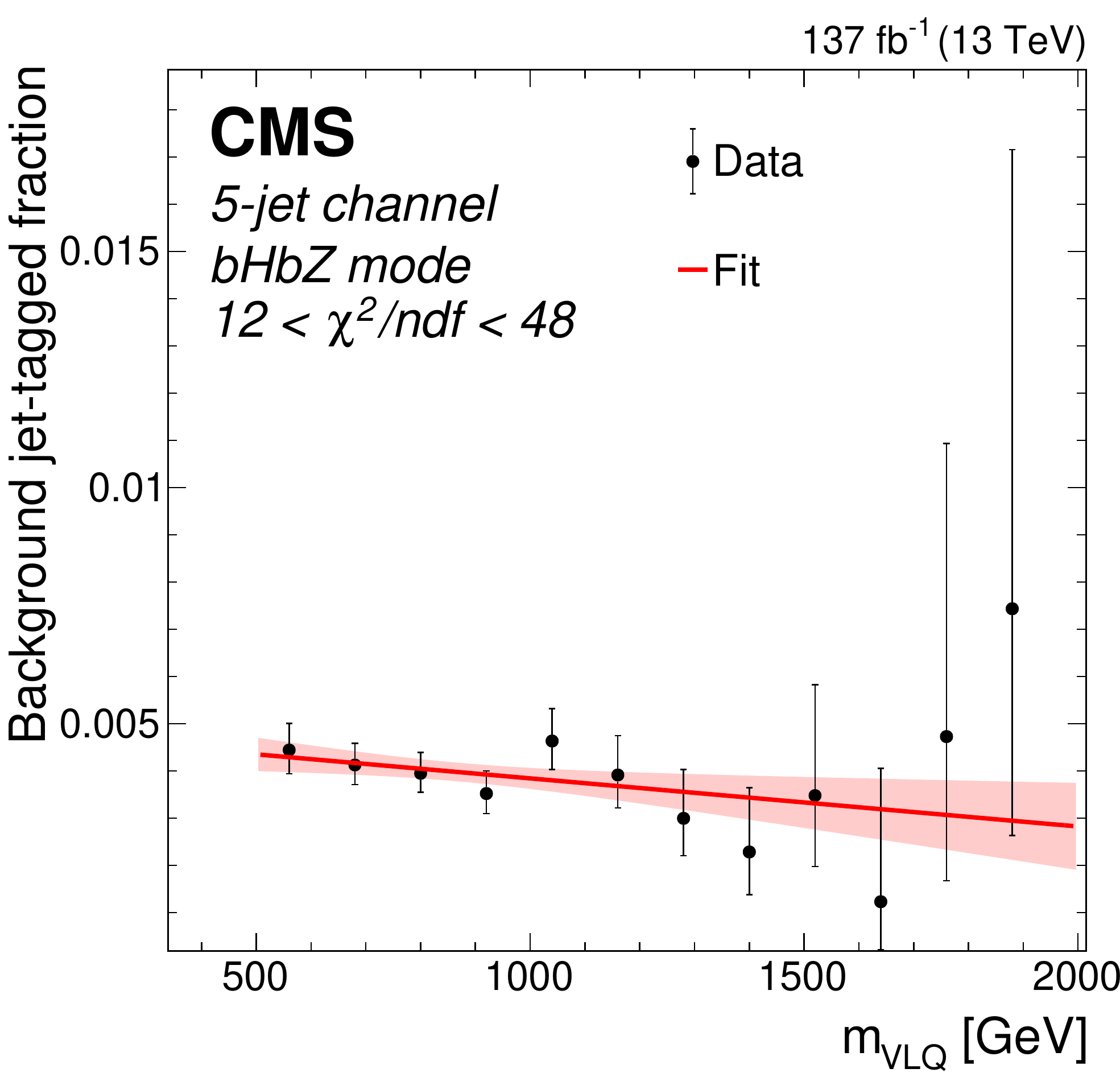}
\includegraphics[width=0.32\textwidth]{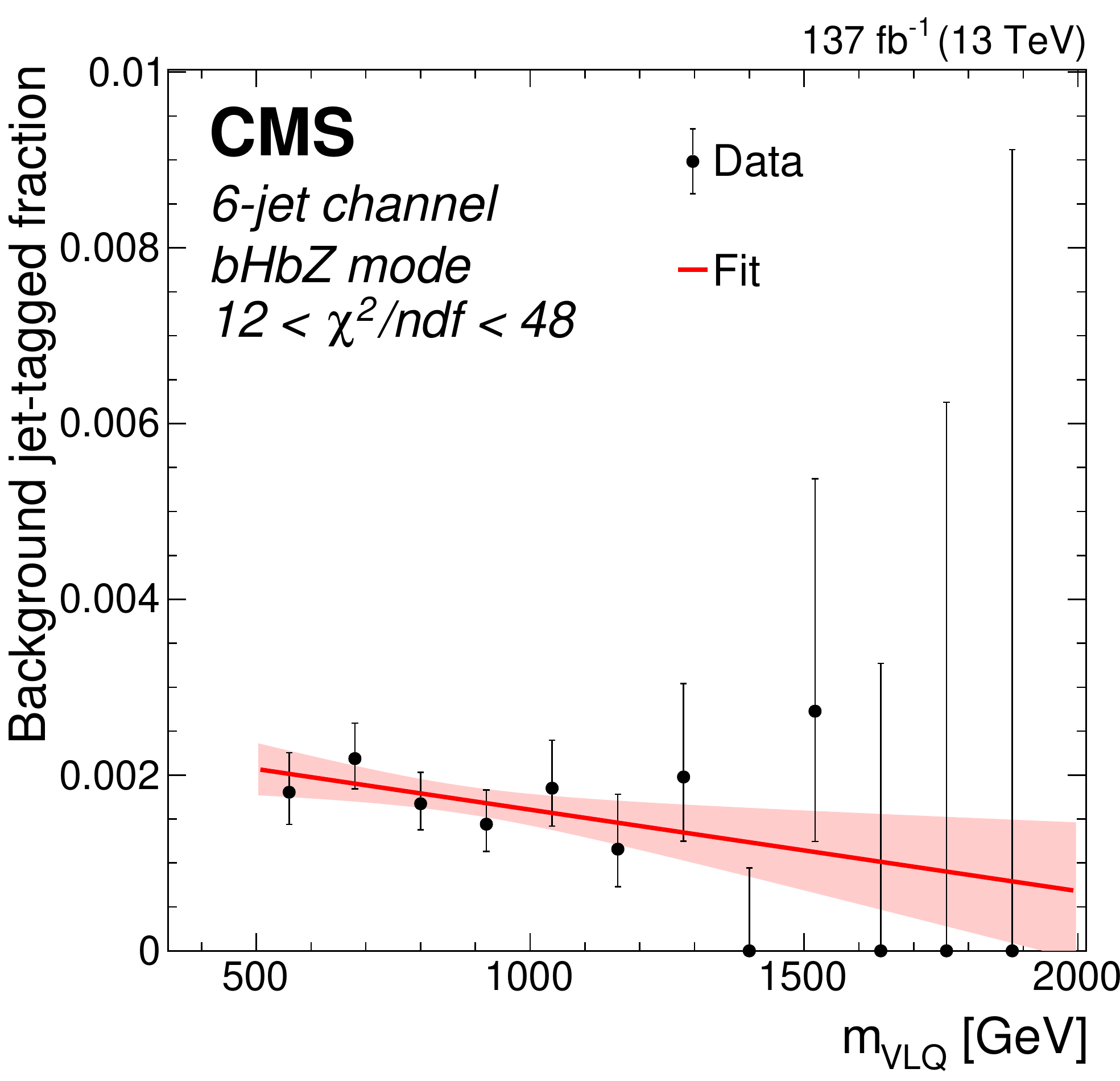}
\includegraphics[width=0.32\textwidth]{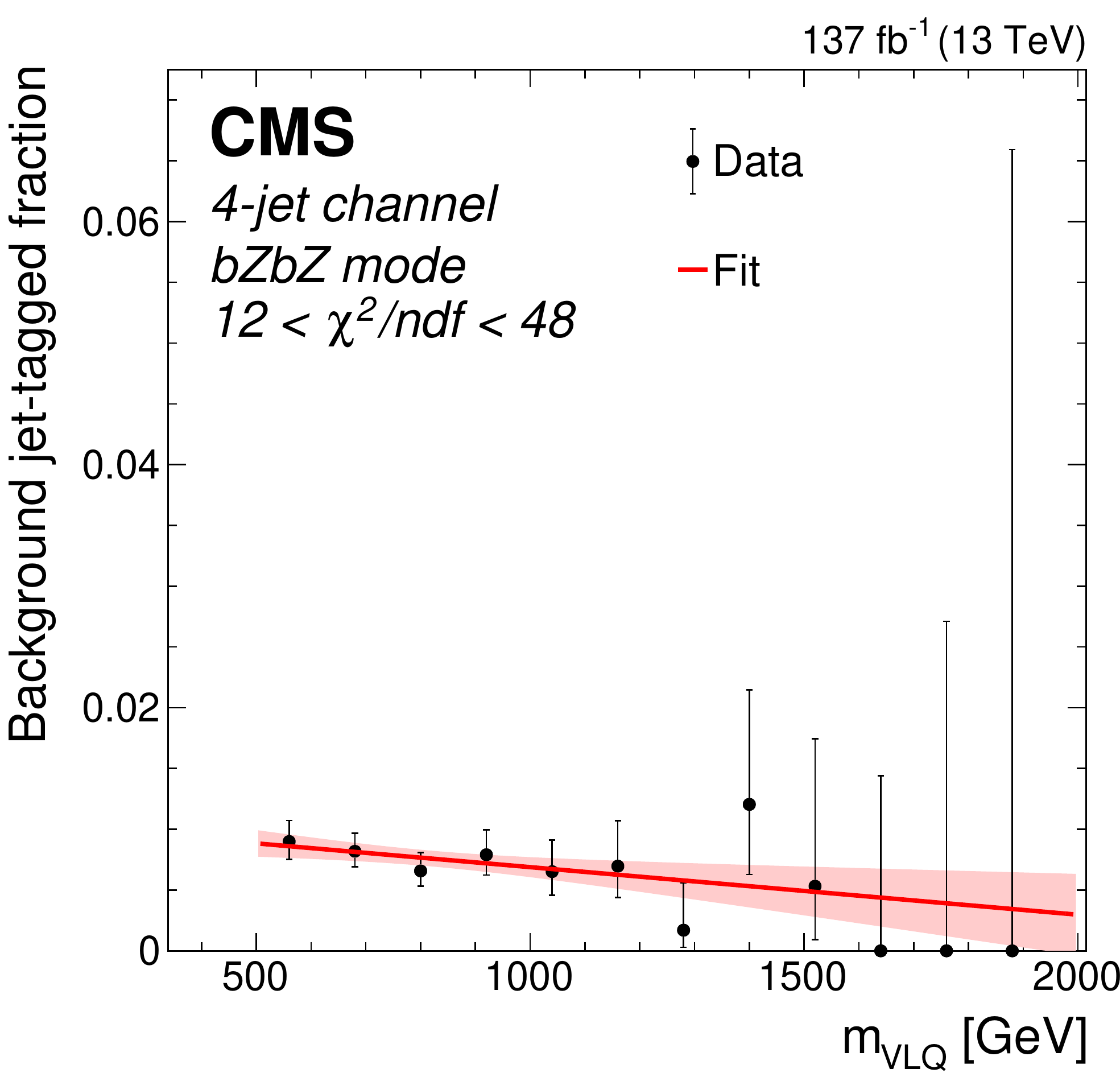}
\includegraphics[width=0.32\textwidth]{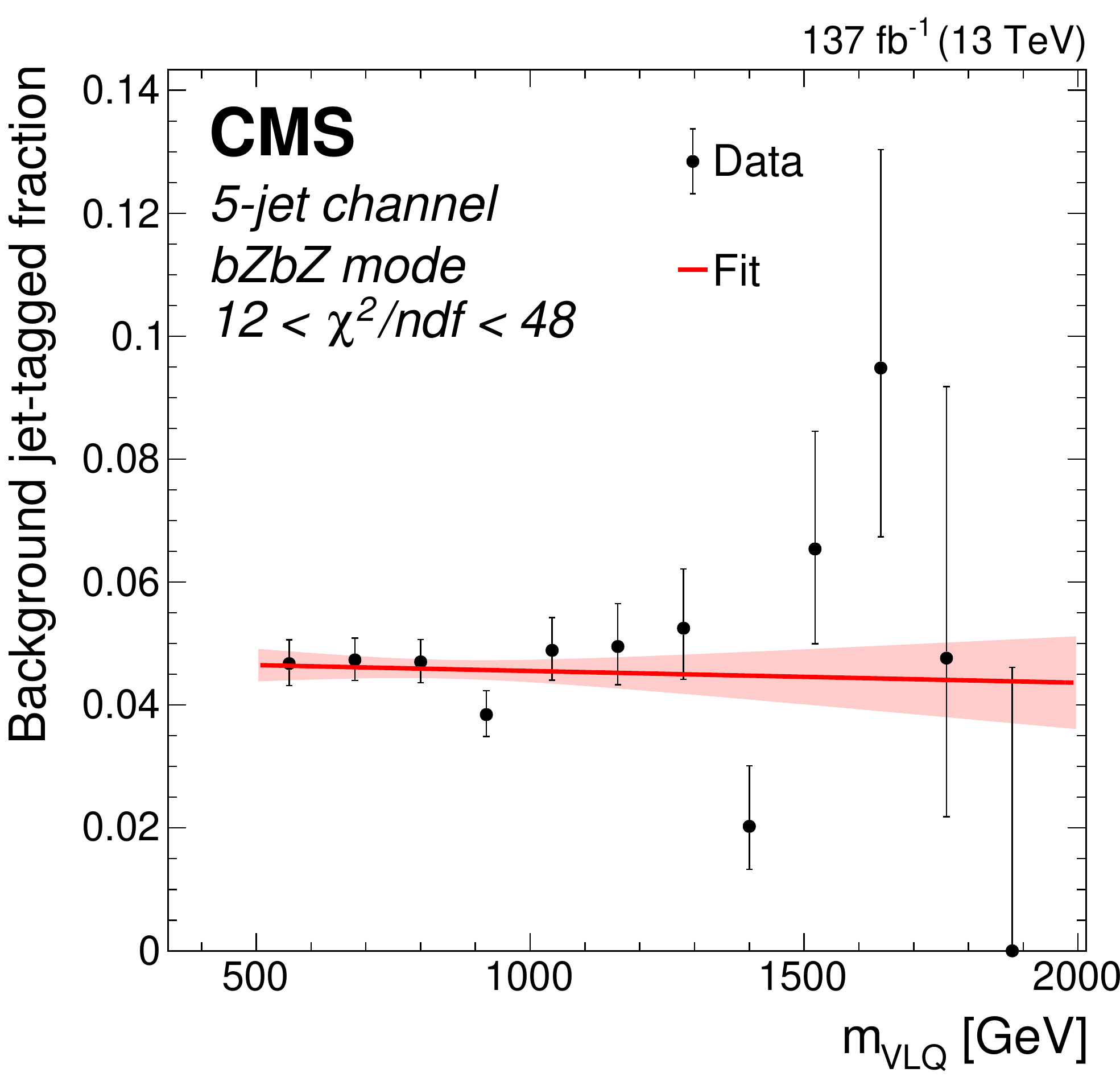}
\includegraphics[width=0.32\textwidth]{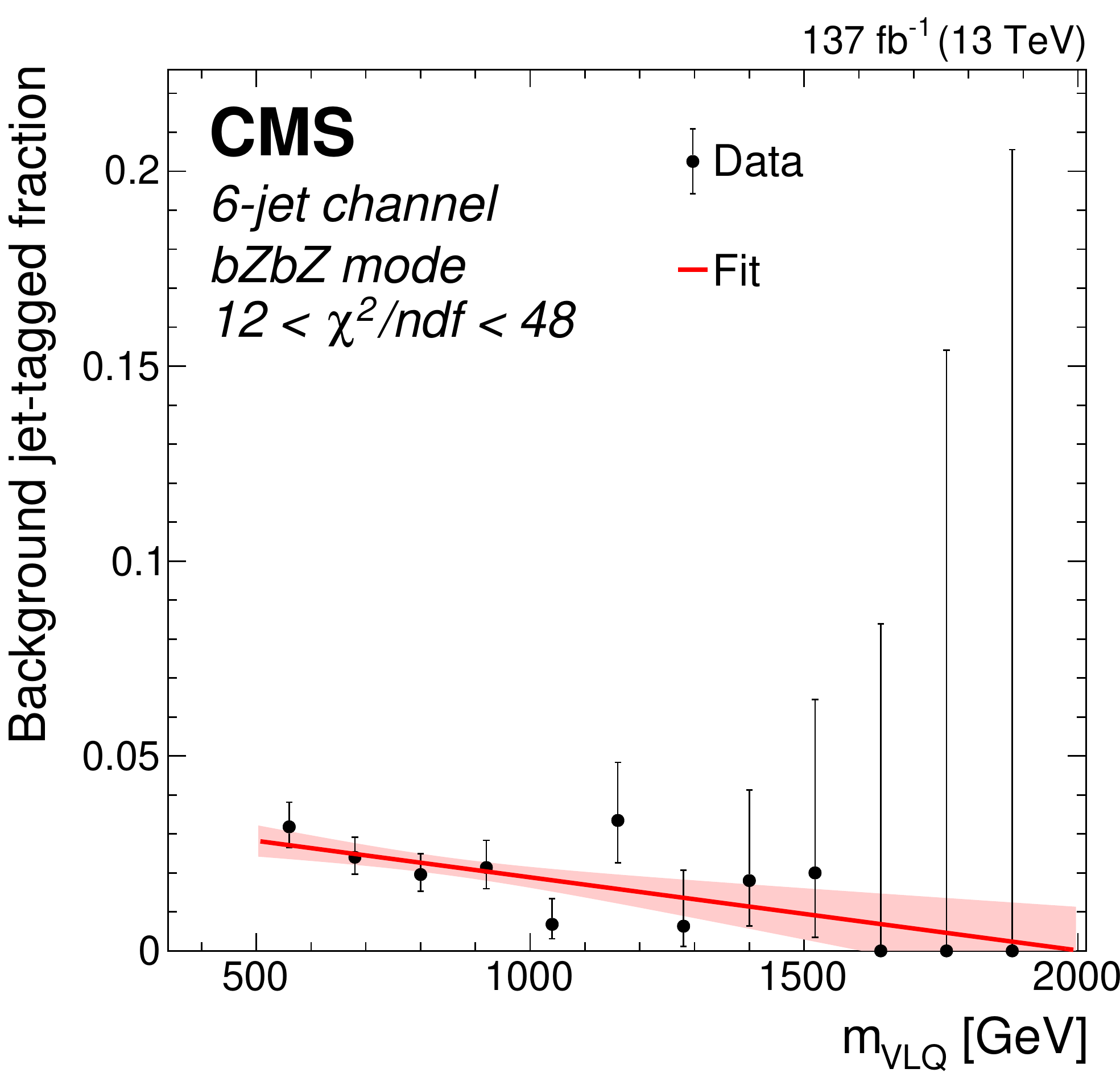}
\caption{Dependence of the BJTF on \mvlq in the control region $12 < \chimodndf < 48$, for 4-jet (left column), 5-jet (center column), and 6-jet (right column) multiplicities, and for the \bhbh (upper row), \bhbz (middle row), and \bzbz (lower row) event modes. The data are shown as black points with vertical error bars, and the linear fit and associated uncertainty are shown as a solid red line and the shaded red band.}
\label{fig:rf_mass_depen}
\end{figure*}

Because the jet tagging efficiency depends on the \pt of the jet, the BJTF might depend on the mass of the VLQ candidate, since events with greater VLQ mass generally have higher \pt jets.  A control region is therefore used to determine the VLQ mass dependence of the BJTF by offsetting the window of the \chimod selection. The signal \chimod region depends on the event mode and multiplicity, as determined by the optimization procedure described in Section~\ref{sec:eventreco}, but in all cases is at most $\chimodndf < 5.5$. A control region is defined by using a region of $12 < \chimodndf < 48$. Figure~\ref{fig:rf_mass_depen} shows the mass dependence of the BJTF for data events in the \chimod control region. A first-order polynomial fit is used to determine the BJTF mass dependence. Another $F$-test shows that there is no improvement for a second-order polynomial fit compared to a first-order one, and also that the first-order polynomial fit performs better than a constant fit. A systematic uncertainty is assigned by comparing the first-order polynomial fit to an exponential fit, and using the average difference of these two fits over the mass range as the uncertainty. This covers the uncertainty due to the choice of the BJTF shape.

\begin{figure*}[htbp]
\centering
\includegraphics[width=0.32\textwidth]{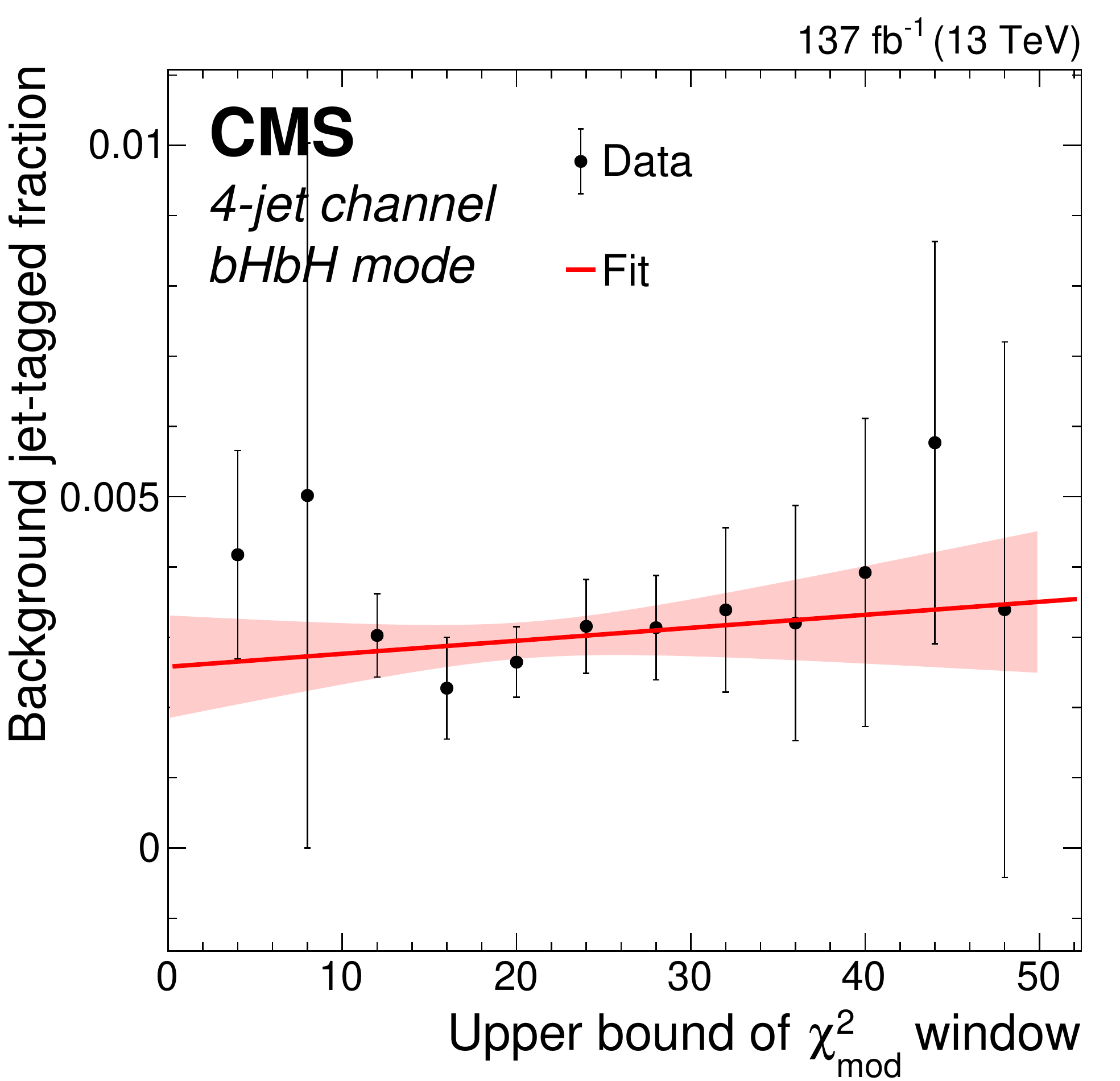}
\includegraphics[width=0.32\textwidth]{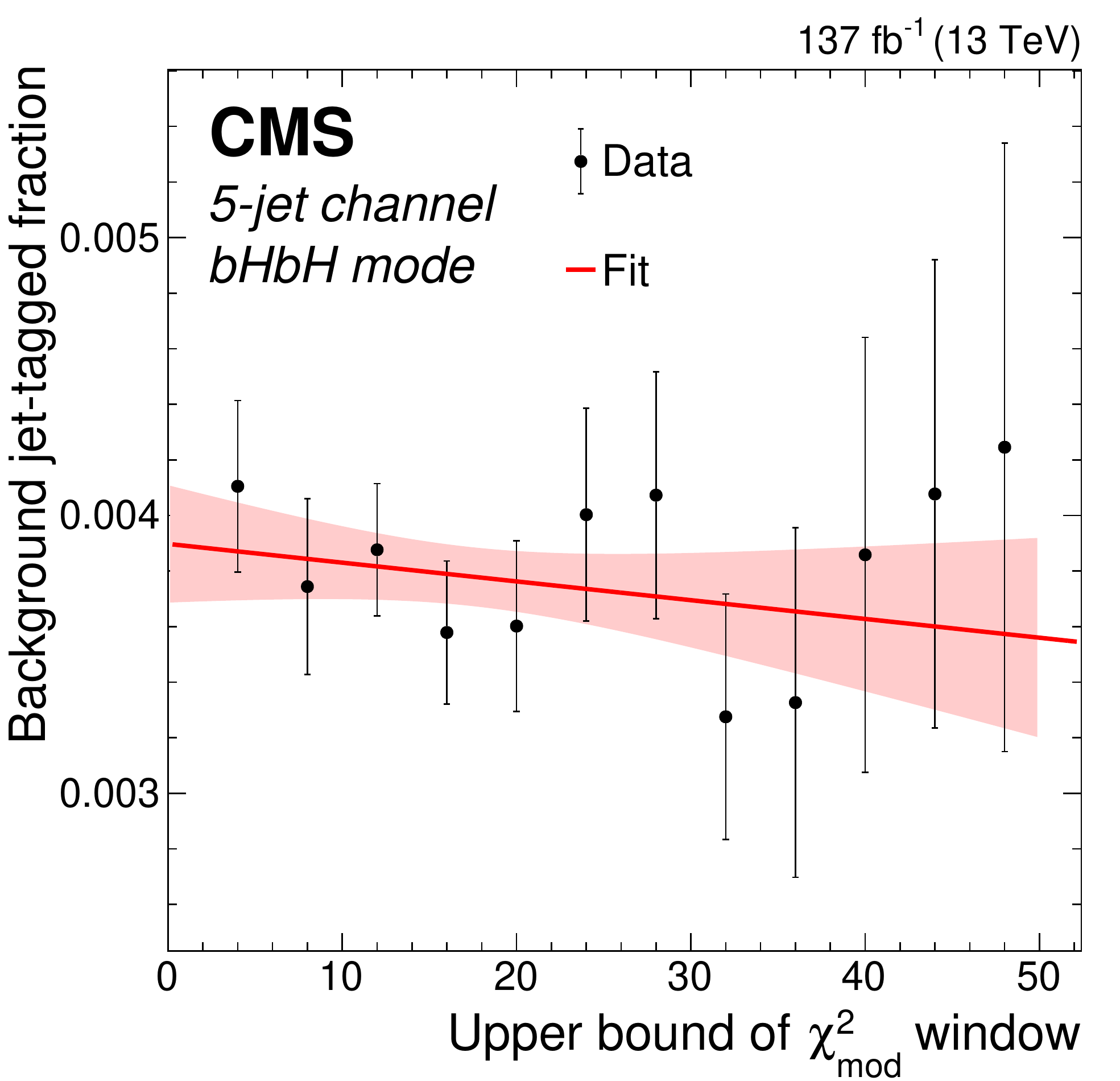}
\includegraphics[width=0.32\textwidth]{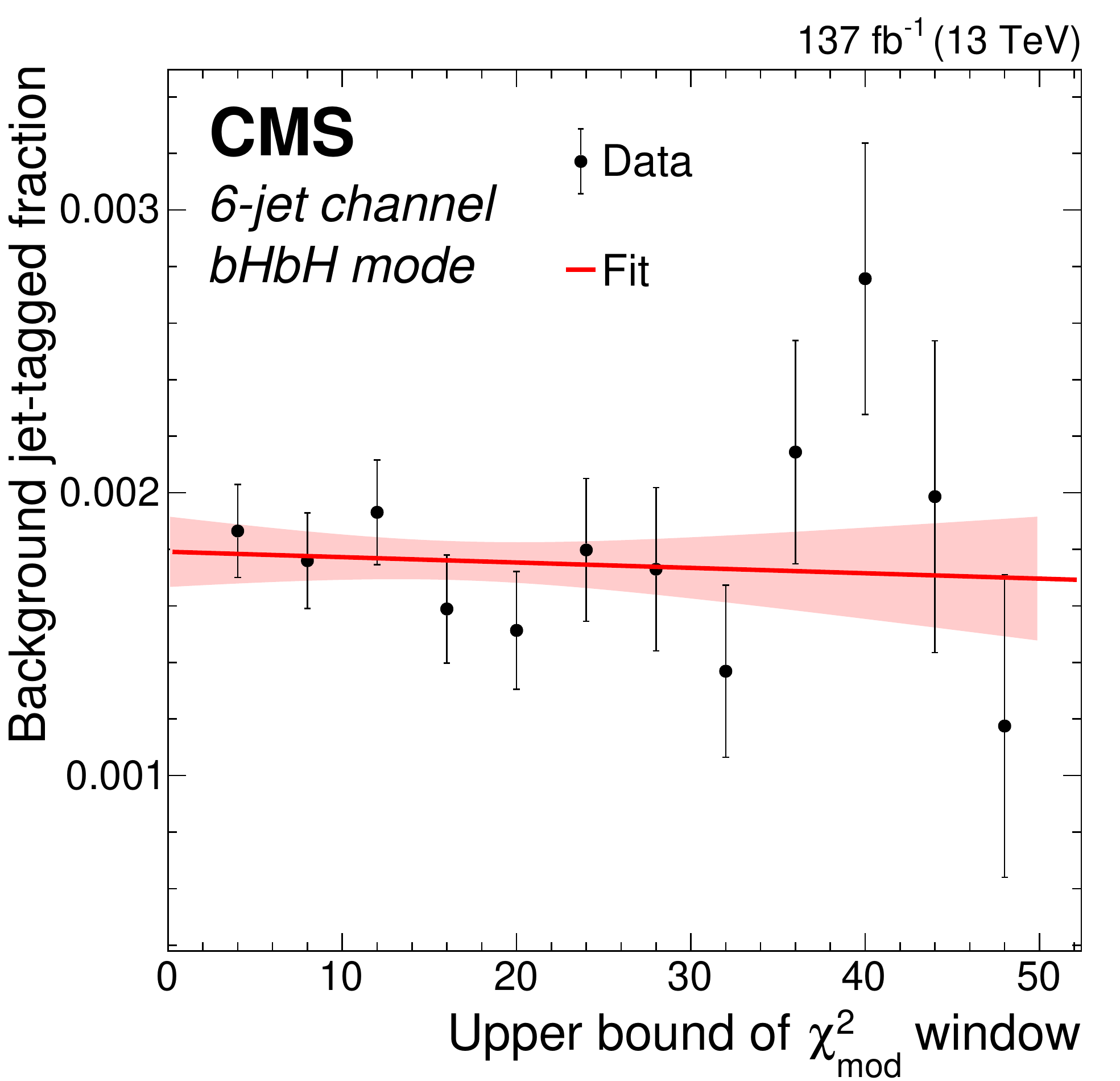}
\includegraphics[width=0.32\textwidth]{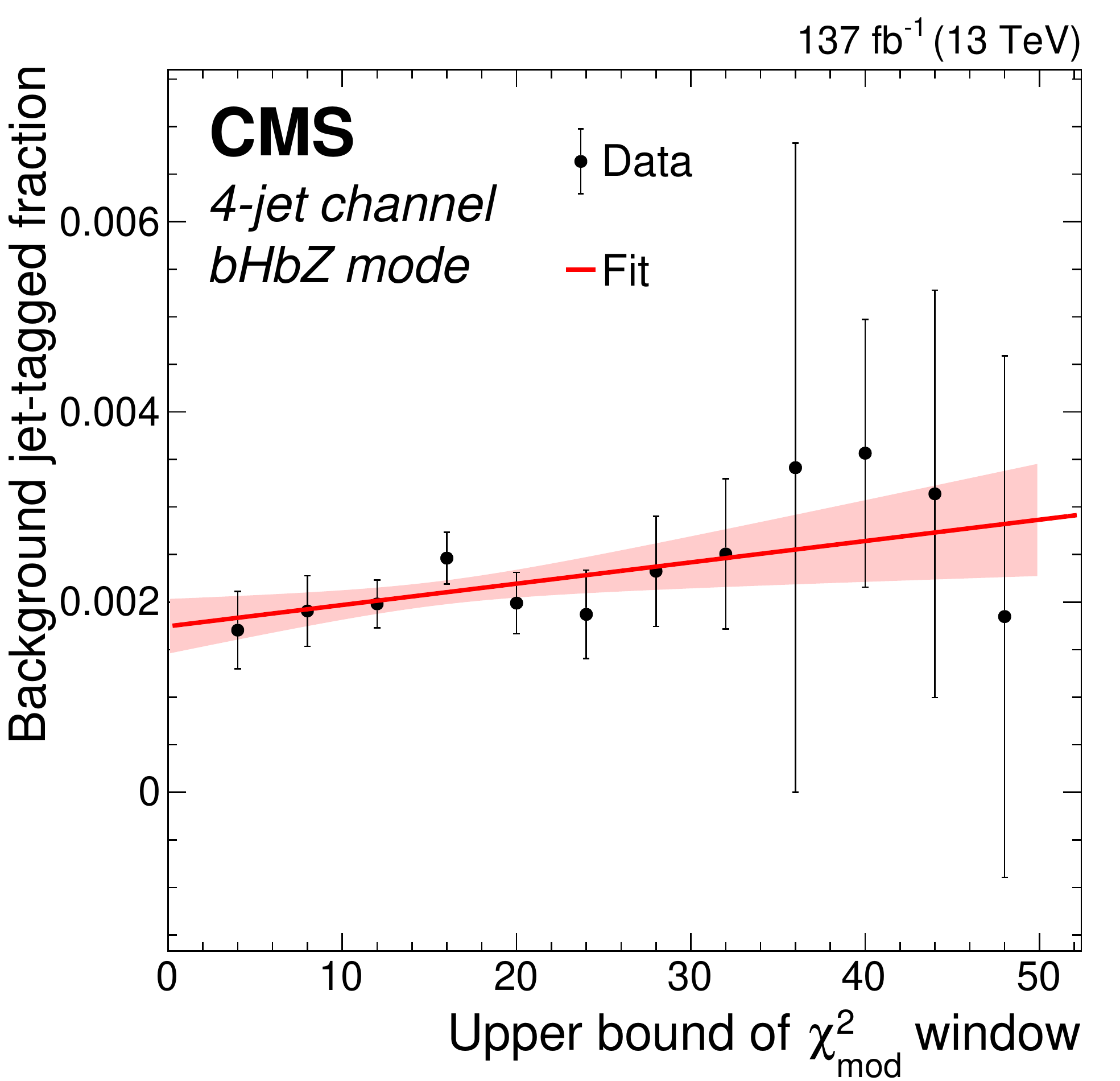}
\includegraphics[width=0.32\textwidth]{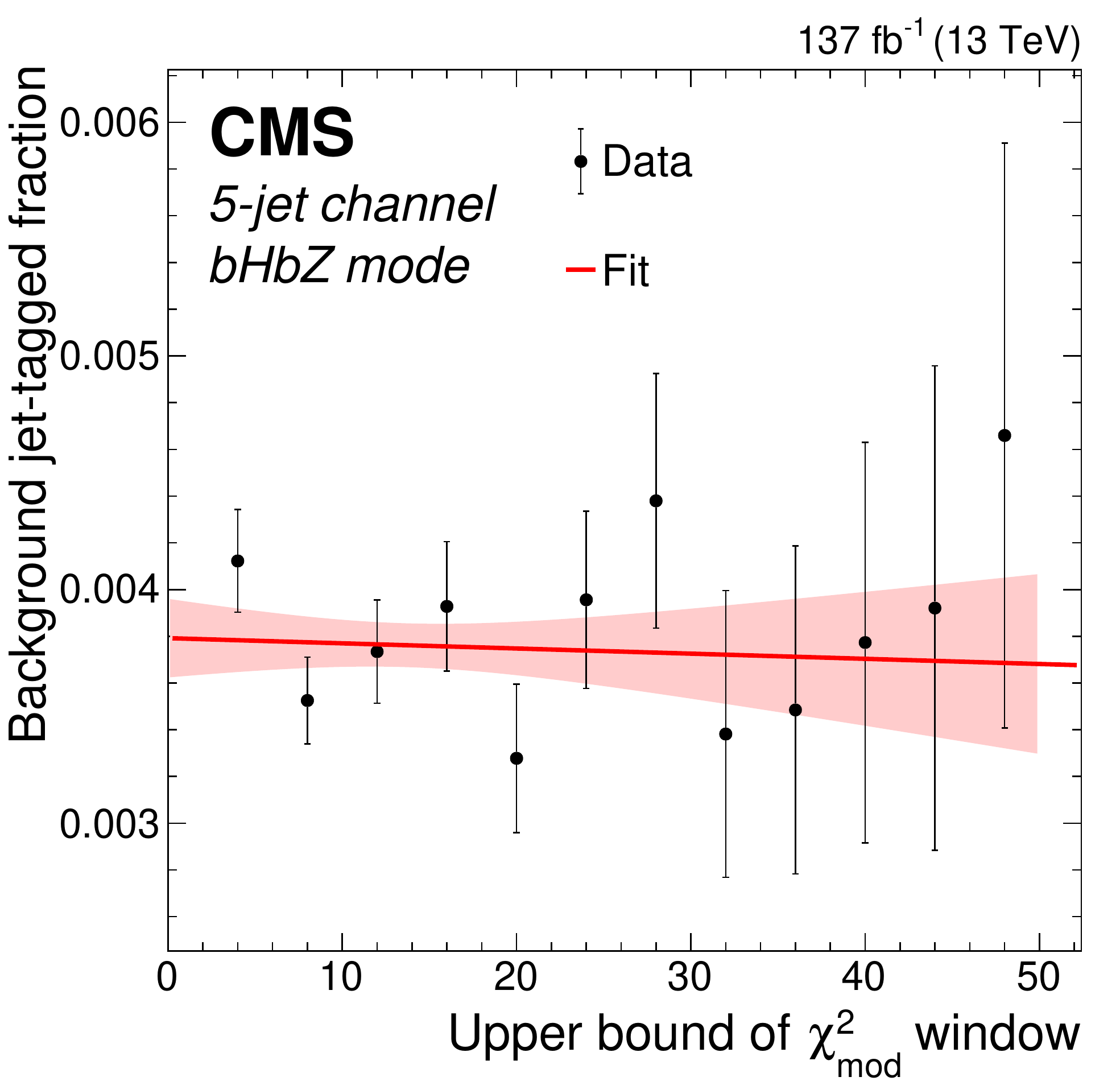}
\includegraphics[width=0.32\textwidth]{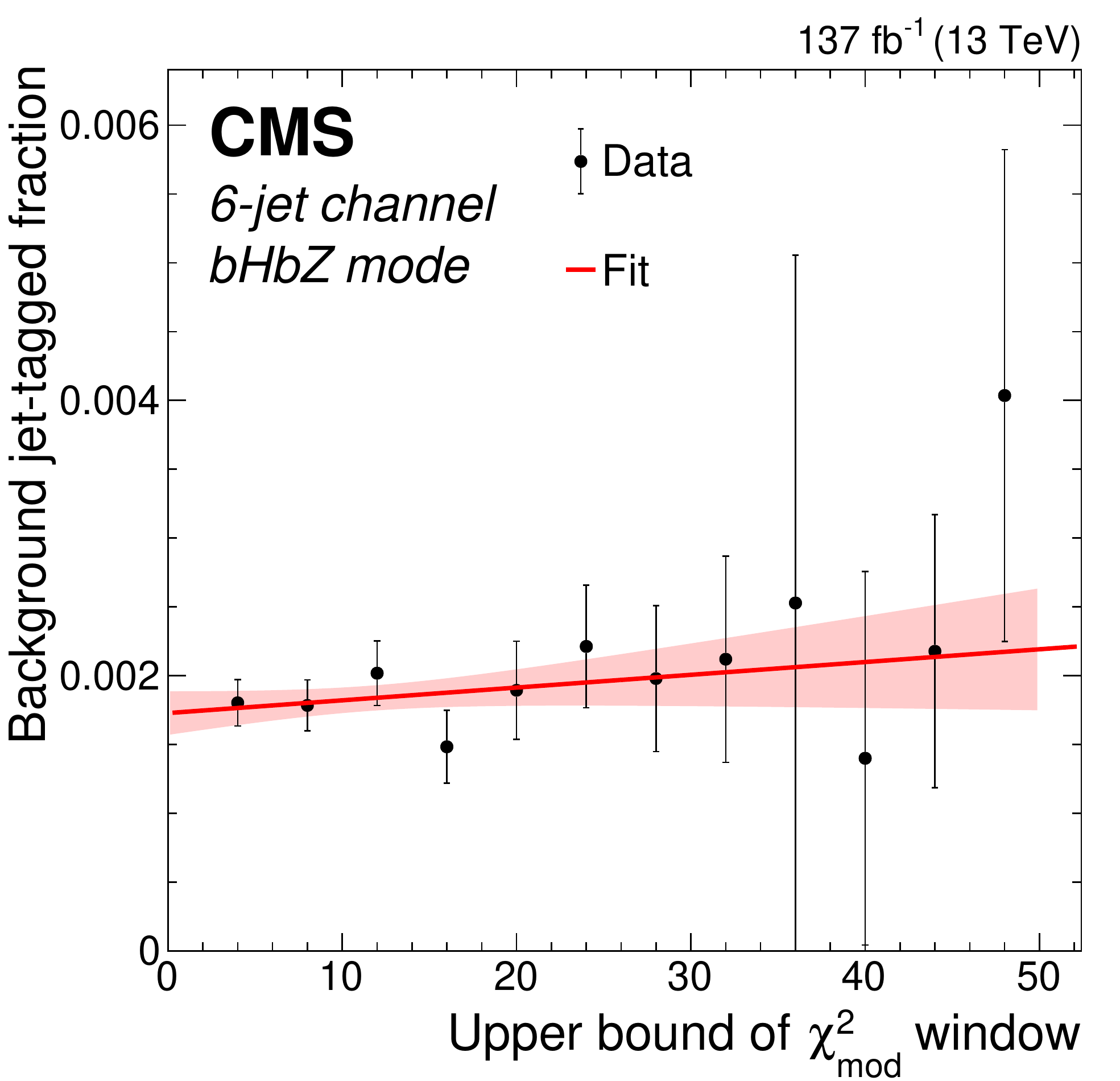}
\includegraphics[width=0.32\textwidth]{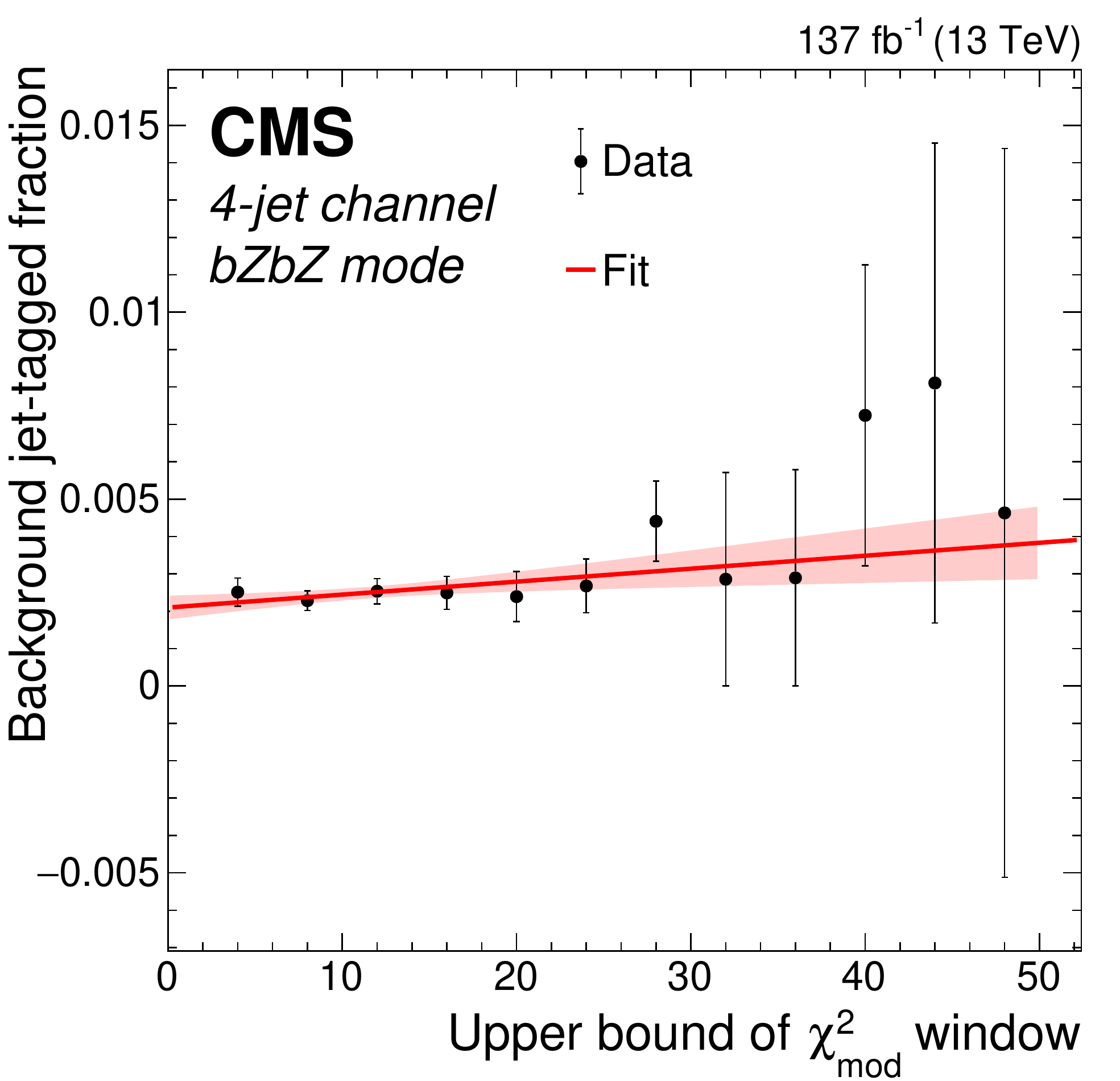}
\includegraphics[width=0.32\textwidth]{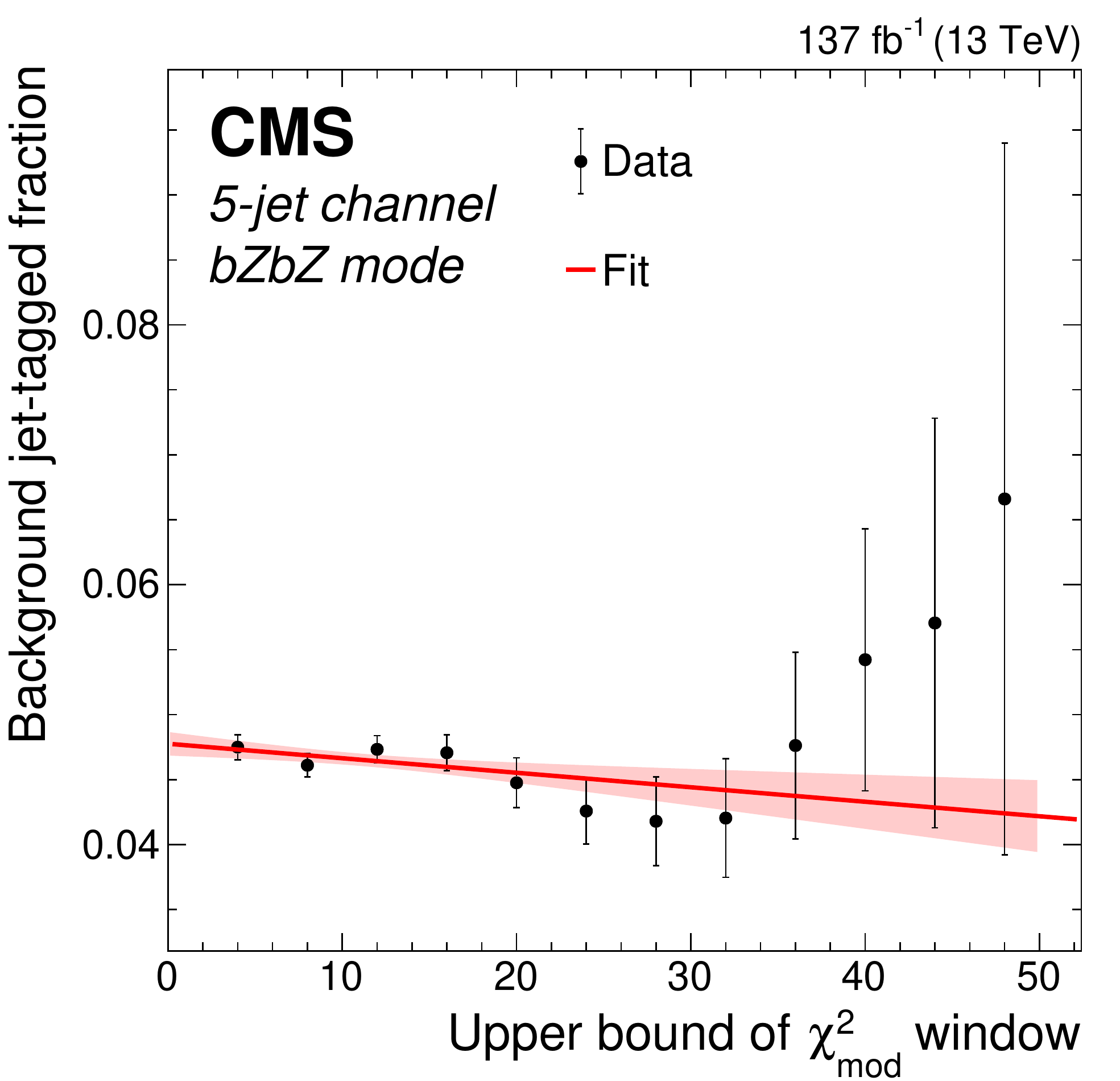}
\includegraphics[width=0.32\textwidth]{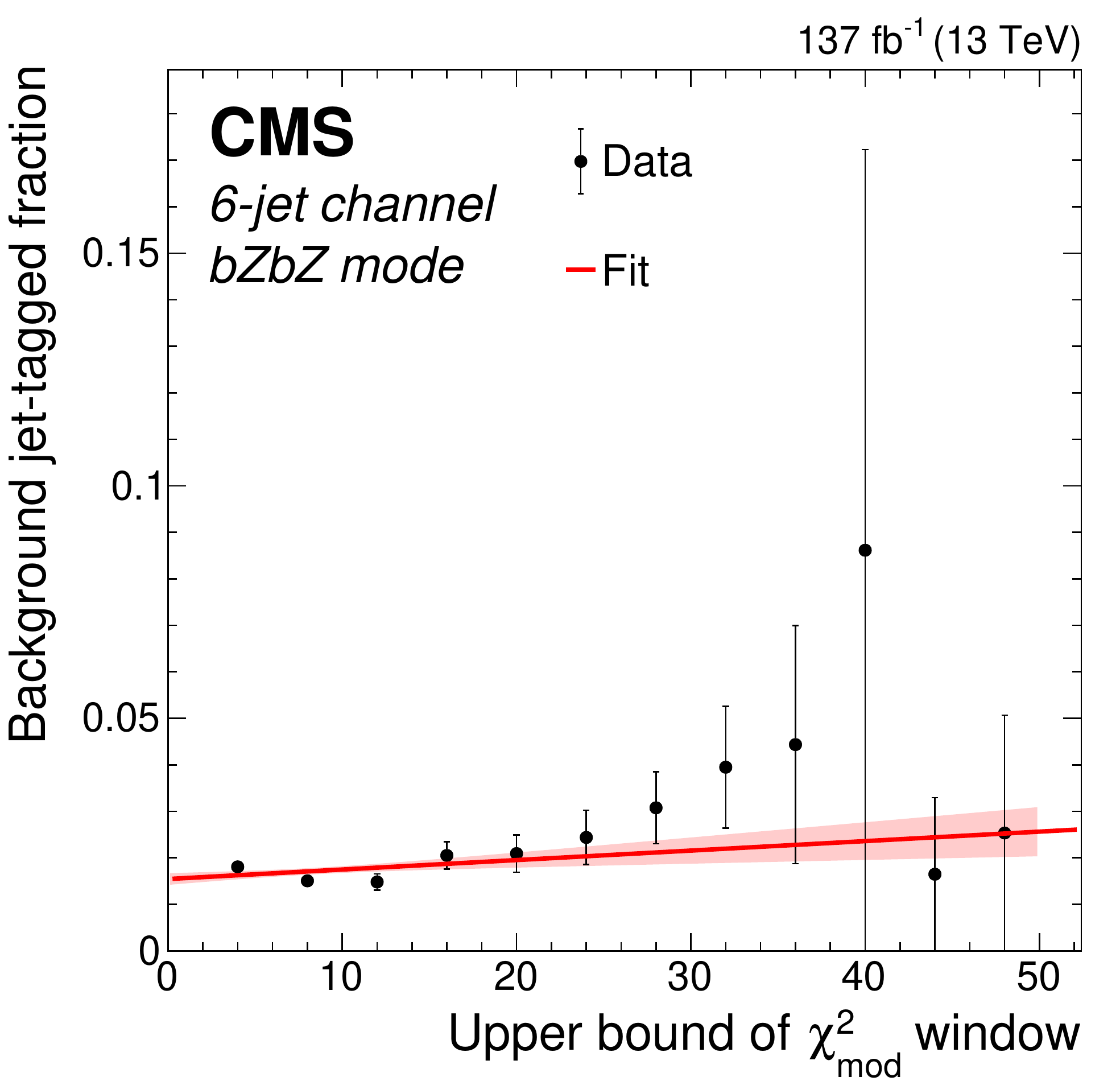}
\caption{Dependence of the BJTF on \chimodndf in the low-mass (500--800\GeV) VLQ region, for 4-jet (left column), 5-jet (center column), and 6-jet (right column) multiplicities, and for the \bhbh (upper row), \bhbz (middle row), and \bzbz (lower row) event modes. The data are shown as black points with vertical error bars, and the linear fit and associated uncertainty are shown as a solid red line and the shaded red band.}
\label{fig:rf_chi2_depen}
\end{figure*}

In order to validate that the control region used has the same BJTF behavior as the signal region, we perform a test where the BJTF in the low VLQ mass range (500--800\GeV) is plotted as a function of \chimodndf, in twelve equally spaced regions for \chimodndf from 0 to 48. This is shown in Fig.~\ref{fig:rf_chi2_depen}. The slope of this plot is consistent with zero, indicating no statistically significant dependence on \chimodndf.

\begin{figure*}[hbtp]
\centering
\includegraphics[width=0.32\textwidth]{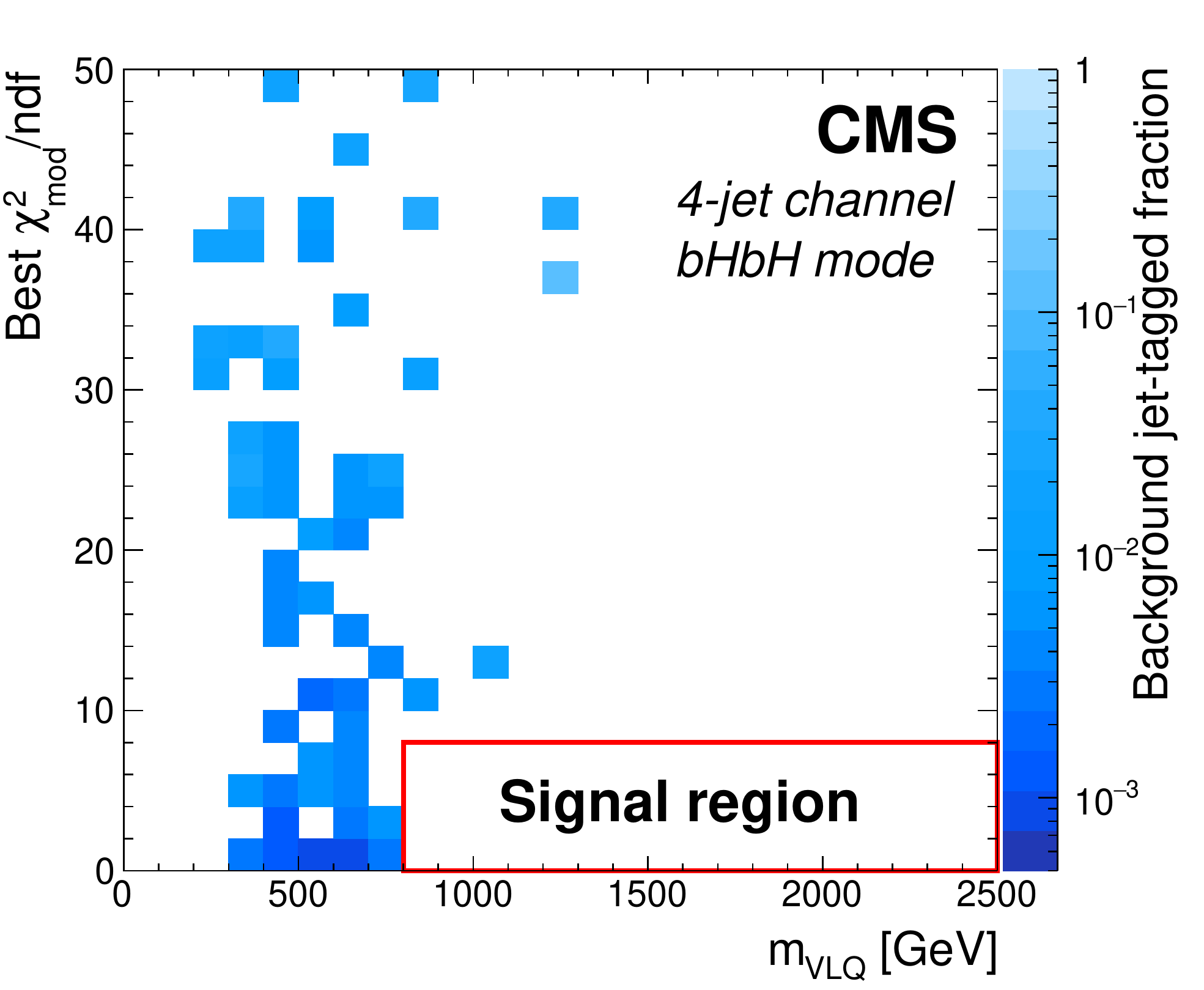}\hfill
\includegraphics[width=0.32\textwidth]{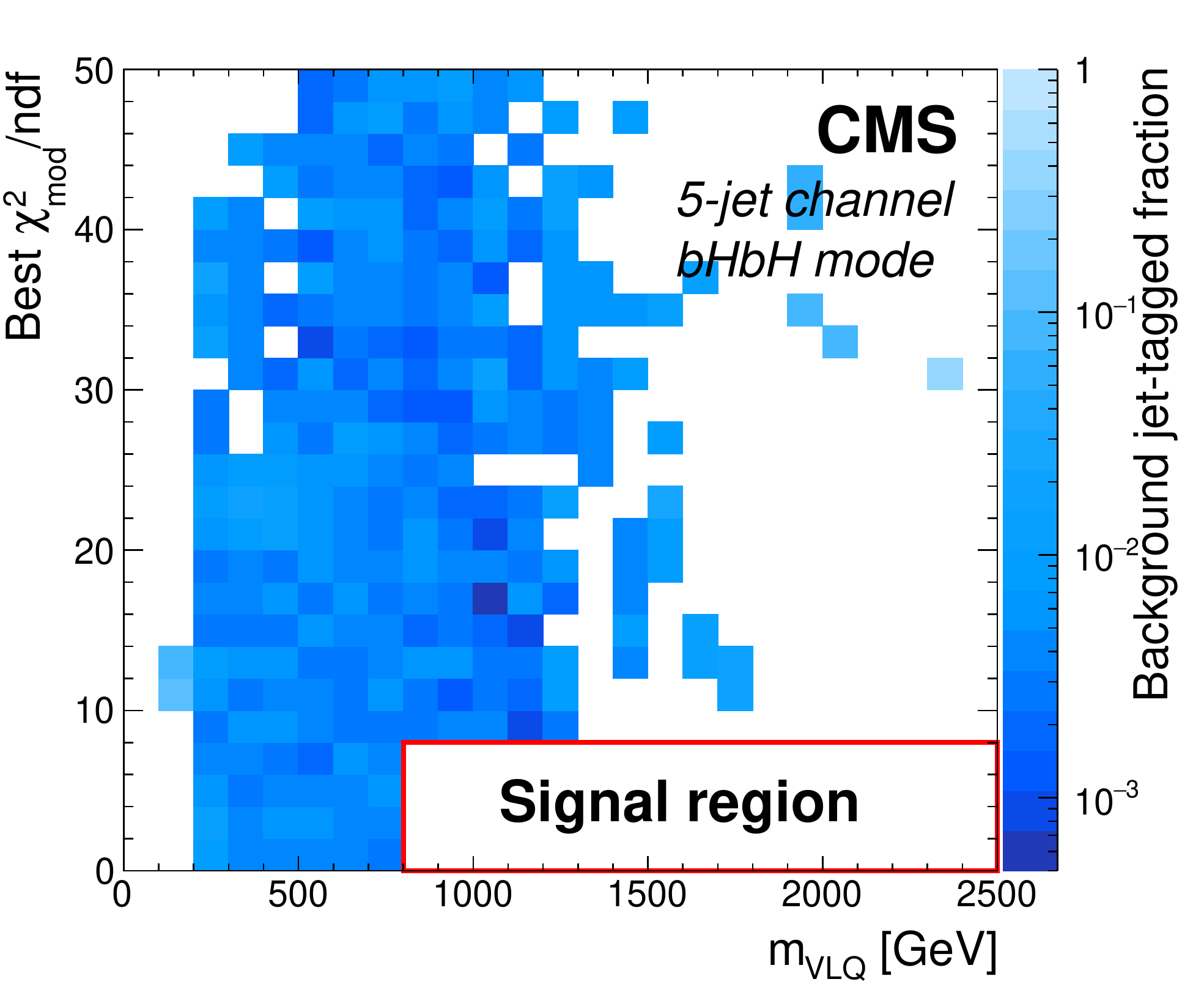}\hfill
\includegraphics[width=0.32\textwidth]{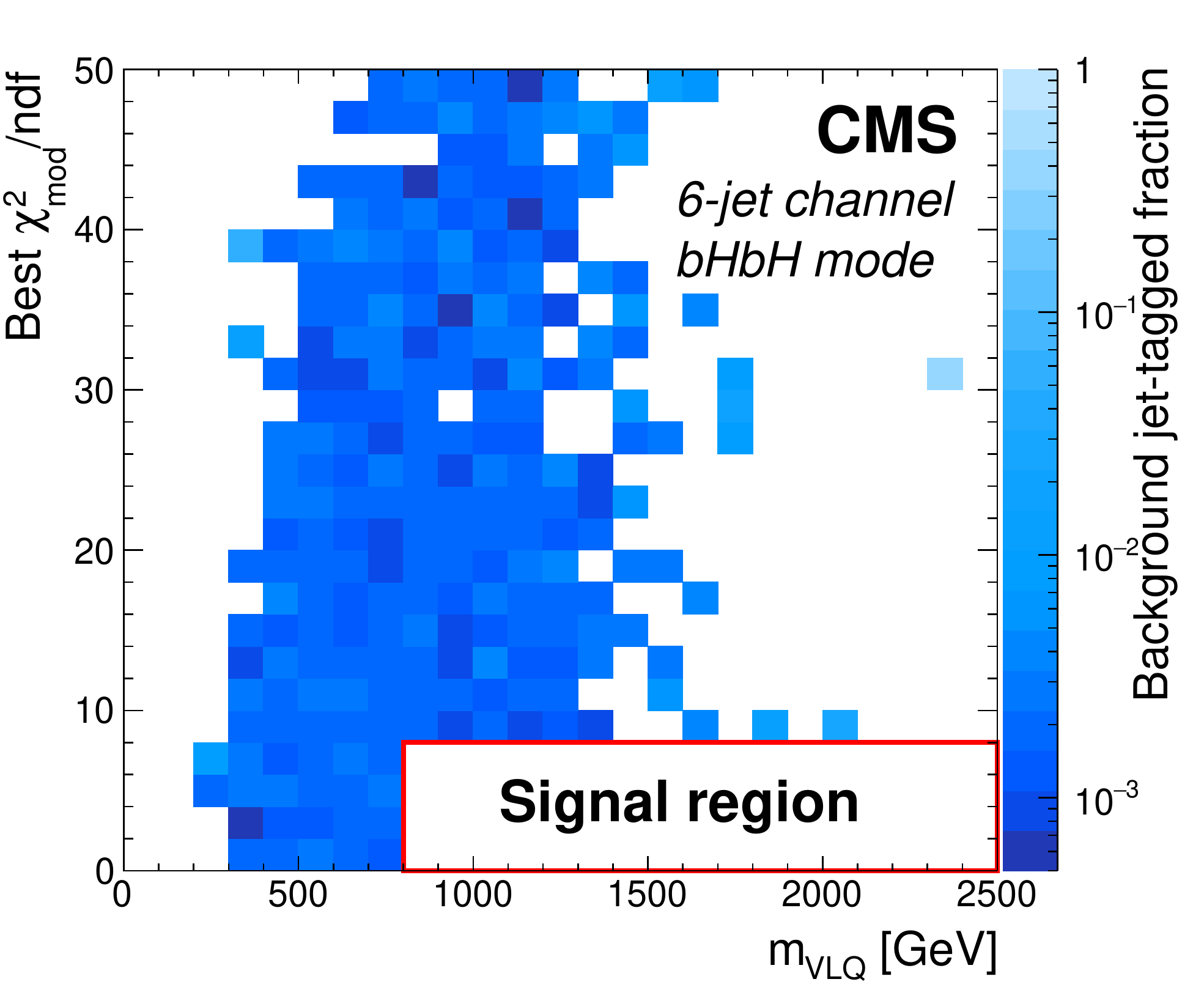}\hfill
\includegraphics[width=0.32\textwidth]{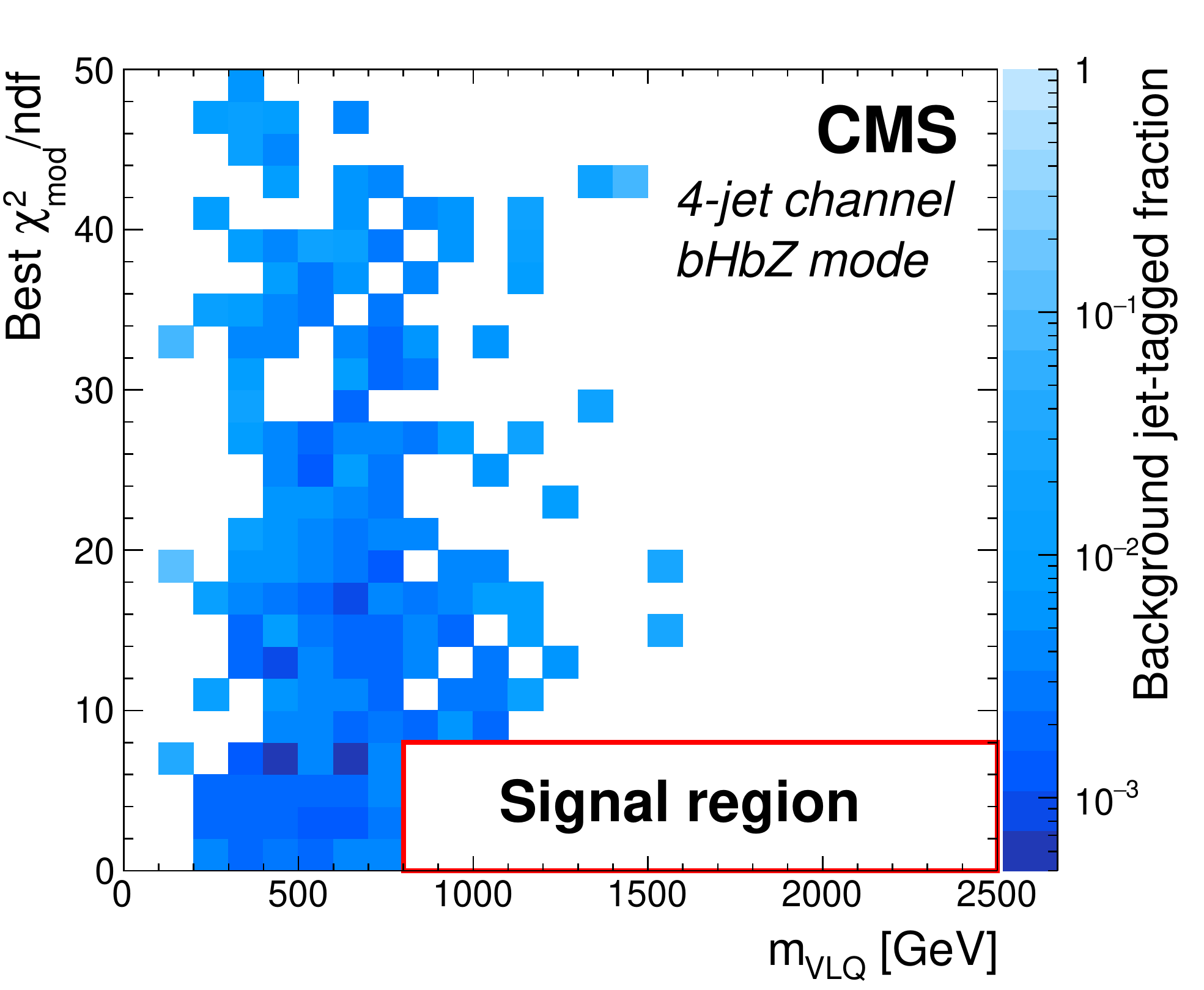}\hfill
\includegraphics[width=0.32\textwidth]{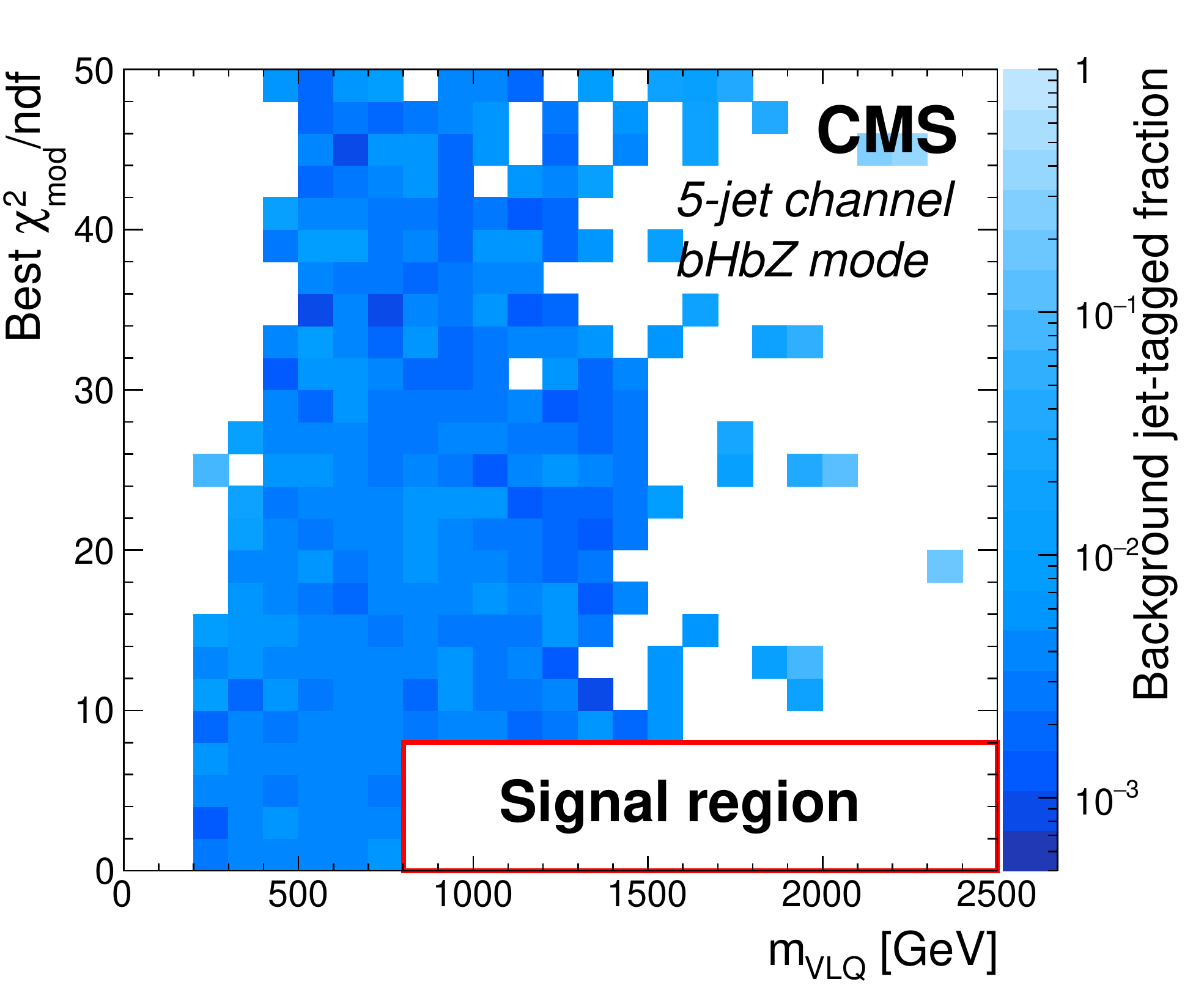}\hfill
\includegraphics[width=0.32\textwidth]{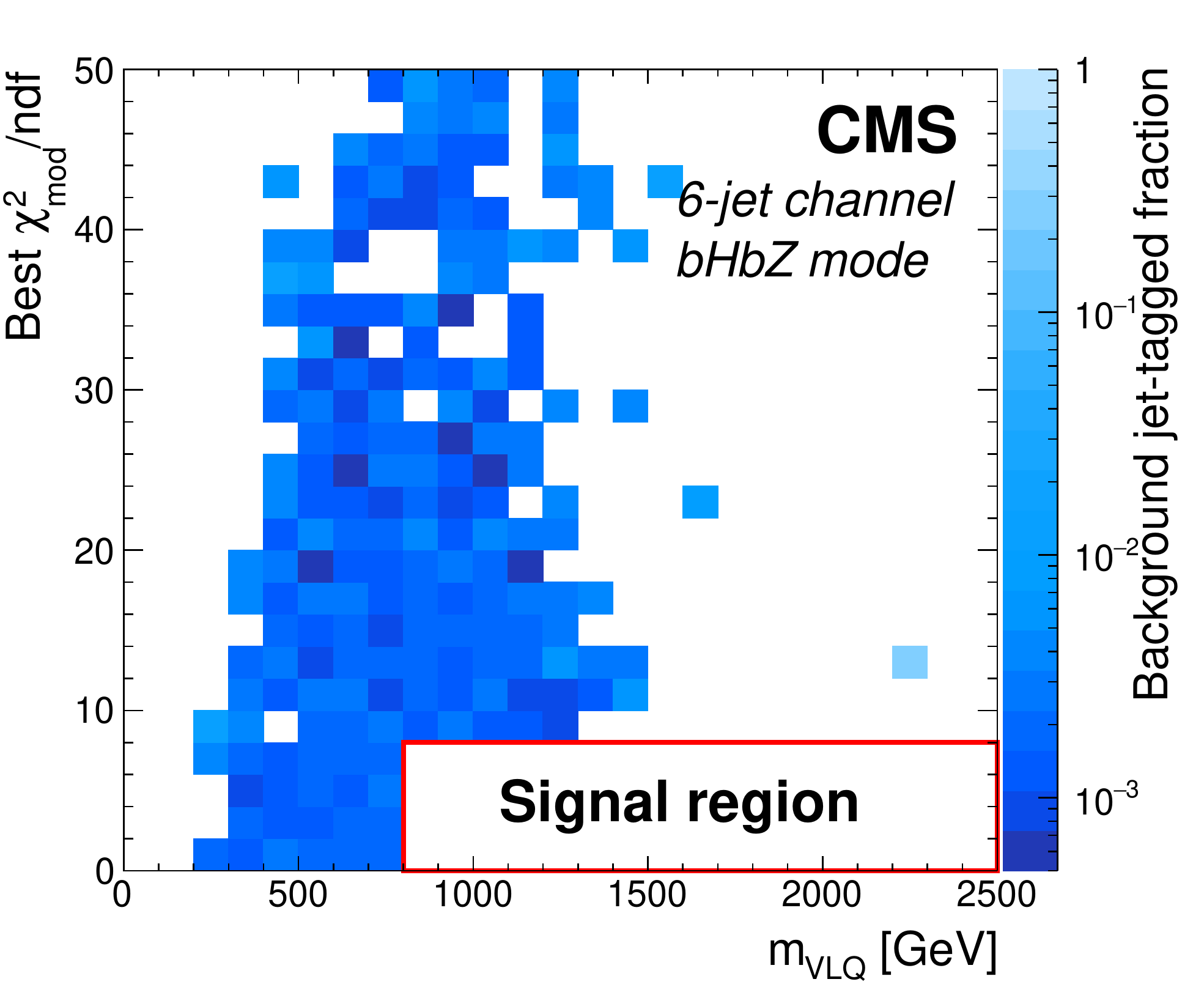}\hfill
\includegraphics[width=0.32\textwidth]{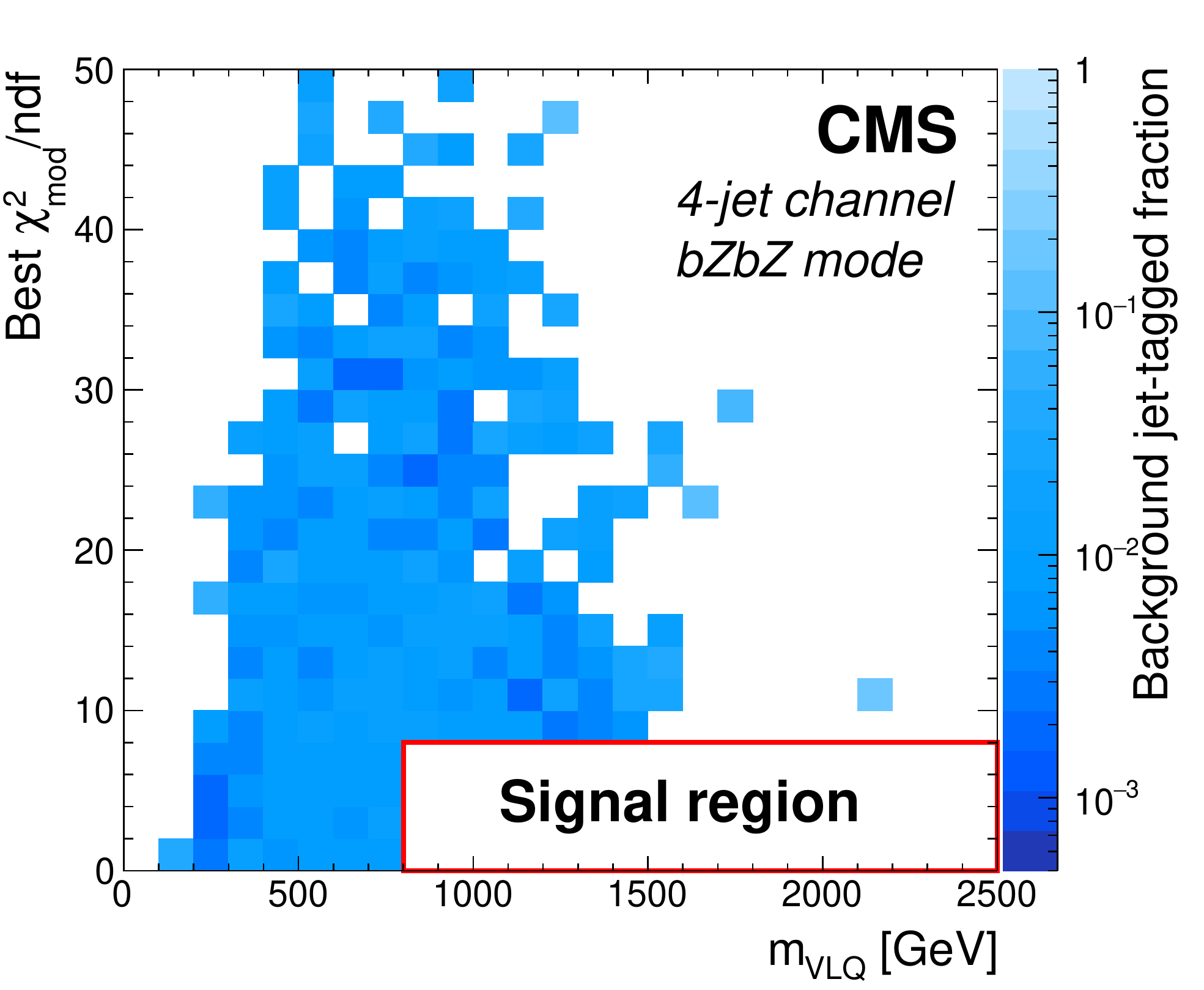}\hfill
\includegraphics[width=0.32\textwidth]{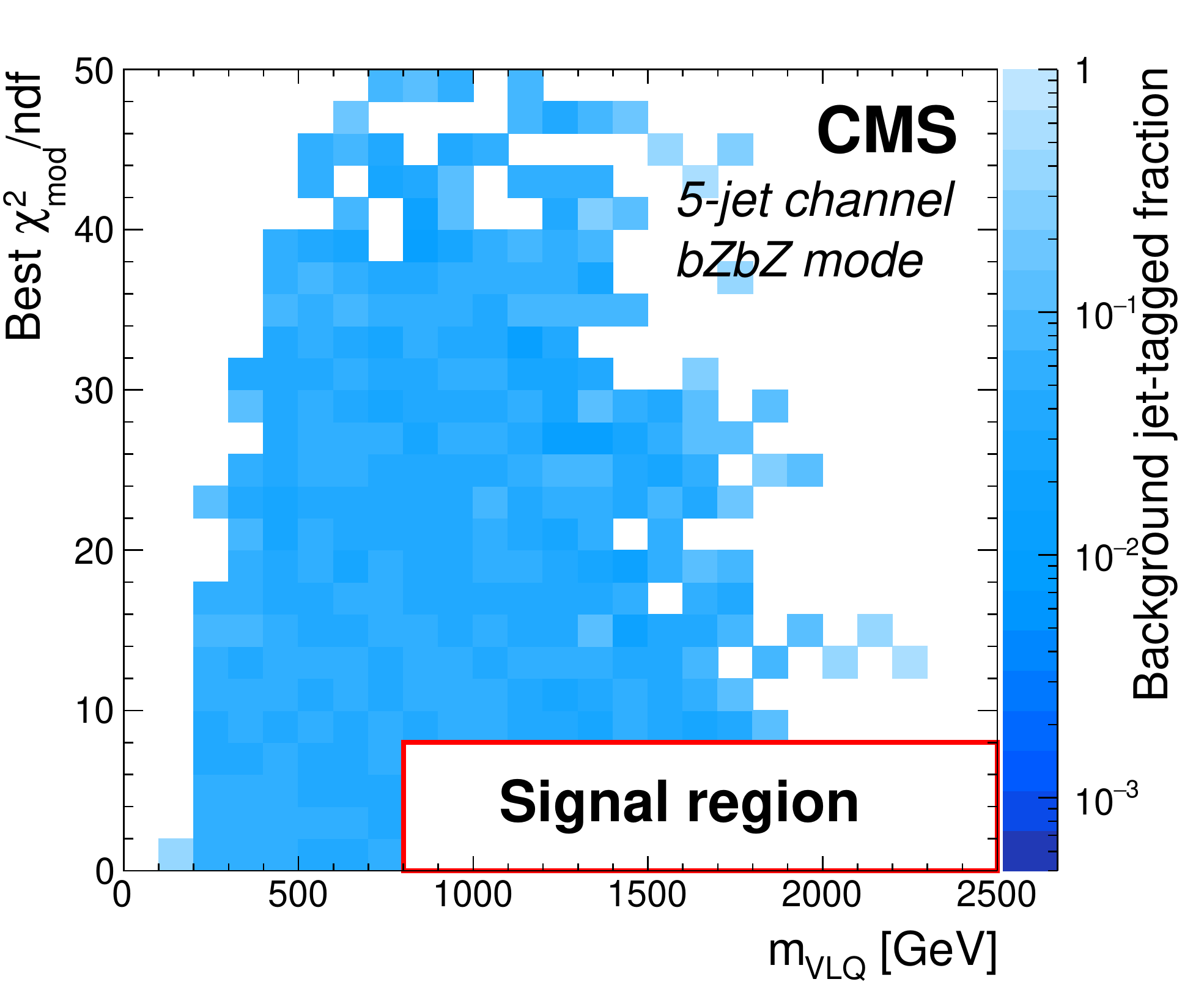}\hfill
\includegraphics[width=0.32\textwidth]{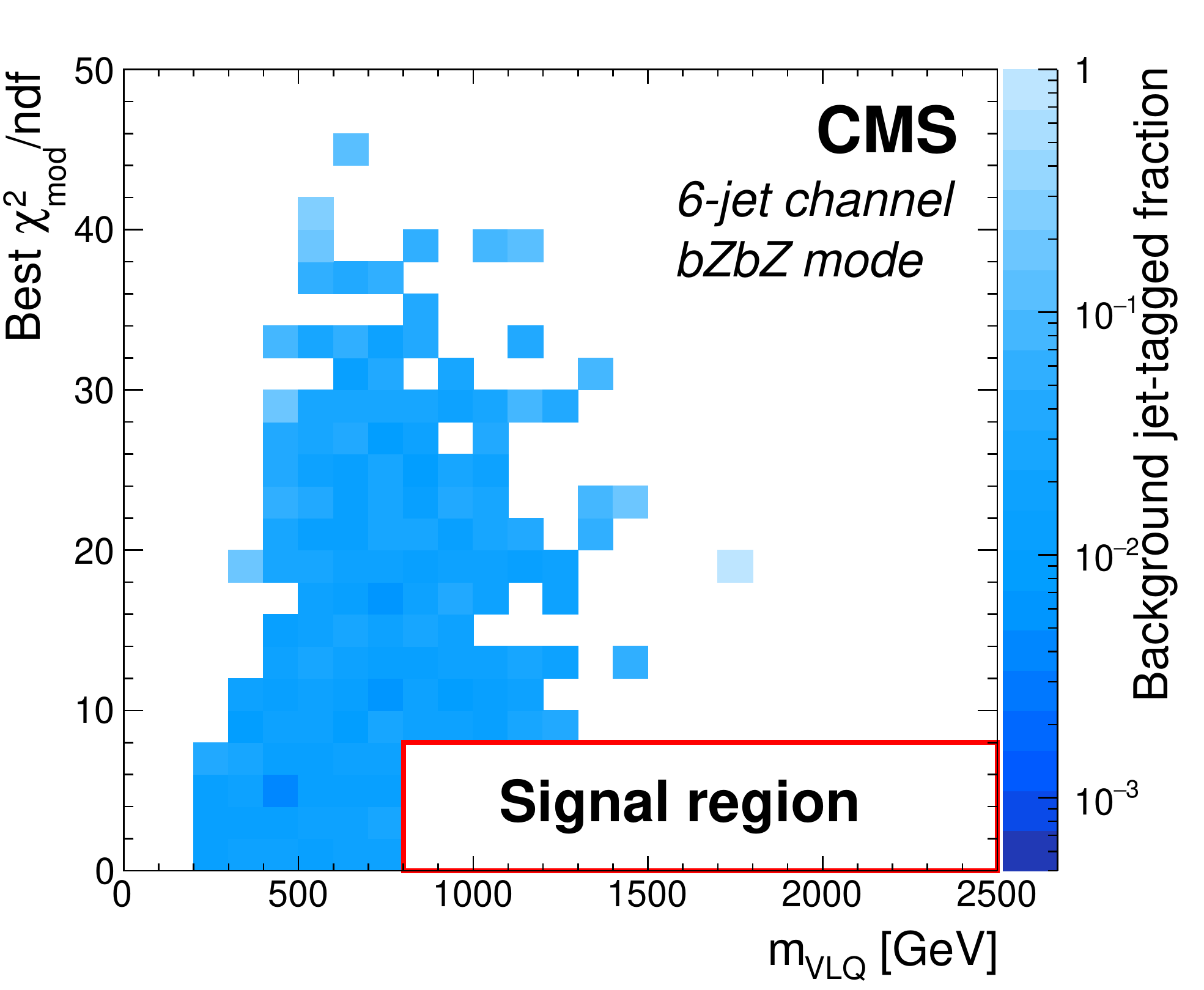}\hfill
\caption{Dependence of the BJTF on \mvlq and the best \chimodndf in data events, for 4-jet (left column), 5-jet (center column), and 6-jet (right column) multiplicities, and for the \bhbh (upper row), \bhbz (middle row), and \bzbz (lower row) event modes. The red box indicates the signal region, which is excluded from these plots.}
\label{fig:rf_temp_bg}
\end{figure*}

\begin{figure*}[hbtp]
\centering
\includegraphics[width=0.32\textwidth]{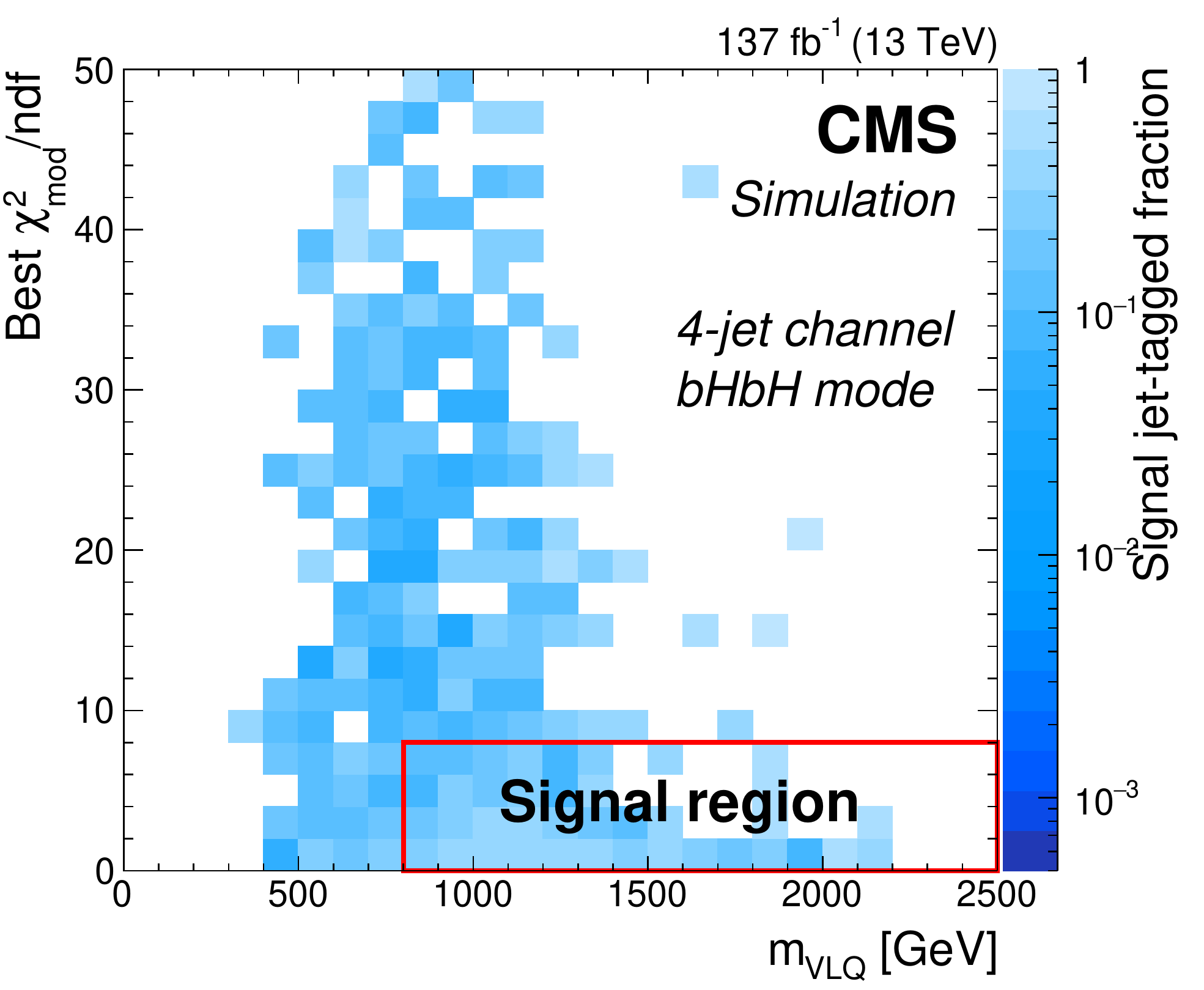}\hfill
\includegraphics[width=0.32\textwidth]{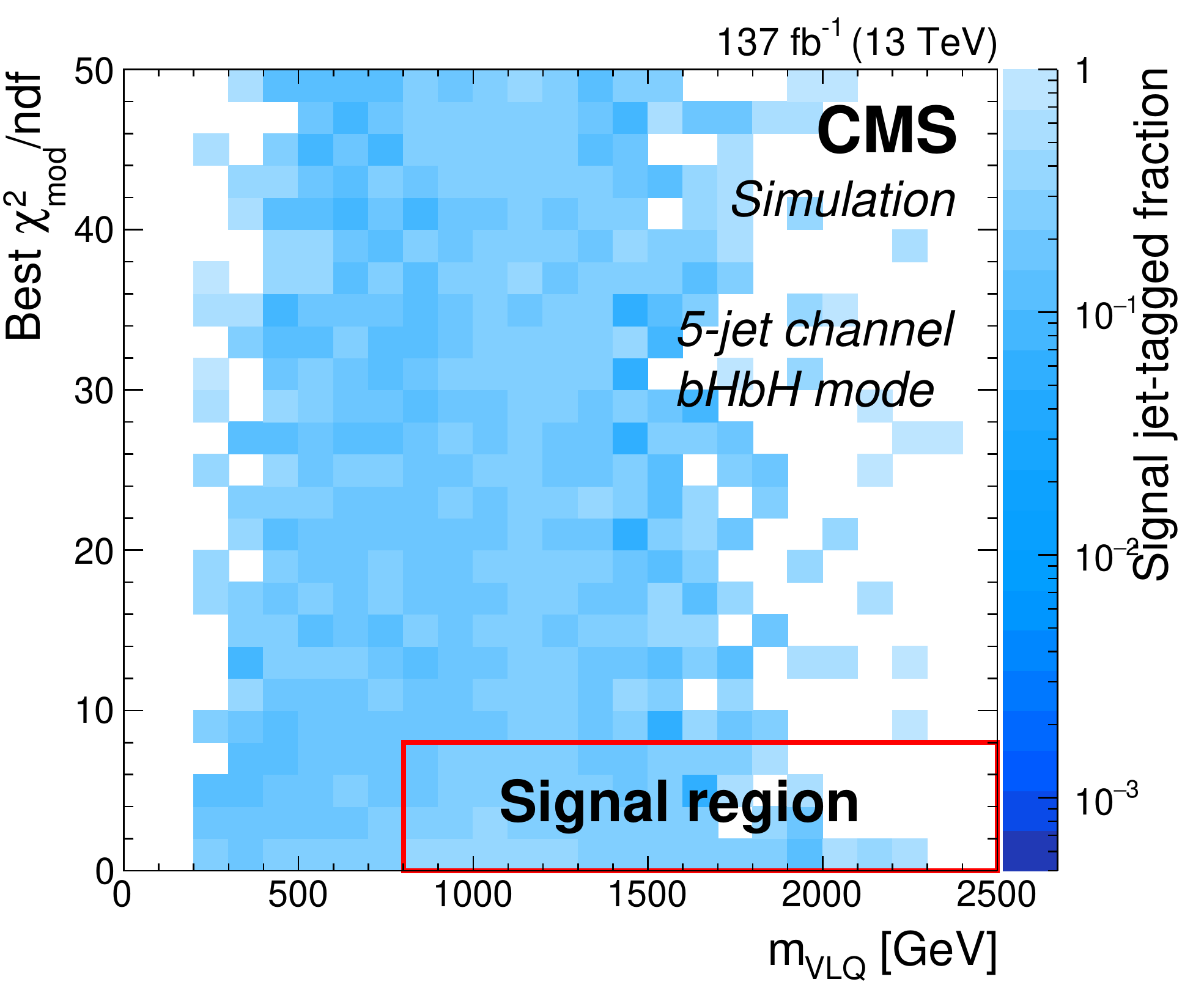}\hfill
\includegraphics[width=0.32\textwidth]{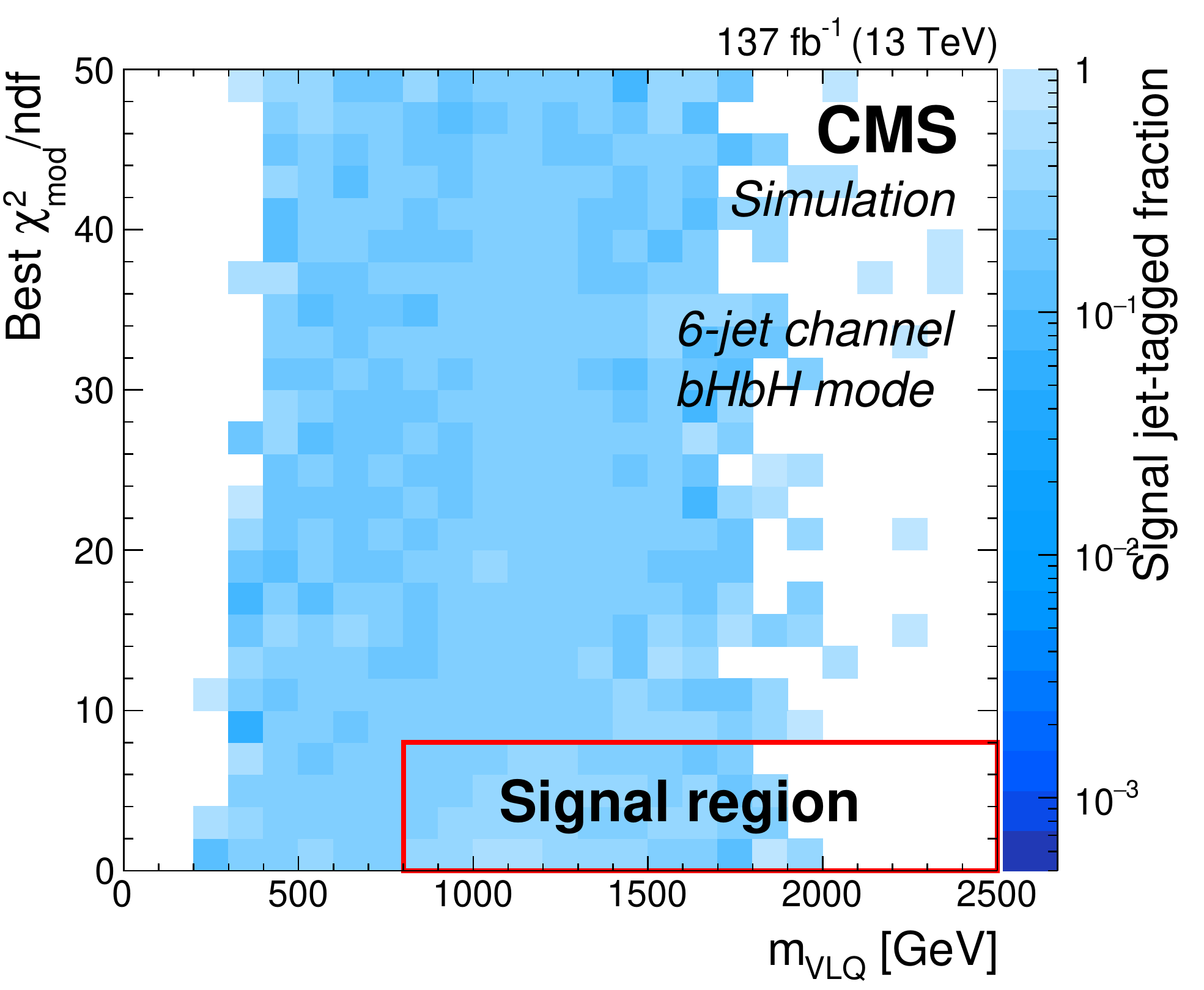}\hfill
\includegraphics[width=0.32\textwidth]{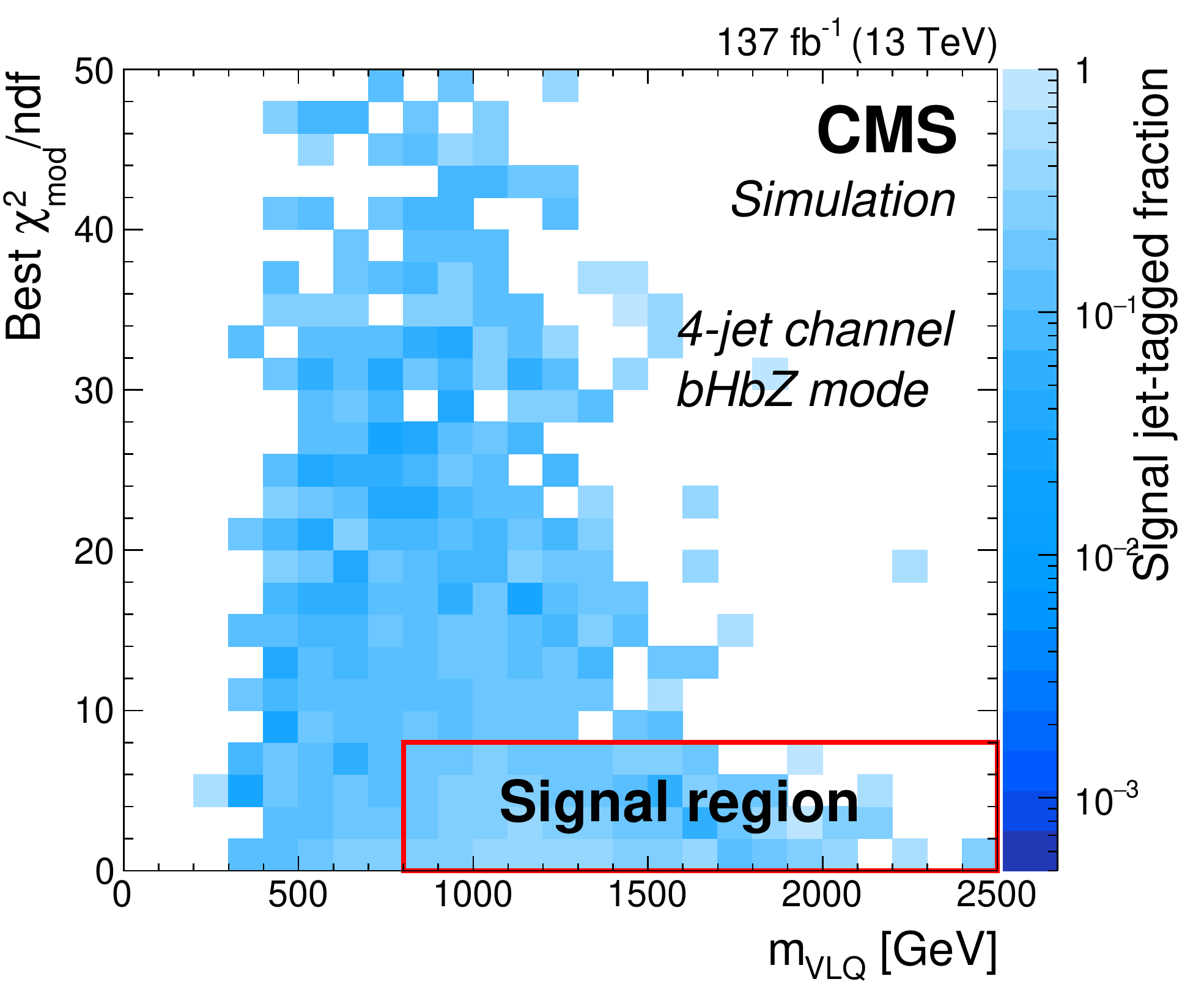}\hfill
\includegraphics[width=0.32\textwidth]{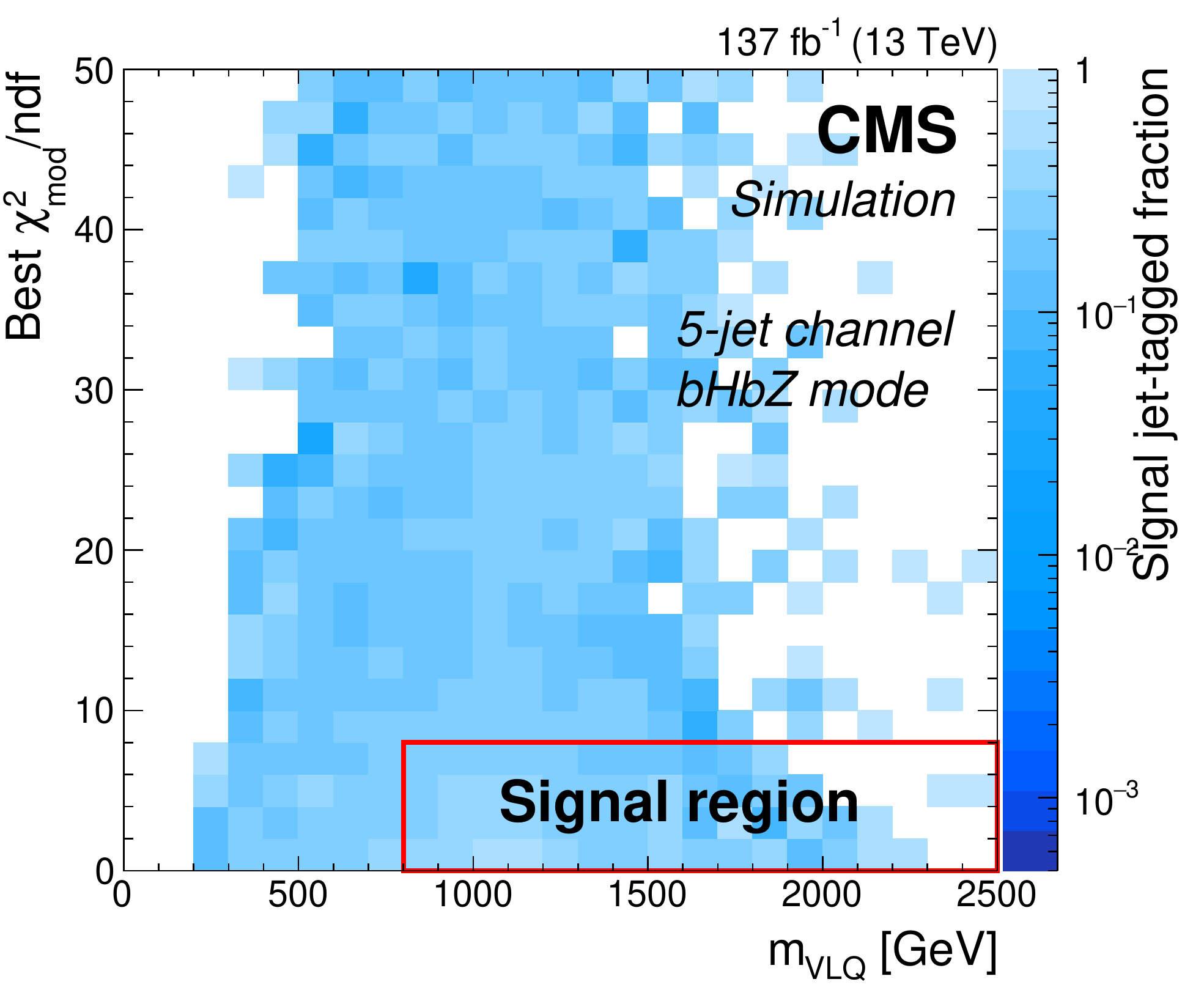}\hfill
\includegraphics[width=0.32\textwidth]{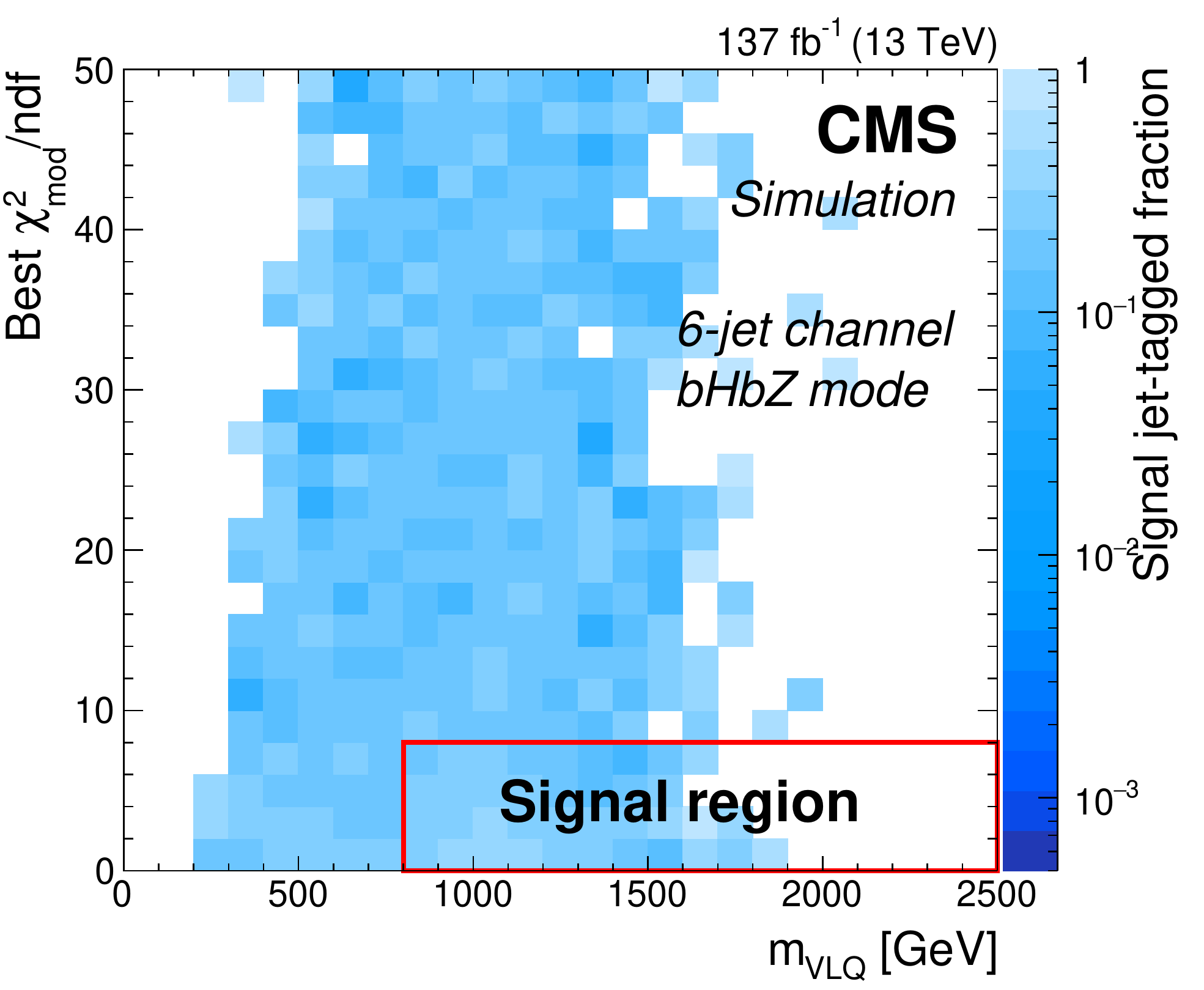}\hfill
\includegraphics[width=0.32\textwidth]{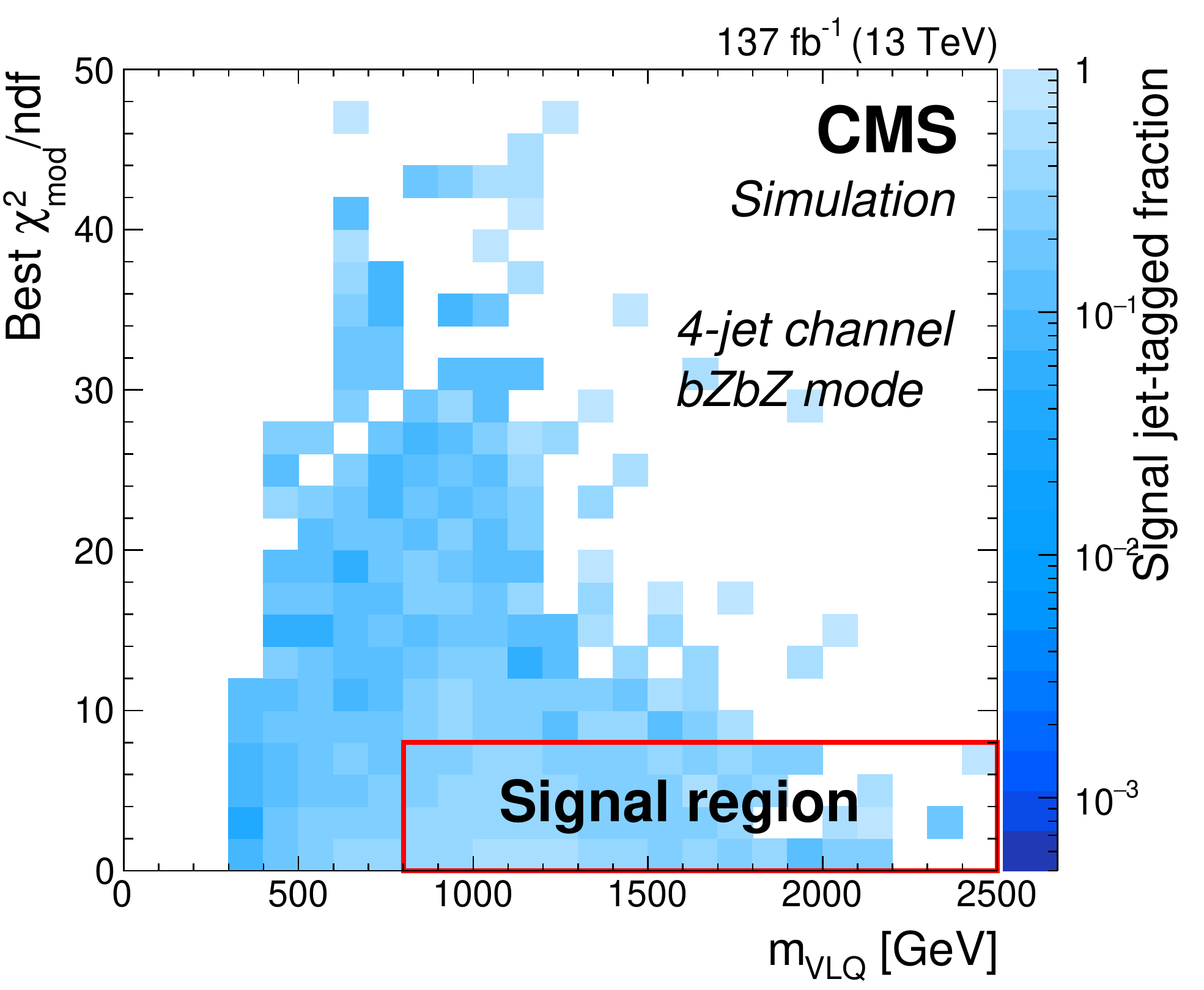}\hfill
\includegraphics[width=0.32\textwidth]{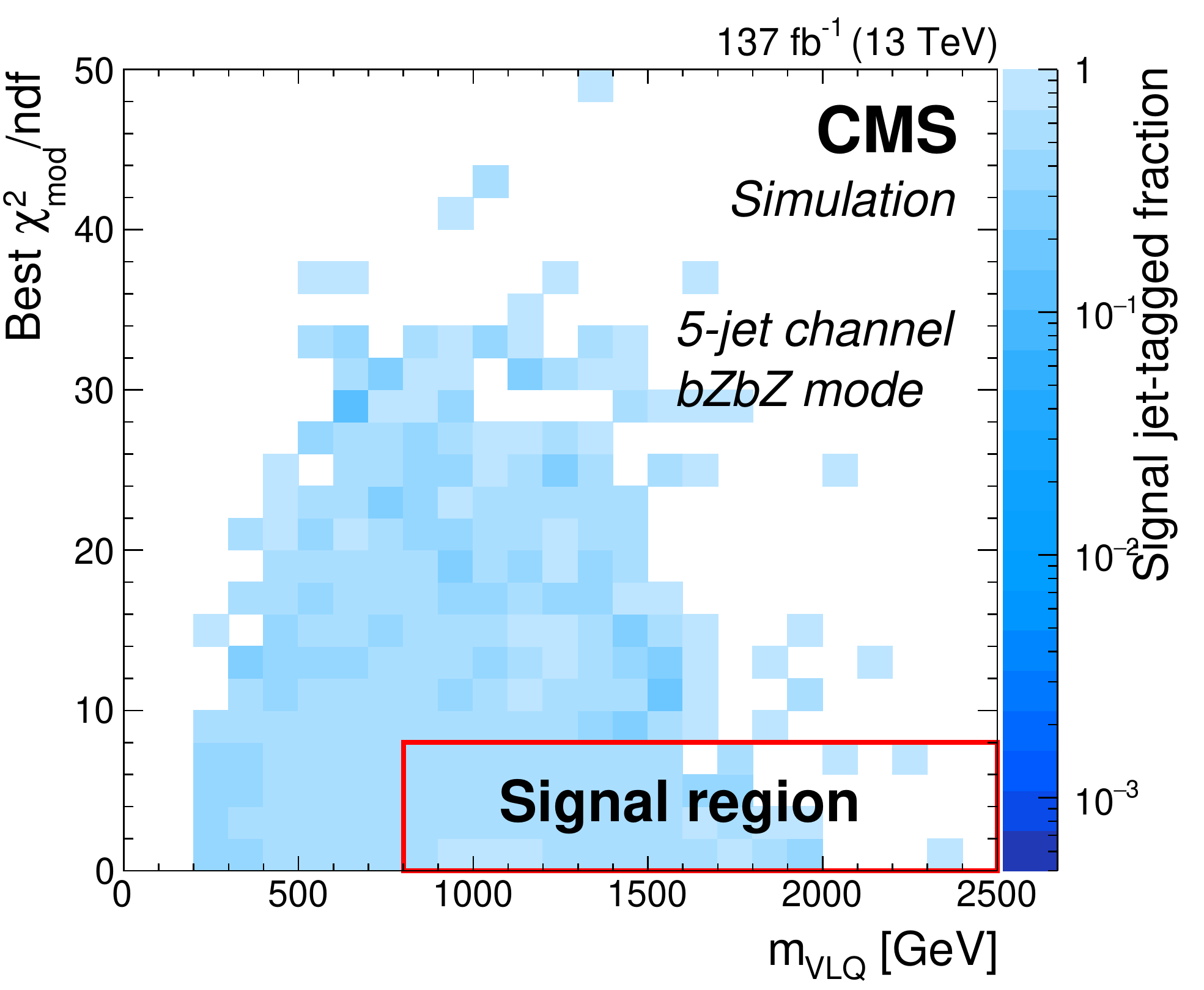}\hfill
\includegraphics[width=0.32\textwidth]{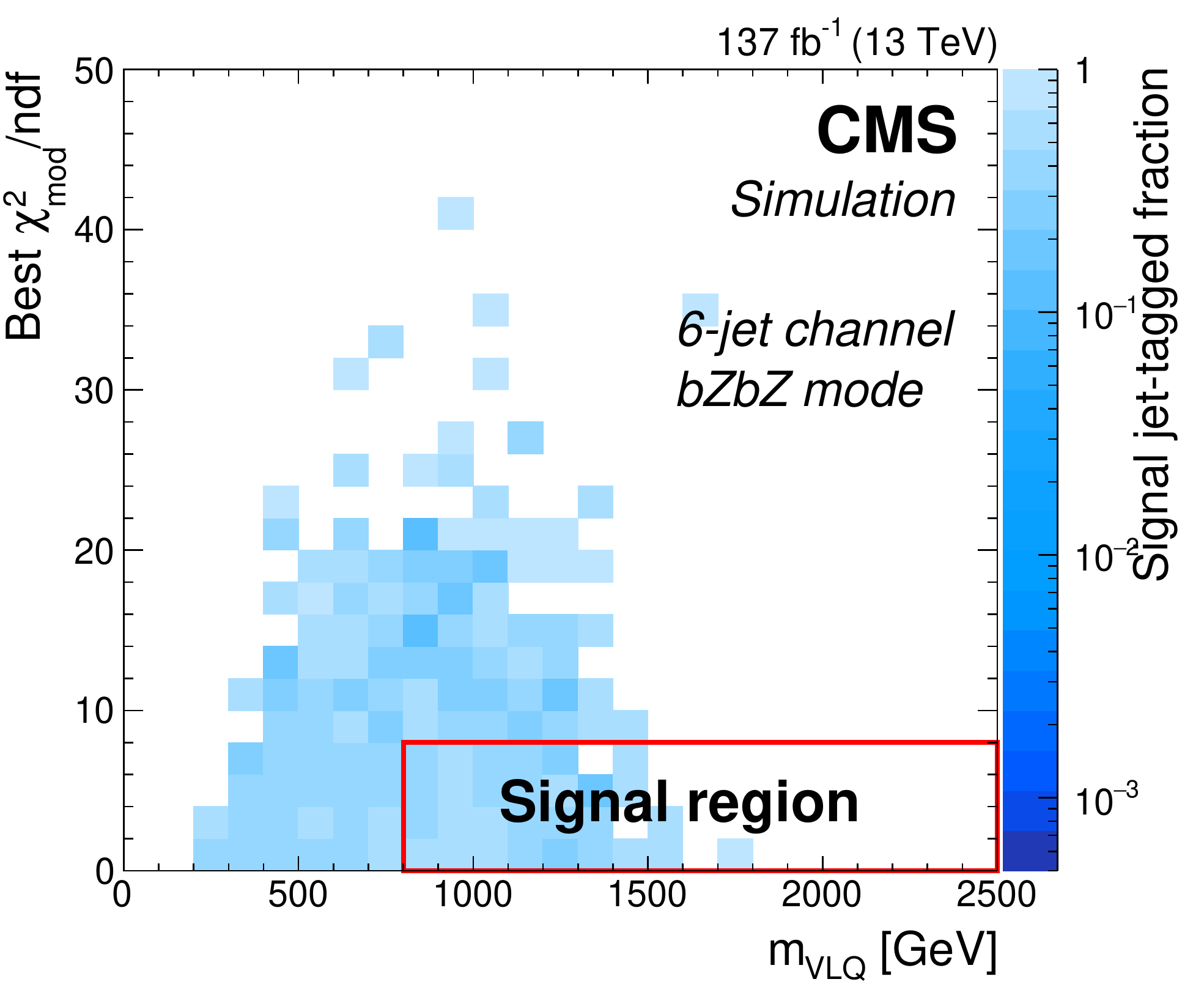}\hfill
\caption{Dependence of the BJTF on \mvlq and the best \chimodndf in simulated VLQ signal events with $m_{\PB} = 1200\GeV$, for 4-jet (left column), 5-jet (center column), and 6-jet (right column) multiplicities, and for the \bhbh (upper row), \bhbz (middle row), and \bzbz (lower row) event modes. The red box indicates the signal region.}
\label{fig:rf_temp_sig}
\end{figure*}

Figure~\ref{fig:rf_temp_bg} shows the two-dimensional dependence of the BJTF in data on \mvlq and \chimodndf, and Fig.~\ref{fig:rf_temp_sig} shows the corresponding distributions for simulated VLQ signal events with a generated VLQ mass of 1200\GeV. The signal region is indicated in these plots by the red rectangle, and is excluded from the data plots.

The final estimate of the number of background events $n_{\mathrm{b}}$ as a function of VLQ mass $m$ is given by the following expression:
\begin{linenomath}
\begin{equation}
\label{eq:background}
n_{\mathrm{b}}(m) = n(m) \varepsilon_0 \frac{\varepsilon(m)}{(\int_{500\GeV}^{800\GeV}\varepsilon(m')\,\rd m')/(300\GeV)},
\end{equation}
\end{linenomath}
where $n(m)$ is the number of candidates as a function of \mvlq before jet tagging for candidates passing the $\chimod$ selection shown in Fig.~\ref{fig:bckg}, $\varepsilon_0$ is the BJTF at low VLQ mass as shown in Table~\ref{table:redfac}, and the last factor accounts for the potential mass dependence of the BJTF, with $\varepsilon(m)$ the distribution of the BJTF as a function of mass, as shown in Fig.~\ref{fig:rf_mass_depen}; the factor of 300\GeV is to normalize over the range considered. The final estimate is also validated by comparison with the region $8 < \chimodndf < 12$ and with simulated events.

\section{Systematic uncertainties}

We consider two types of systematic uncertainties, those that are common to all event modes and jet multiplicities, and those that depend on the particular channel. The uncertainties in the first category are listed in Table~\ref{table:systematic_common}; these are the integrated luminosity, trigger efficiency, and the choice of fit function for the dependence of the BJTF on VLQ mass. The integrated luminosities of the 2016, 2017, and 2018 data-taking periods are individually known with uncertainties in the range 2.3--2.5\%~\cite{CMS-PAS-LUM-17-001,CMS-PAS-LUM-17-004,CMS-PAS-LUM-18-002}, while the total 2016--2018 integrated luminosity has an uncertainty of 1.8\%, the improvement in precision reflecting the (uncorrelated) time evolution of some systematic effects. The uncertainty associated with the choice of fit function for the \mvlq dependence of the BJTF is determined by finding the average difference between the fit functions used for each mode and multiplicity combination, and using the maximum one as the common uncertainty for all modes and multiplicities. Table~\ref{table:systematic_common} also indicates whether an uncertainty affects the signal efficiency or the background estimate.

\begin{table}[hbtp!]
\centering
\topcaption{Systematic uncertainties common to all three event modes and all three jet multiplicities. All uncertainties listed here are rate uncertainties, meaning they affect only the normalization.}
\label{table:systematic_common} 
\vspace*{-0.2cm}
\begin{scotch}{lcc}
Type                   & Signal/Background & Uncertainty \\
\hline
Integrated luminosity  & Signal                   & 1.8\% \\
Trigger efficiency     & Signal                   & 0.02\% \\
Choice of fit function & Background               & 4.9\% \\
\end{scotch}
\end{table}

The uncertainties that depend on event mode and jet multiplicity are those due to the background estimation, jet tag scale factors, jet energy resolution and scale, choice of PDF, and pileup. 

There are several sources of uncertainty in the background estimation, corresponding to the three terms in Eq.~\ref{eq:background}. The first uncertainty arises from the exponential fit to the distribution $n(m)$, the number of events as a function of mass before jet tagging is applied. The uncertainties in the fit parameters $p_0$ and $p_1$ are used to determine the uncertainty in the fit value for a given mass. The second is the uncertainty in the BJTF determined for low-mass VLQ candidates, $\varepsilon_0$, as shown in Table~\ref{table:redfac}. Finally, the third uncertainty arises from the third term, to account for a potential mass dependence of the BJTF, and is obtained from the uncertainties in the fit parameters, as in the first case.

The efficiencies for jet tagging are measured in simulated events and then corrected to data events using a data-to-simulation scale factor. The uncertainty in this scale factor is propagated to the signal reconstruction efficiency by varying the scale factors within their uncertainties~\cite{Sirunyan:2017ezt}. The uncertainties due to the scale factors for jet energy scale and resolution~\cite{Khachatryan:2016kdb} are determined similarly. The uncertainty due to the choice of PDF weighting is calculated from a set of 100 weights selected from the NNPDF3.0 distribution, following the prescription in Ref.~\cite{Butterworth:2015oua}. The pileup uncertainties are due to a 4.6\% systematic uncertainty in the $\Pp\Pp$ inelastic cross section.

Table~\ref{table:systematic_mode} summarizes these uncertainties, and indicates whether they affect the signal efficiency or the background estimate, and whether the uncertainty affects the overall rate or the shape of the mass distribution. For the PDF systematic uncertainties, the values refer only to the event acceptance rate. There is in addition an uncertainty in the VLQ pair production cross section. This uncertainty depends only weakly on VLQ mass and an average value of 6\% is used for all masses~\cite{Sirunyan:2018omb}.

\begin{table*}[htbp!]
  \centering
  \topcaption{Table of systematic uncertainties for each event mode and jet multiplicity. The reported values indicate the uncertainty in the event yield in a $\pm$75\GeV window about the signal peak for a generated signal mass $m_{\PB} = 1600\GeV$.}
  \begin{scotch}{lccccc}
 Type                  & Signal/Background & Rate/Shape & 4 jets & 5 jets  & 6 jets \\
\hline \\[-1.5ex]
\multicolumn{6}{c}{\textit{\bhbh event mode}} \\
[\cmsTabSkip]
 Background fit $p_0$  & Background        & Shape      & 59\% & 14\% & 13\%  \\
 Background fit $p_1$  & Background        & Shape      & 78\% & 18\% & 16\%  \\
 BJTF $m$ dependence $p_0$ & Background    & Shape      & 1.3\%  & 5.9\%  & 4.5\%  \\
 BJTF $m$ dependence $p_1$ & Background    & Shape      & 19\% & 25\% & 17\%  \\
 Low-mass BJTF         & Background        & Rate       & 34\% & 9.7\%  & 11\%  \\
 Jet tag scale factors & Signal            & Shape      & 16\% & 15\% & 17\%  \\
 Jet energy scale      & Signal            & Shape      & 4.0\% & 5.3\%   & 6.4\%  \\
 Jet energy resolution & Signal            & Shape      & 2.4\%  & 1.5\%  & 1.6\%   \\
 Pileup                & Signal            & Shape      & 28\% & 28\% & 27\%  \\
 PDF                   & Signal            & Rate       & 1.5\%  & 1.5\%  & 1.5\%  \\
[\cmsTabSkip]
\multicolumn{6}{c}{\textit{\bhbz event mode}} \\
[\cmsTabSkip]
 Background fit $p_0$  & Background        & Shape      & 21\% & 12\% & 10\%  \\
 Background fit $p_1$  & Background        & Shape      & 21\% & 14\% & 12\%  \\
 BJTF $m$ dependence $p_0$ & Background    & Shape      & 2.1\%  & 7.7\%  & 3.5\%  \\
 BJTF $m$ dependence $p_1$ & Background    & Shape      & 21\% & 30\% & 27\%  \\
 Low-mass BJTF       & Background        & Rate       & 22\% & 7.7\%  & 11\%  \\
 Jet tag scale factors & Signal            & Shape      & 15\% & 13\% & 17\%  \\
 Jet energy scale      & Signal            & Shape      & 4.9\%  & 5.7\%  & 5.1\%  \\
 Jet energy resolution & Signal            & Shape      & 1.8\%  & 2.7\%  & 3.2\%  \\
 Pileup                & Signal            & Shape      & 33\% & 28\% & 21\%  \\
 PDF                   & Signal            & Rate       & 1.6\% & 1.5\% & 1.5\%  \\
[\cmsTabSkip]
\multicolumn{6}{c}{\textit{\bzbz event mode}} \\
[\cmsTabSkip]
 Background fit $p_0$  & Background        & Shape      & 26\% & 17\% & 24\%  \\
 Background fit $p_1$  & Background        & Shape      & 28\% & 21\% & 32\%  \\
 BJTF $m$ dependence $p_0$ & Background    & Shape      & 3.7\%  & 0.6\%  & 11\% \\
 BJTF $m$ dependence $p_1$ & Background    & Shape      & 15\% & 7.8\%  & 21\%  \\
 Low-mass BJTF       & Background        & Rate       & 16\% & 19\% & 25\%  \\
 Jet tag scale factors & Signal            & Shape      & 8.9\%  & 8.0\%  & 11\%  \\
 Jet energy scale      & Signal            & Shape      & 4.0\% & 2.9\% & 1.6\%  \\
 Jet energy resolution & Signal            & Shape      & 2.5\%  & 2.5\%  & 3.2\%  \\
 Pileup                & Signal            & Shape      & 28\%  & 28\%  & 10\%  \\
 PDF                   & Signal            & Rate       & 1.5\%  & 1.5\%  & 1.5\%  \\
  \end{scotch}
  \label{table:systematic_mode}
\end{table*}

\section{Results}
\label{sec:results}

\begin{figure*}[htbp]
\centering
\includegraphics[width=0.32\textwidth]{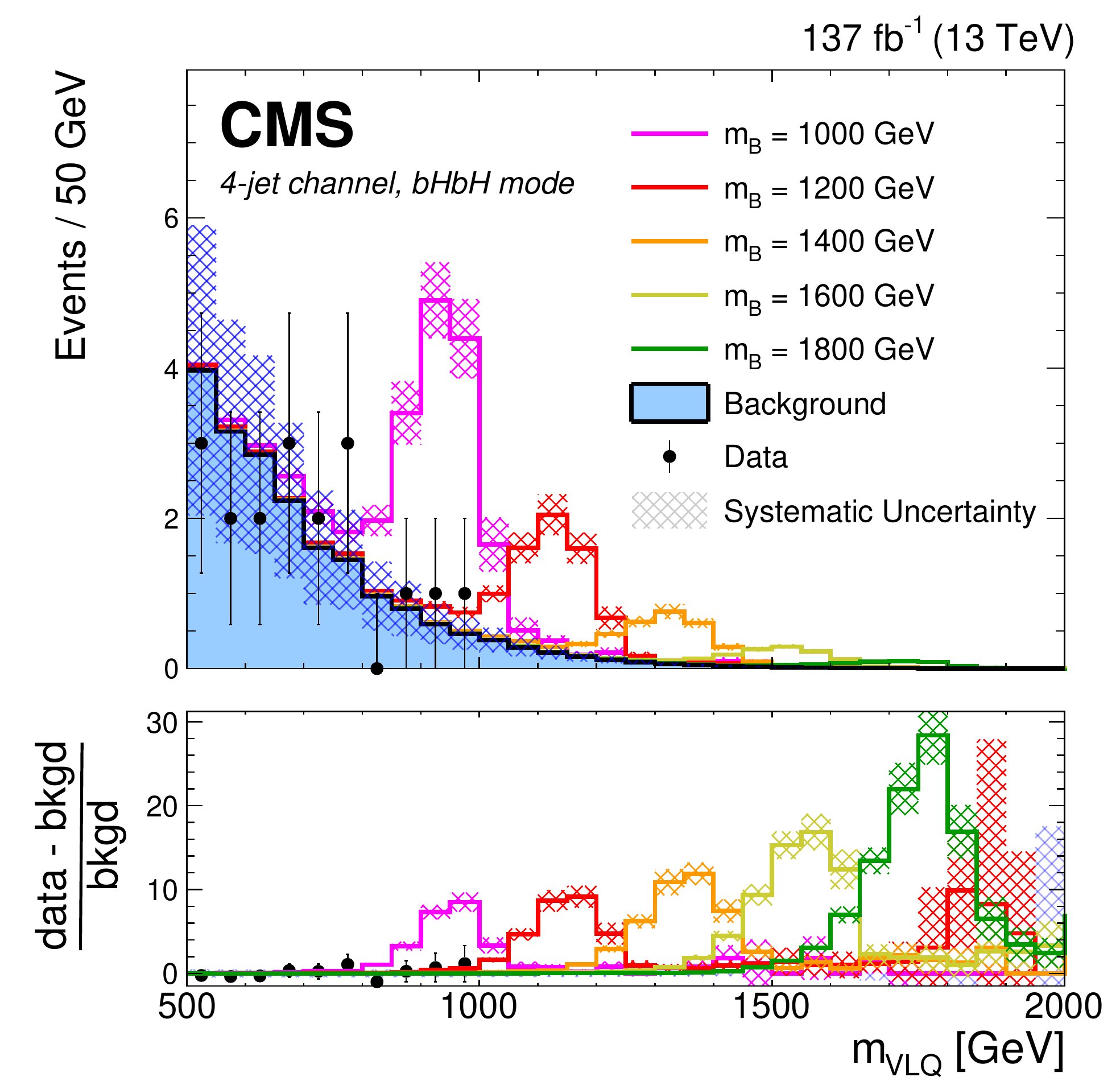}
\includegraphics[width=0.32\textwidth]{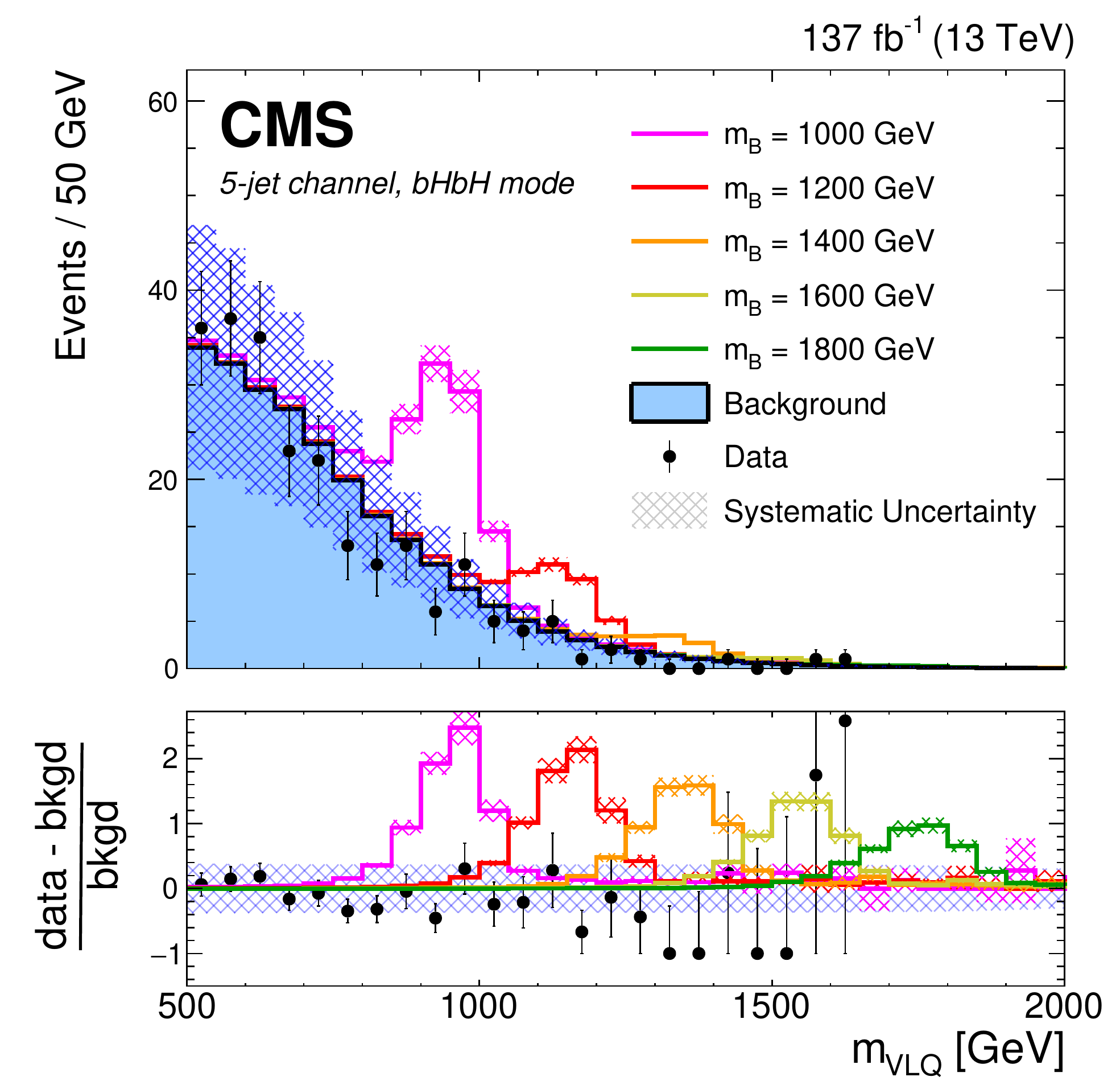}
\includegraphics[width=0.32\textwidth]{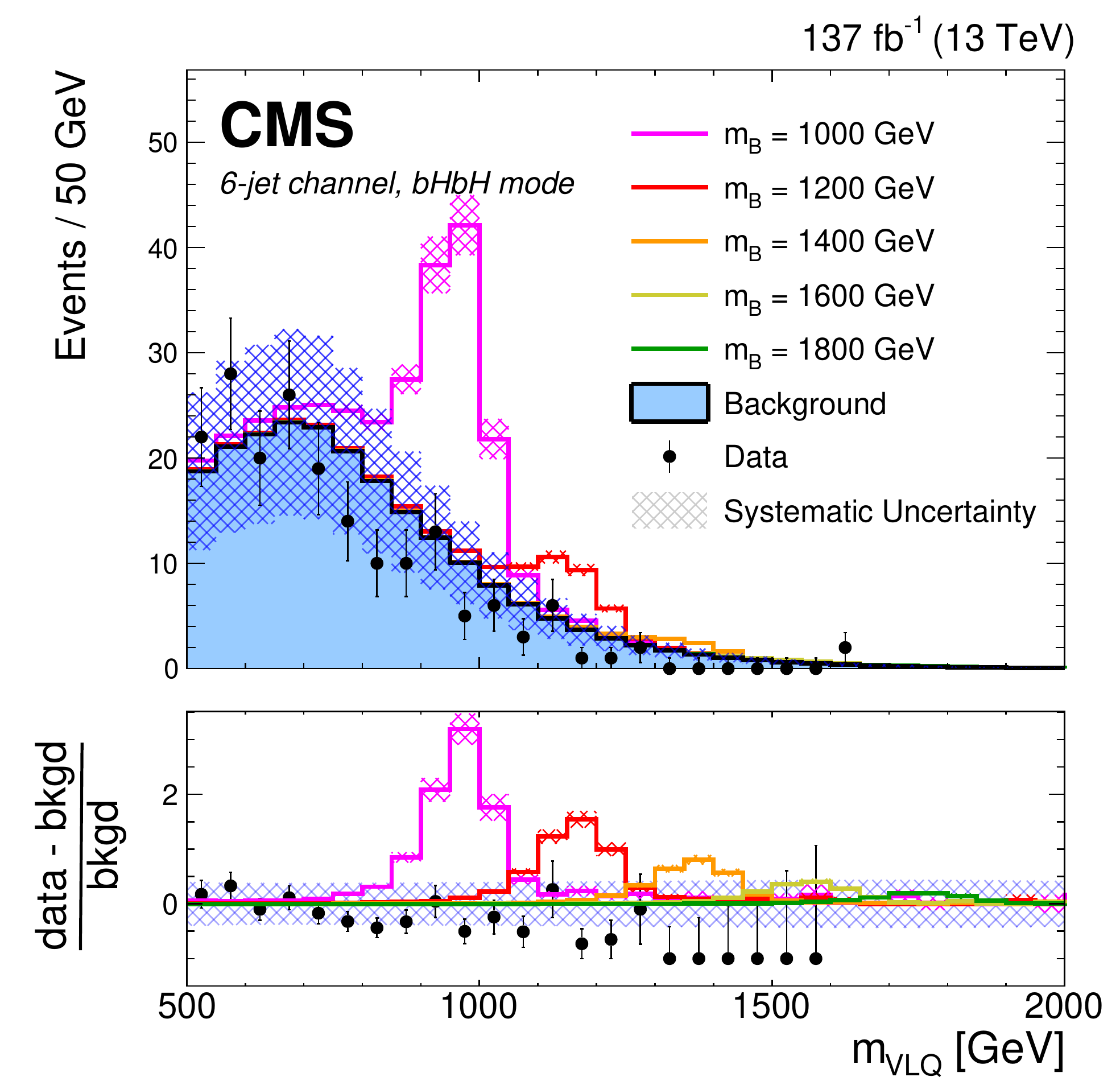}
\includegraphics[width=0.32\textwidth]{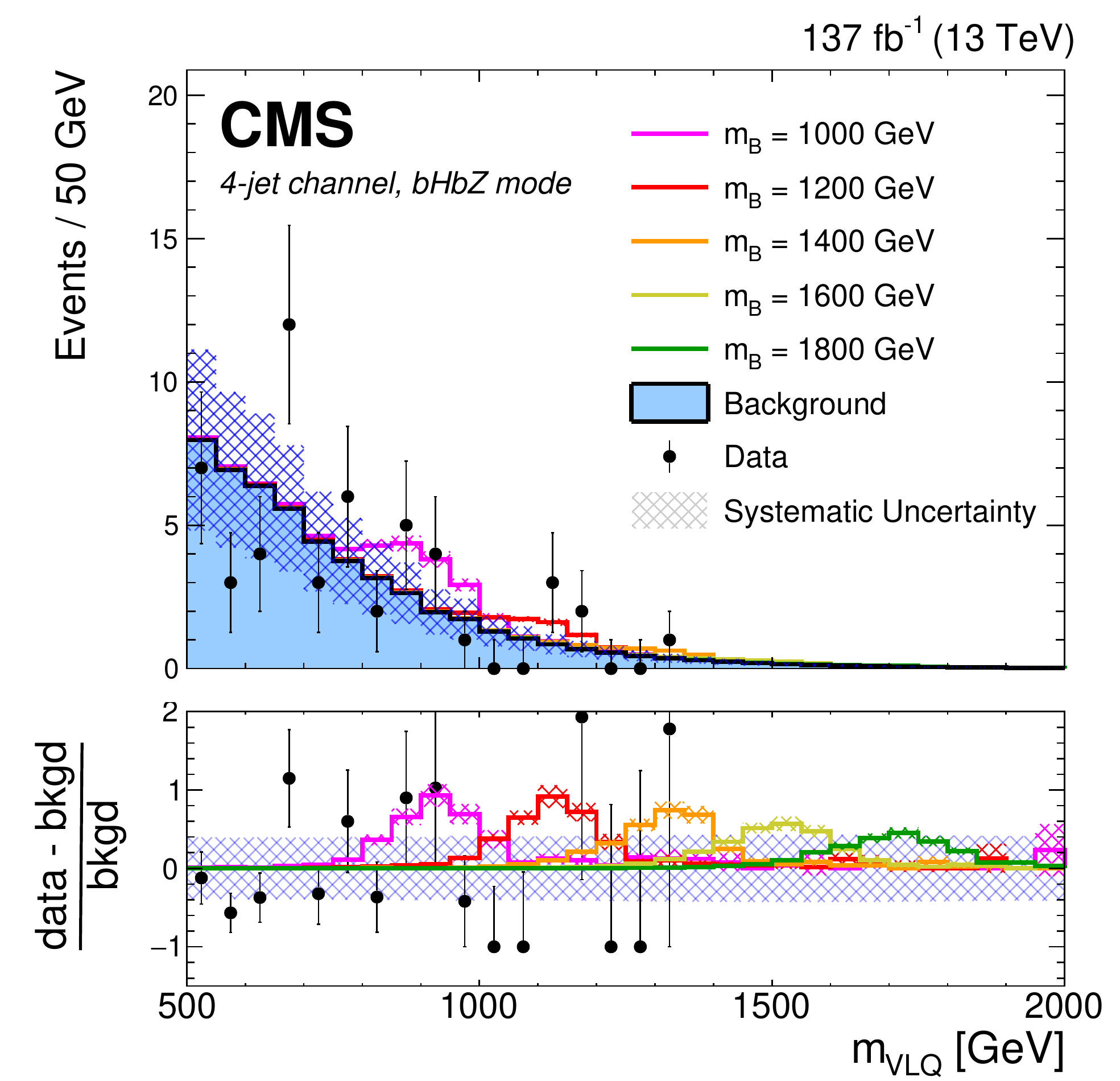}
\includegraphics[width=0.32\textwidth]{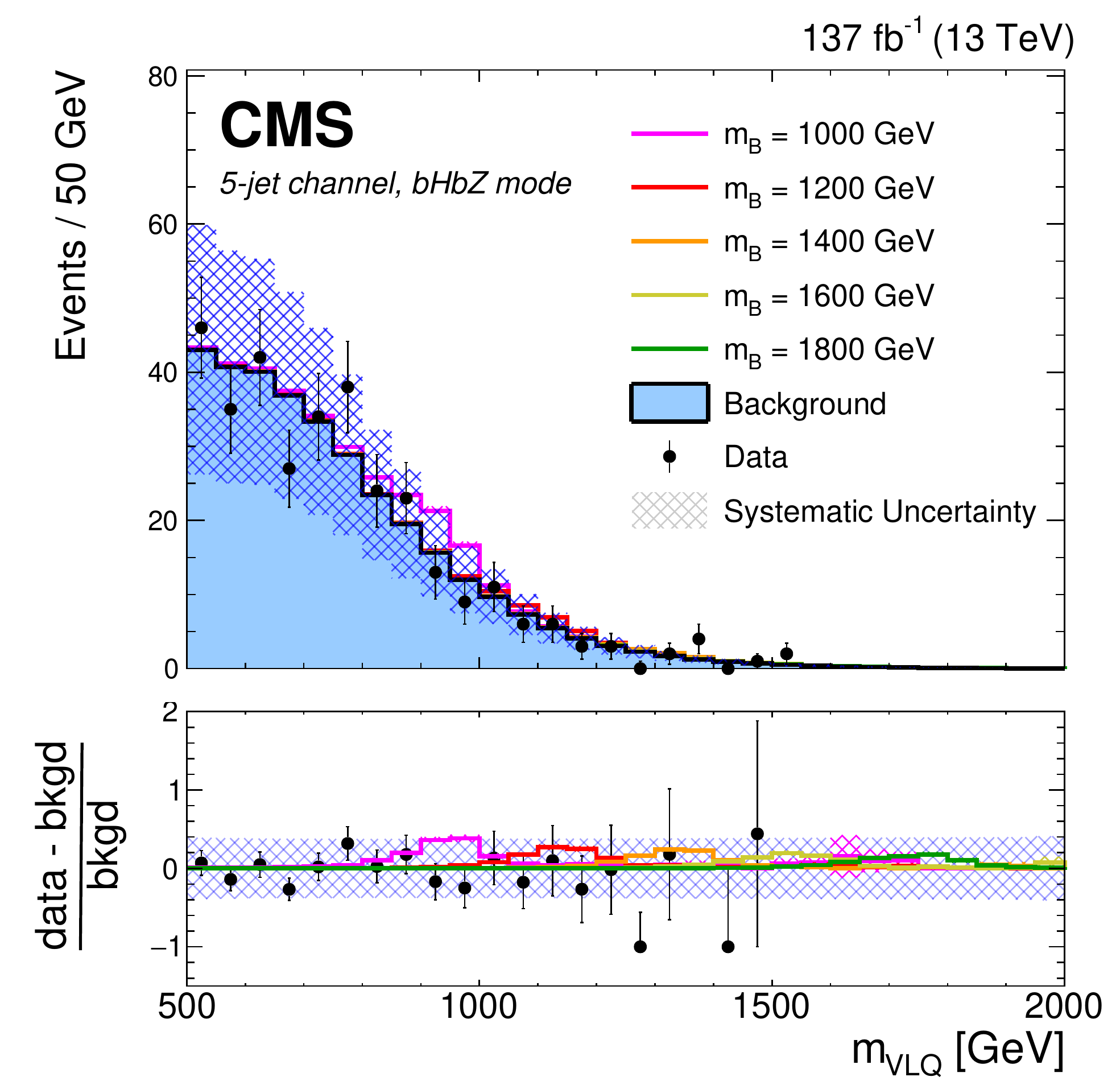}
\includegraphics[width=0.32\textwidth]{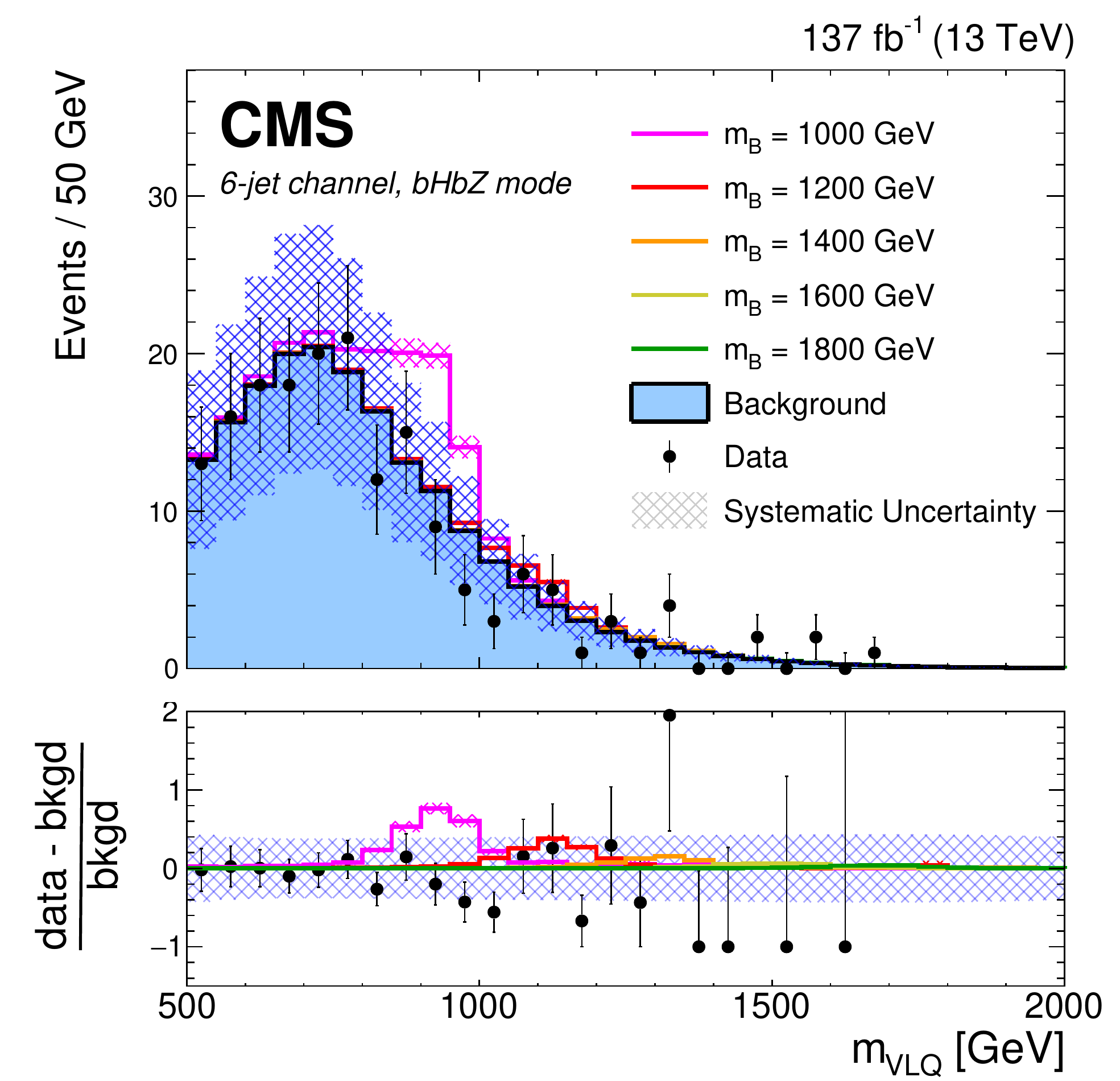}
\includegraphics[width=0.32\textwidth]{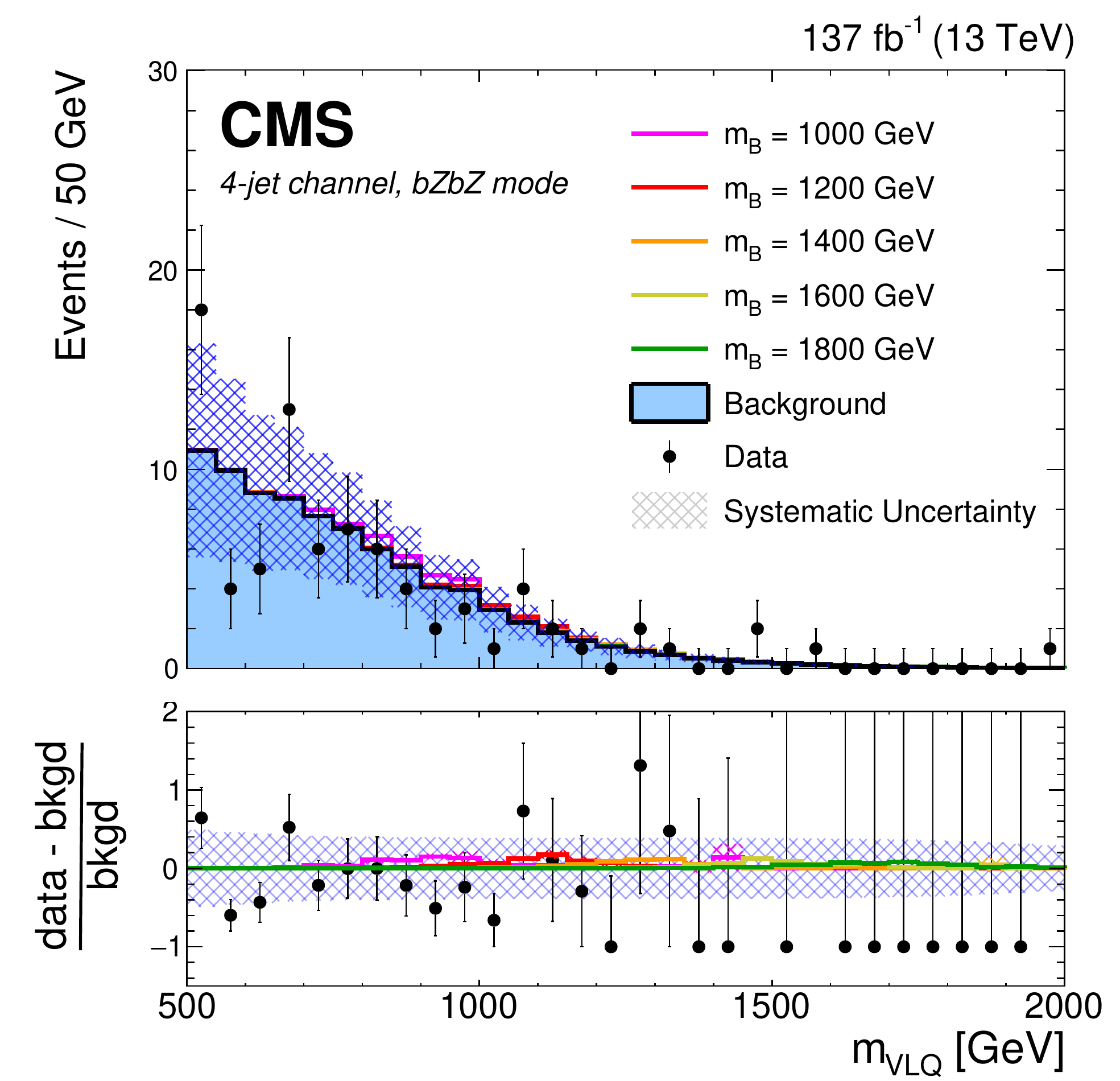}
\includegraphics[width=0.32\textwidth]{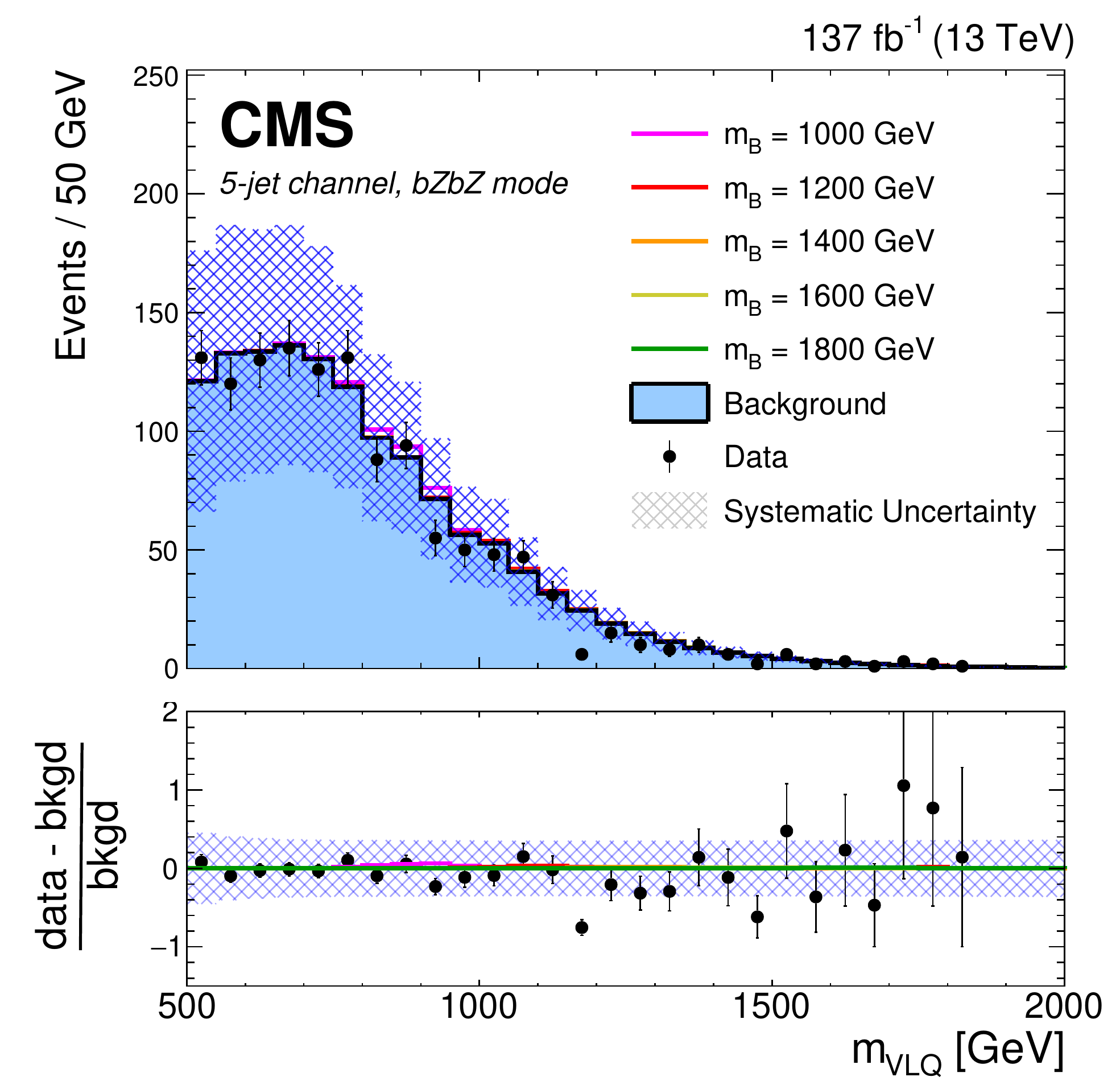}
\includegraphics[width=0.32\textwidth]{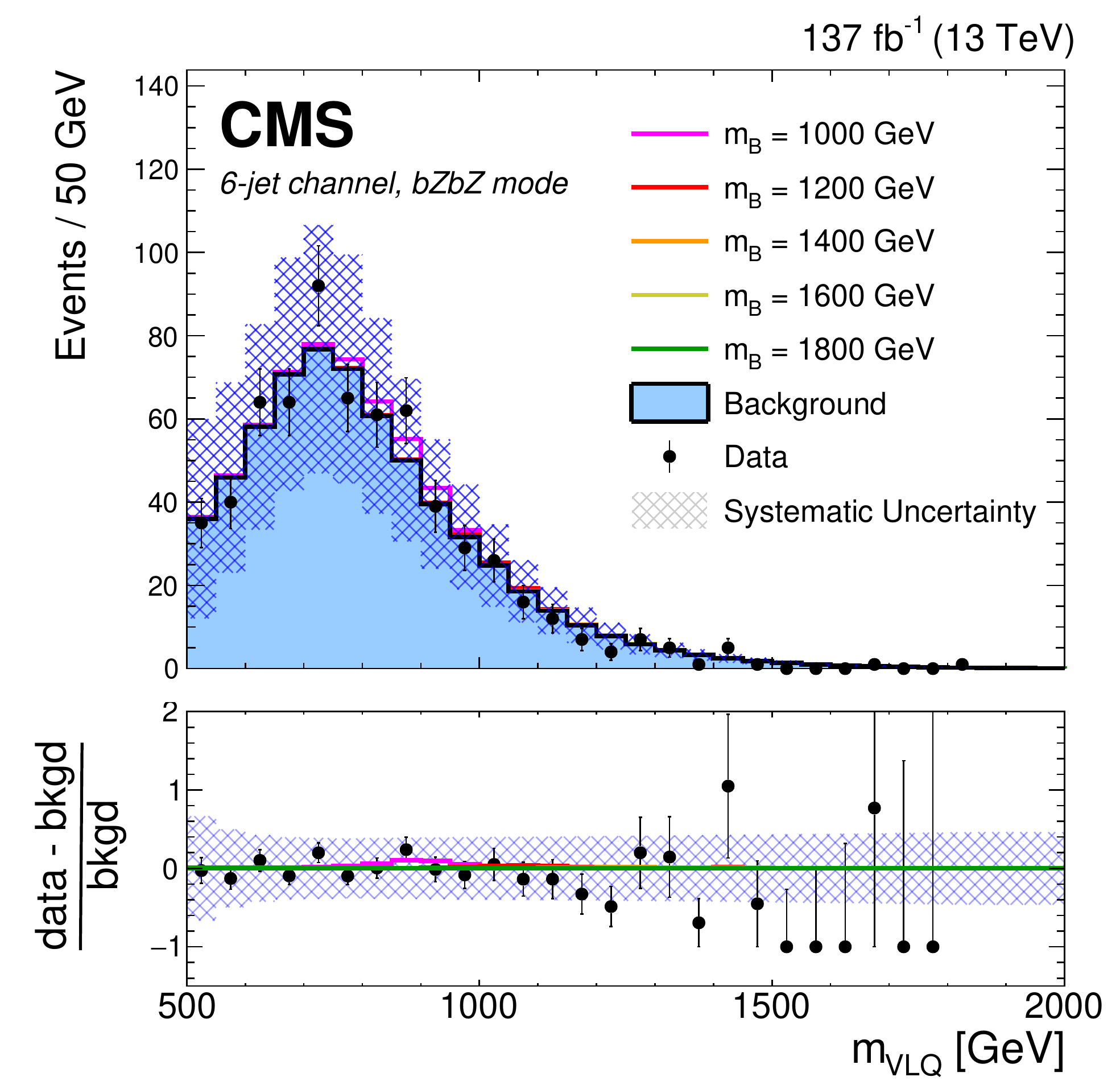}
\caption{Data (black points), expected background (solid blue histogram), and expected background plus a VLQ signal for different VLQ masses (colored lines), for 4-jet (left column), 5-jet (center column), and 6-jet (right column) multiplicities and for \bhbh (upper row), \bhbz (middle row), and \bzbz (lower row) event modes. For the signal, $\BrBbH = 100\%$ is assumed. The hatched regions for the background and background plus signal distributions indicate the systematic uncertainties. All three data-taking years are combined.}
\label{fig:signal_background}
\end{figure*}

Figure~\ref{fig:signal_background} shows the distribution of the reconstructed VLQ mass, after the optimized selections described in Section~\ref{sec:eventreco} have been applied, for data, the expected background, and for simulated signal events with a VLQ mass of 1200, 1400, 1600, and 1800\GeV and $\BrBbH = 100\%$. The signal distributions are normalized to the expected number of events as determined by the VLQ production cross section. No statistically significant excess of data over the background expectation is observed; the largest difference across all nine mass points and branching fraction scenarios is slightly less than $2\sigma$.

\begin{figure*}[hbtp!]
\centering
\includegraphics[width=0.8\textwidth]{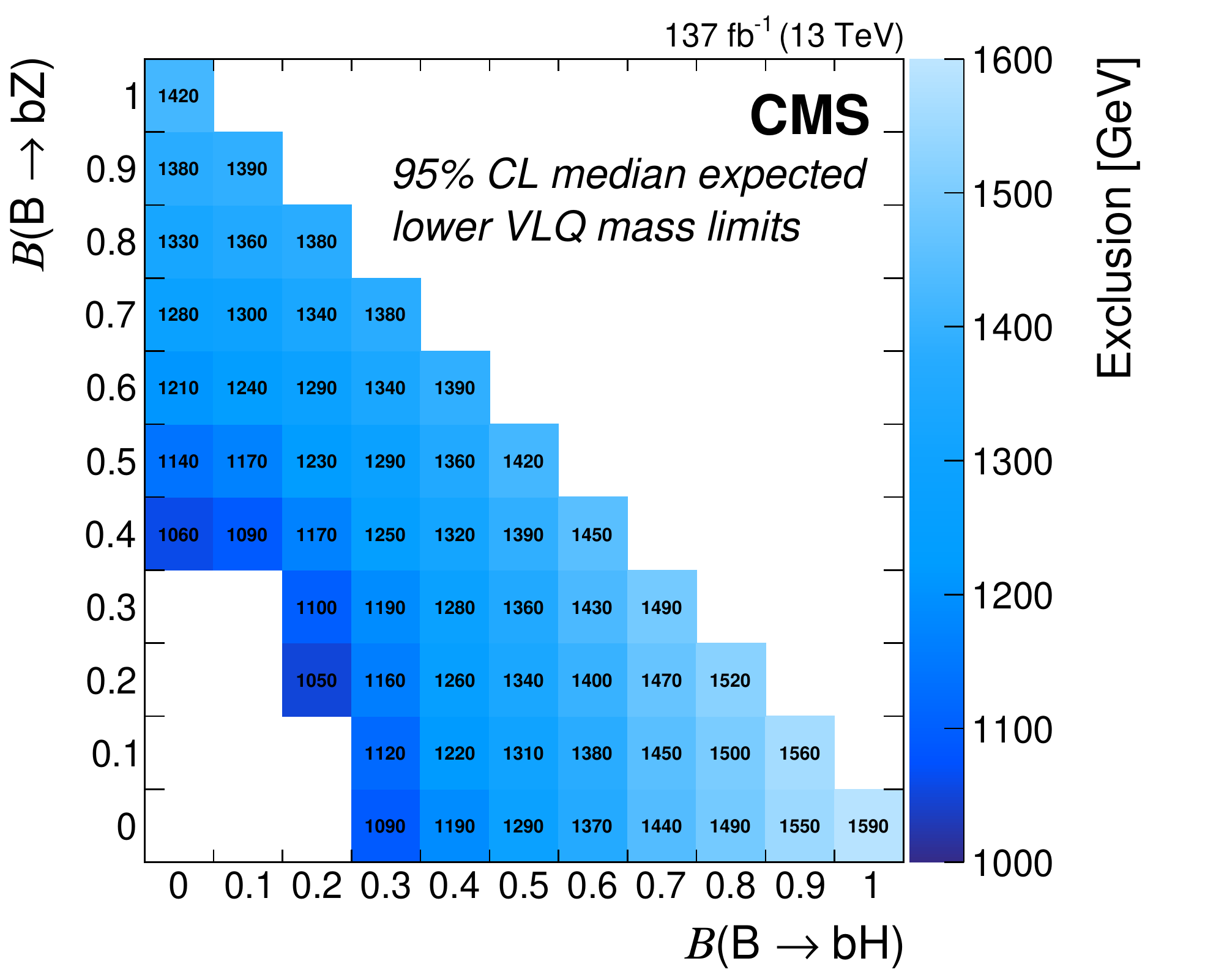}
\includegraphics[width=0.8\textwidth]{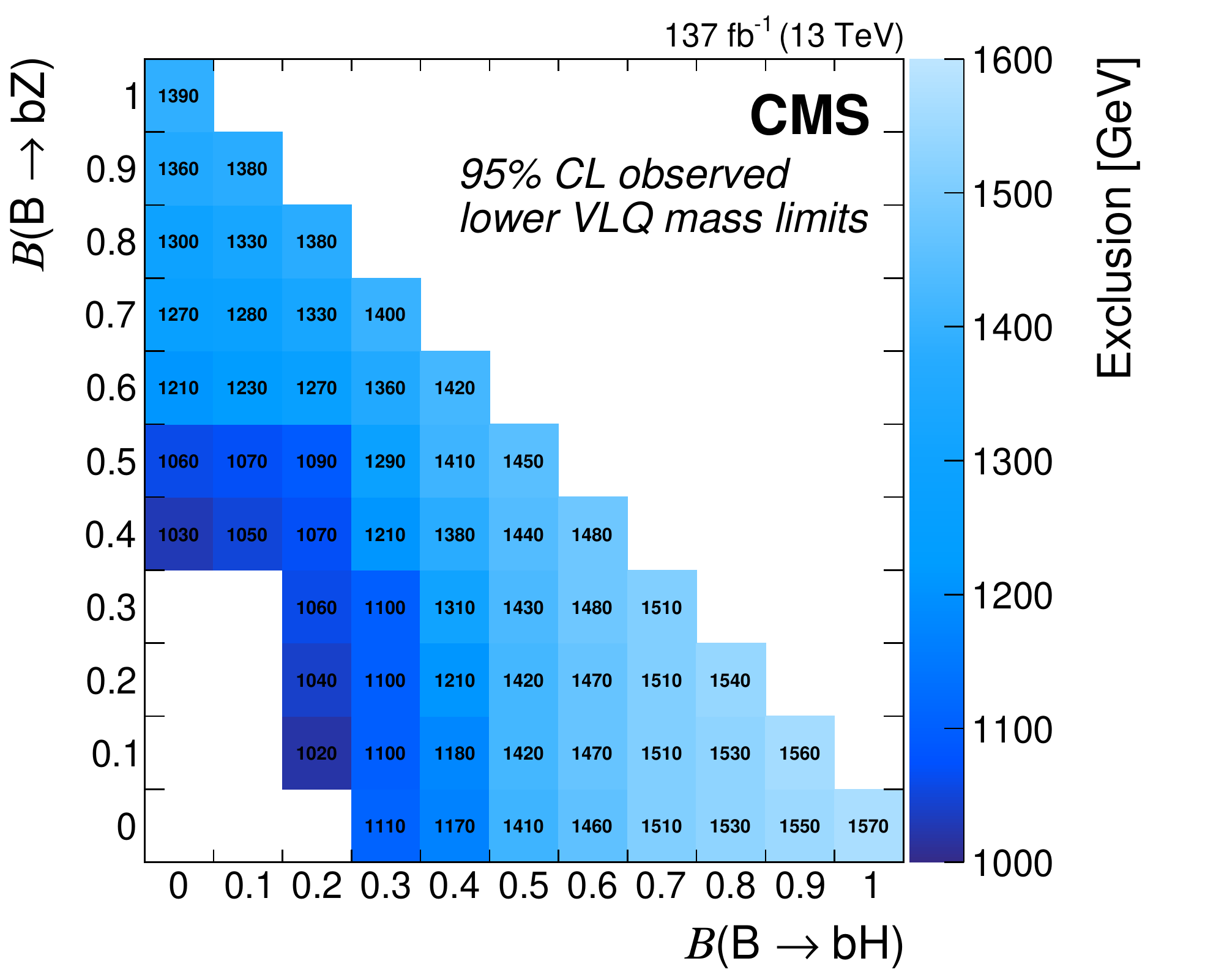}
\caption{Expected (upper) and observed (lower) limits on the VLQ mass at 95\% \CL as a function of the branching fractions \BrBbH and \BrBbZ.}
\label{fig:triangle_limits}
\end{figure*}

We proceed to set exclusion limits on the VLQ mass as a function of the branching fractions. The signal extraction procedure is based on a binned maximum likelihood fit. All systematic uncertainties are incorporated into the fit as nuisance parameters, where the effect of each systematic uncertainty is included as a lognormal probability distribution per bin. A limit at 95\% \CL is calculated using the \CLs method~\cite{CLS2,CLS1} using the profile likelihood test statistic~\cite{Cowan:2010js} with the asymptotic limit approximation. Figure~\ref{fig:triangle_limits} shows the final expected and observed limits on the VLQ mass as a function of \BrBbH and \BrBbZ, after all of the individual jet multiplicities and event modes have been combined. Points for which the exclusion limit is less than 1000\GeV are not shown.

\begin{figure}[hbtp!]
\centering
\includegraphics[width=\masslimitswidth]{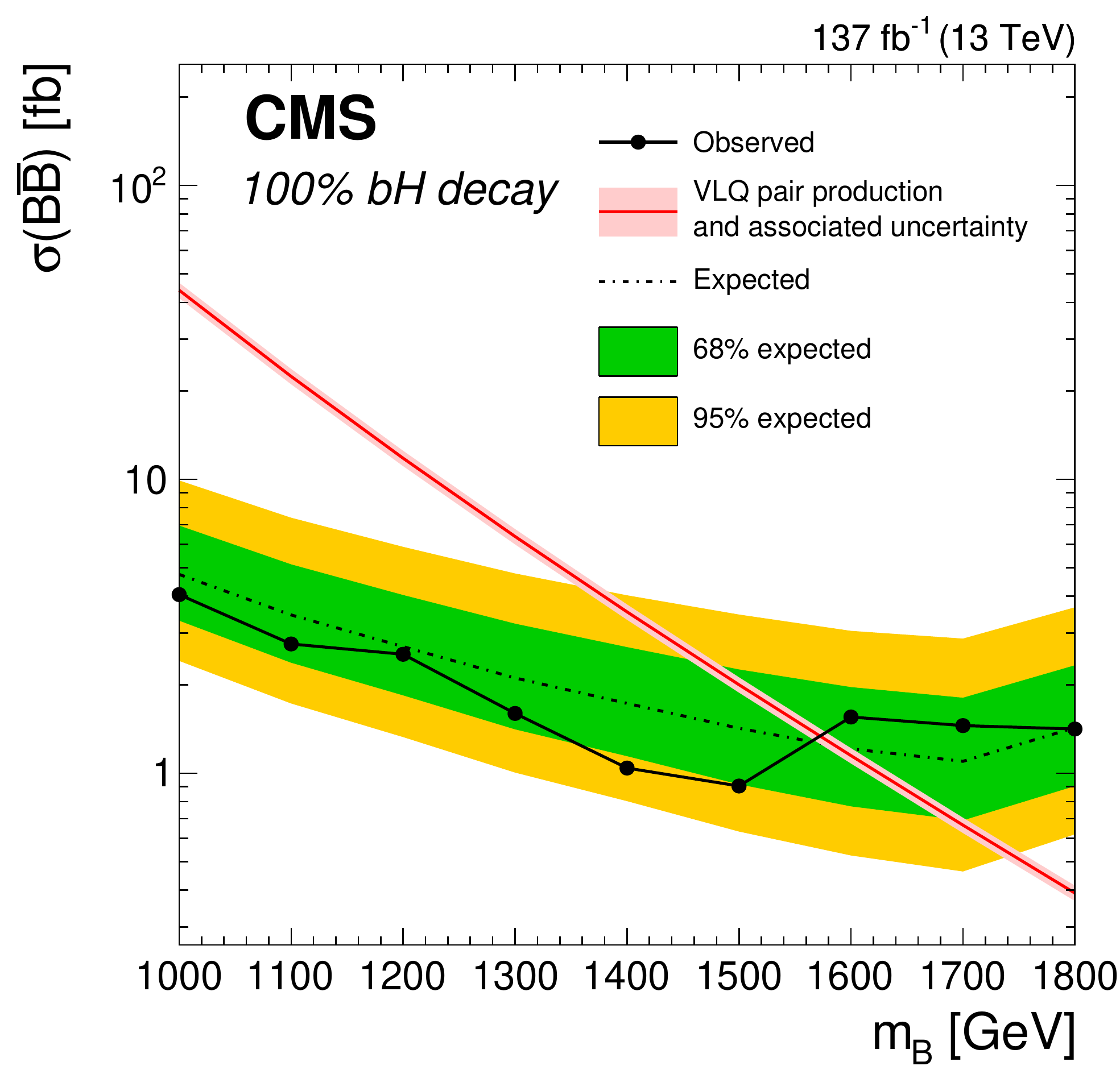}
\includegraphics[width=\masslimitswidth]{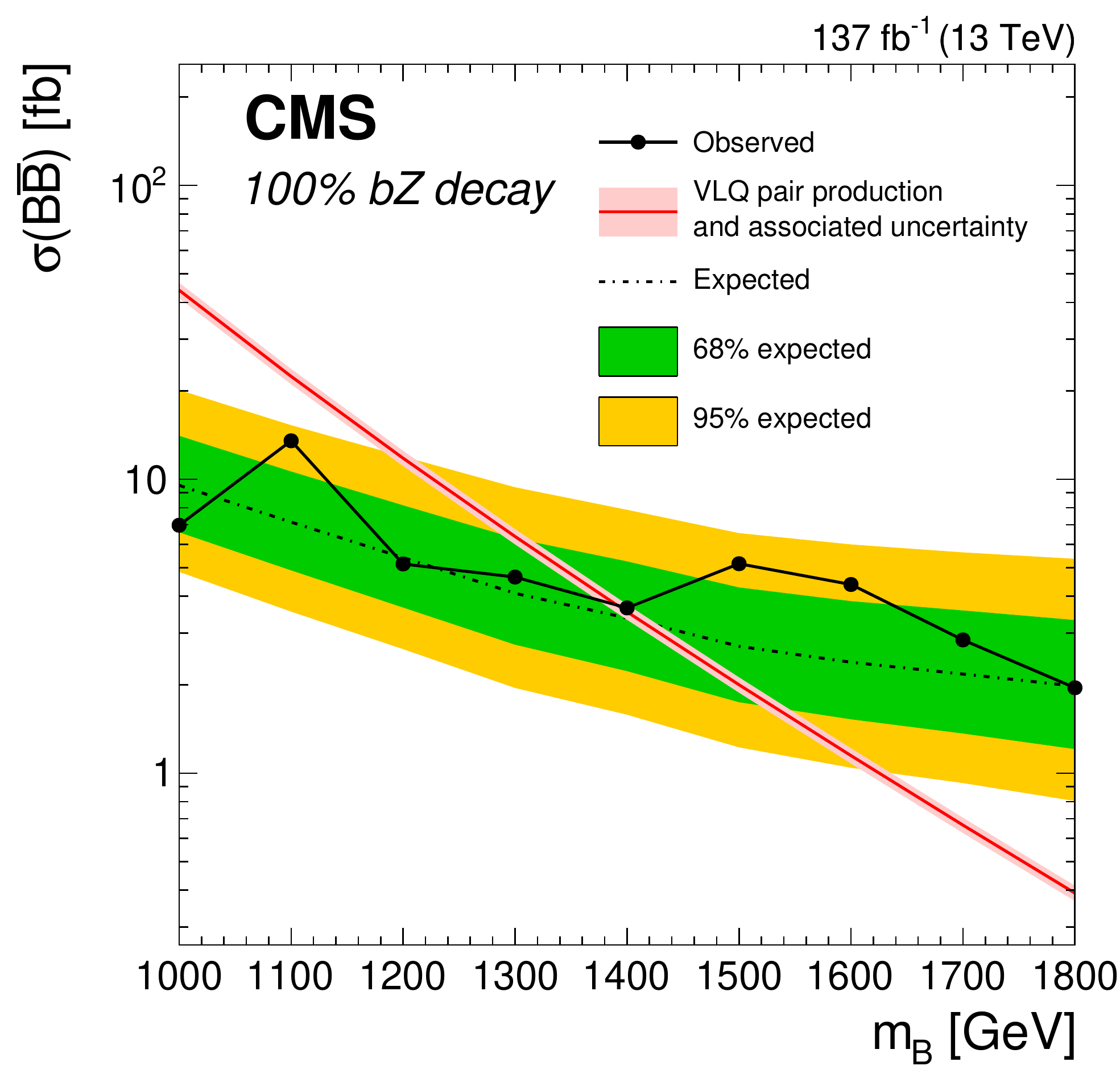}
\includegraphics[width=\masslimitswidth]{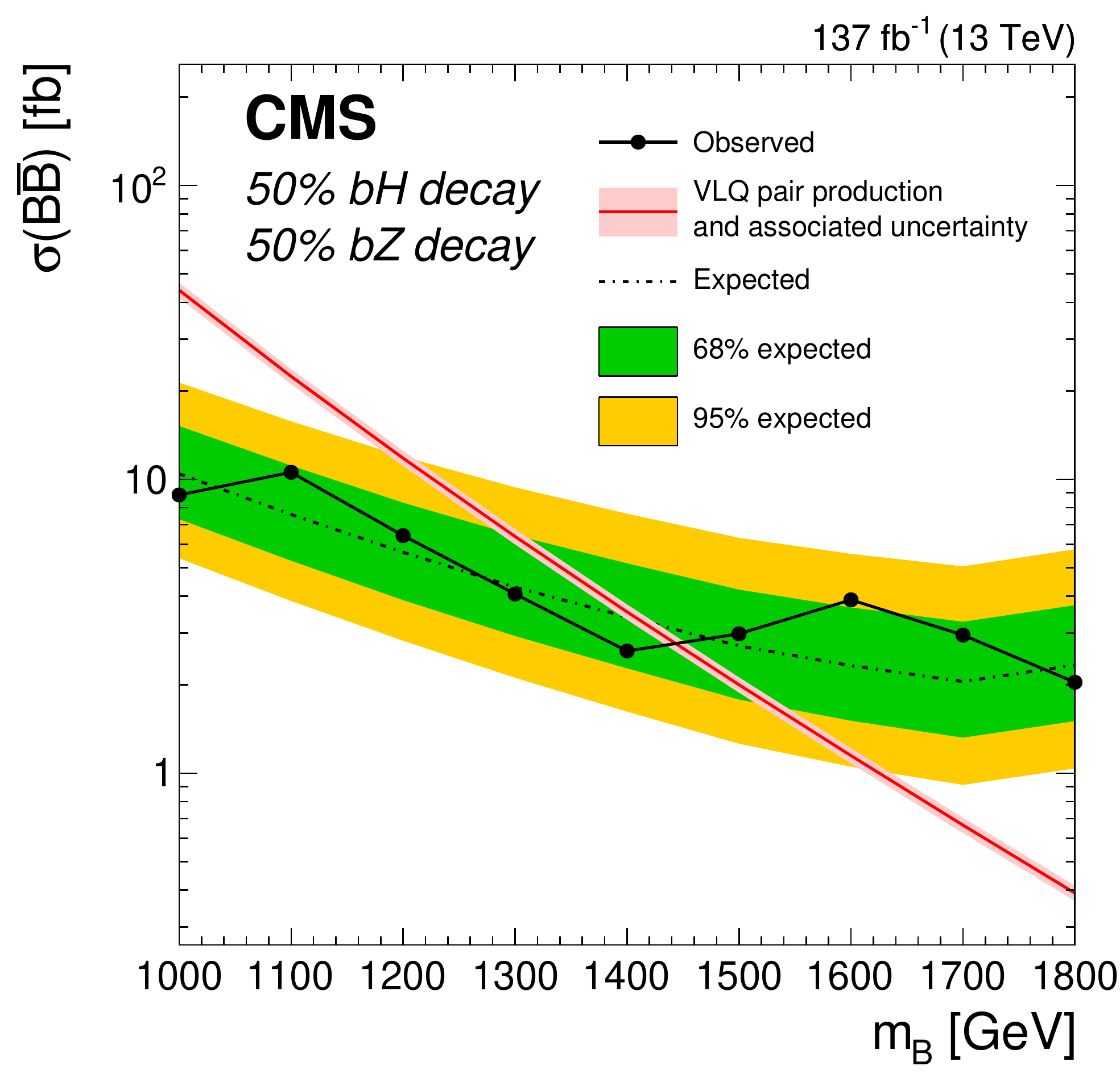}
\caption{The 95\% confidence limit on the cross section for VLQ pair production as a function of VLQ mass for three branching fraction hypotheses: $\BrBbH = 100\%$ (\masslimitsone), $\BrBbZ = 100\%$ (\masslimitstwo), and $\BrBbH = \BrBbZ = 50\%$ (\masslimitsthree). The solid black line indicates the observed limit and the dashed line indicates the expected limit with 1 sigma (green band) and 2 sigma (yellow band) uncertainties. The theoretical cross section and its uncertainty are shown as the red line and pale red band; the band is only slightly visible outside the line.}
\label{fig:exclusion_bHbH}
\end{figure}

Figure~\ref{fig:exclusion_bHbH} shows the expected limits at 95\% \CL on the cross section of VLQ pair production as a function of VLQ mass assuming three different branching fraction combinations: $\BrBbH = 100\%$, $\BrBbZ = 100\%$, and $\BrBbH = \BrBbZ = 50\%$. The observed limits at 95\% \CL are: 1570\GeV in the 100\% $\PQb\PH$ case, 1390\GeV in the 100\% $\PQb\PZ$ case, and 1450\GeV in the 50\% $\PQb\PH$ plus 50\% $\PQb\PZ$ case. In the fully \BbH and \BbZ modes, as well as the mixed \bhbz mode, where this analysis is most sensitive, these limits represent significant improvements over previously published VLQ limits (1010, 1070, and 1025\GeV respectively), extending the existing limits by several hundred \GeVns. These improvements can be attributed to the use of the \chimodndf method, which allows the hadronic final state to be fully reconstructed, as well as to the increased size of the data sample.

\section{Summary}

This paper describes a search for bottom-type, vector-like quark (VLQ) pair production in data collected by the CMS detector in 2016--2018 at $\sqrt{s} = 13\TeV$, where the VLQ \PB decays into a \PQb or \PAQb quark and either a Higgs boson \PH or a \PZ boson. The analysis targets the fully hadronic \BbH and \BbZ decays by tagging jets and using a modified \chisq metric to reconstruct the event. Different jet multiplicity categories were used to account for the fact that Higgs or \PZ boson decays can produce either two distinct jets or, if highly Lorentz boosted, a single merged jet. Backgrounds were estimated from a region of low VLQ mass and extrapolated into the signal region using a modified \chisq control region. Limits were set on the VLQ mass at 95\% confidence level as a function of the branching fractions for \BbH and \BbZ. Compared to previous measurements~\cite{Sirunyan:2019sza,Aaboud:2018wxv}, limits on the \PB VLQ mass have been increased from 1010 to 1570\GeV in the $\BrBbH = 100\%$ case, from 1070 to 1390\GeV in the $\BrBbZ = 100\%$ case, and from 1025 to 1450\GeV in the $\BrBbH = \BrBbZ = 50\%$ case.

\begin{acknowledgments}
  We congratulate our colleagues in the CERN accelerator departments for the excellent performance of the LHC and thank the technical and administrative staffs at CERN and at other CMS institutes for their contributions to the success of the CMS effort. In addition, we gratefully acknowledge the computing centers and personnel of the Worldwide LHC Computing Grid for delivering so effectively the computing infrastructure essential to our analyses. Finally, we acknowledge the enduring support for the construction and operation of the LHC and the CMS detector provided by the following funding agencies: BMBWF and FWF (Austria); FNRS and FWO (Belgium); CNPq, CAPES, FAPERJ, FAPERGS, and FAPESP (Brazil); MES (Bulgaria); CERN; CAS, MoST, and NSFC (China); COLCIENCIAS (Colombia); MSES and CSF (Croatia); RIF (Cyprus); SENESCYT (Ecuador); MoER, ERC IUT, PUT and ERDF (Estonia); Academy of Finland, MEC, and HIP (Finland); CEA and CNRS/IN2P3 (France); BMBF, DFG, and HGF (Germany); GSRT (Greece); NKFIA (Hungary); DAE and DST (India); IPM (Iran); SFI (Ireland); INFN (Italy); MSIP and NRF (Republic of Korea); MES (Latvia); LAS (Lithuania); MOE and UM (Malaysia); BUAP, CINVESTAV, CONACYT, LNS, SEP, and UASLP-FAI (Mexico); MOS (Montenegro); MBIE (New Zealand); PAEC (Pakistan); MSHE and NSC (Poland); FCT (Portugal); JINR (Dubna); MON, RosAtom, RAS, RFBR, and NRC KI (Russia); MESTD (Serbia); SEIDI, CPAN, PCTI, and FEDER (Spain); MOSTR (Sri Lanka); Swiss Funding Agencies (Switzerland); MST (Taipei); ThEPCenter, IPST, STAR, and NSTDA (Thailand); TUBITAK and TAEK (Turkey); NASU (Ukraine); STFC (United Kingdom); DOE and NSF (USA).
   
  \hyphenation{Rachada-pisek} Individuals have received support from the Marie-Curie program and the European Research Council and Horizon 2020 Grant, contract Nos.\ 675440, 752730, and 765710 (European Union); the Leventis Foundation; the A.P.\ Sloan Foundation; the Alexander von Humboldt Foundation; the Belgian Federal Science Policy Office; the Fonds pour la Formation \`a la Recherche dans l'Industrie et dans l'Agriculture (FRIA-Belgium); the Agentschap voor Innovatie door Wetenschap en Technologie (IWT-Belgium); the F.R.S.-FNRS and FWO (Belgium) under the ``Excellence of Science -- EOS" -- be.h project n.\ 30820817; the Beijing Municipal Science \& Technology Commission, No. Z191100007219010; the Ministry of Education, Youth and Sports (MEYS) of the Czech Republic; the Deutsche Forschungsgemeinschaft (DFG) under Germany's Excellence Strategy -- EXC 2121 ``Quantum Universe" -- 390833306; the Lend\"ulet (``Momentum") Program and the J\'anos Bolyai Research Scholarship of the Hungarian Academy of Sciences, the New National Excellence Program \'UNKP, the NKFIA research grants 123842, 123959, 124845, 124850, 125105, 128713, 128786, and 129058 (Hungary); the Council of Science and Industrial Research, India; the HOMING PLUS program of the Foundation for Polish Science, cofinanced from European Union, Regional Development Fund, the Mobility Plus program of the Ministry of Science and Higher Education, the National Science Center (Poland), contracts Harmonia 2014/14/M/ST2/00428, Opus 2014/13/B/ST2/02543, 2014/15/B/ST2/03998, and 2015/19/B/ST2/02861, Sonata-bis 2012/07/E/ST2/01406; the National Priorities Research Program by Qatar National Research Fund; the Ministry of Science and Higher Education, project no. 02.a03.21.0005 (Russia); the Programa Estatal de Fomento de la Investigaci{\'o}n Cient{\'i}fica y T{\'e}cnica de Excelencia Mar\'{\i}a de Maeztu, grant MDM-2015-0509 and the Programa Severo Ochoa del Principado de Asturias; the Thalis and Aristeia programs cofinanced by EU-ESF and the Greek NSRF; the Rachadapisek Sompot Fund for Postdoctoral Fellowship, Chulalongkorn University and the Chulalongkorn Academic into Its 2nd Century Project Advancement Project (Thailand); the Kavli Foundation; the Nvidia Corporation; the SuperMicro Corporation; the Welch Foundation, contract C-1845; and the Weston Havens Foundation (USA). 
\end{acknowledgments}
     
\bibliography{auto_generated}
\cleardoublepage \appendix\section{The CMS Collaboration \label{app:collab}}\begin{sloppypar}\hyphenpenalty=5000\widowpenalty=500\clubpenalty=5000\input{B2G-19-005-authorlist.tex}\end{sloppypar}
\end{document}

%% file: B2G-19-005-authorlist.tex
\vskip\cmsinstskip
\textbf{Yerevan Physics Institute, Yerevan, Armenia}\\*[0pt]
A.M.~Sirunyan$^{\textrm{\dag}}$, A.~Tumasyan
\vskip\cmsinstskip
\textbf{Institut f\"{u}r Hochenergiephysik, Wien, Austria}\\*[0pt]
W.~Adam, T.~Bergauer, M.~Dragicevic, J.~Er\"{o}, A.~Escalante~Del~Valle, R.~Fr\"{u}hwirth\cmsAuthorMark{1}, M.~Jeitler\cmsAuthorMark{1}, N.~Krammer, L.~Lechner, D.~Liko, T.~Madlener, I.~Mikulec, F.M.~Pitters, N.~Rad, J.~Schieck\cmsAuthorMark{1}, R.~Sch\"{o}fbeck, M.~Spanring, S.~Templ, W.~Waltenberger, C.-E.~Wulz\cmsAuthorMark{1}, M.~Zarucki
% ...
\vskip\cmsinstskip
\dag: Deceased\\
1:  Also at Vienna University of Technology, Vienna, Austria\\
% ...


%% file: B2G-19-005_temp.bbl
\providecommand{\href}[2]{#2}\begingroup\raggedright\begin{thebibliography}{10}%
\makeatletter
\providecommand{\hrefCMSnoop }[0]{\@secondoftwo}%
\makeatother
\providecommand{\doi}{\texttt{doi:}\begingroup \urlstyle{tt}\Url}

\bibitem{tHooft:1979rat}
\hrefCMSnoop {}{G.~'t~Hooft, ``{Naturalness, chiral symmetry, and spontaneous
  chiral symmetry breaking}'',} \textit{ NATO Sci. Ser. B} \textbf{ 59} (1980)
  135,
\href{http://dx.doi.org/10.1007/978-1-4684-7571-5_9}{\doi{10.1007/978-1-4684-7571-5_9}}.

\bibitem{Wess:1973kz}
\hrefCMSnoop {}{J.~Wess and B.~Zumino, ``A {Lagrangian} model invariant under
  supergauge transformations'',} \textit{ Phys. Lett. B} \textbf{ 49} (1974)
  52,
\href{http://dx.doi.org/10.1016/0370-2693(74)90578-4}{\doi{10.1016/0370-2693(74)90578-4}}.

\bibitem{Fayet:1976cr}
\hrefCMSnoop {}{P.~Fayet and S.~Ferrara, ``{Supersymmetry}'',} \textit{ Phys.
  Rept.} \textbf{ 32} (1977) 249,
\href{http://dx.doi.org/10.1016/0370-1573(77)90066-7}{\doi{10.1016/0370-1573(77)90066-7}}.

\bibitem{Georgi:1974yw}
\hrefCMSnoop {}{H.~Georgi and A.~Pais, ``Calculability and naturalness in gauge
  theories'',} \textit{ Phys. Rev. D} \textbf{ 10} (1974) 539,
\href{http://dx.doi.org/10.1103/PhysRevD.10.539}{\doi{10.1103/PhysRevD.10.539}}.

\bibitem{Kaplan:1983sm}
\hrefCMSnoop {}{D.~B. Kaplan, H.~Georgi, and S.~Dimopoulos, ``Composite {Higgs}
  scalars'',} \textit{ Phys. Lett. B} \textbf{ 136} (1984) 187,
  \href{http://dx.doi.org/10.1016/0370-2693(84)91178-X}{\doi{10.1016/0370-2693(84)91178-X}}.

\bibitem{Agashe:2004rs}
\hrefCMSnoop {}{K.~Agashe, R.~Contino, and A.~Pomarol, ``The minimal composite
  {Higgs} model'',} \textit{ Nucl. Phys. B} \textbf{ 719} (2005) 165,
  \href{http://dx.doi.org/10.1016/j.nuclphysb.2005.04.035}{\doi{10.1016/j.nuclphysb.2005.04.035}},
  \href{http://www.arXiv.org/abs/hep-ph/0412089}{\texttt{arXiv:hep-ph/0412089}}.

\bibitem{ArkaniHamed:2001nc}
\hrefCMSnoop {}{N.~Arkani-Hamed, A.~G. Cohen, and H.~Georgi, ``Electroweak
  symmetry breaking from dimensional deconstruction'',} \textit{ Phys. Lett. B}
  \textbf{ 513} (2001) 232,
  \href{http://dx.doi.org/10.1016/S0370-2693(01)00741-9}{\doi{10.1016/S0370-2693(01)00741-9}},
\href{http://www.arXiv.org/abs/hep-ph/0105239}{\texttt{arXiv:hep-ph/0105239}}.

\bibitem{ArkaniHamed:2002qy}
\hrefCMSnoop {}{N.~Arkani-Hamed, A.~G. Cohen, E.~Katz, and A.~E. Nelson, ``The
  littlest {Higgs}'',} \textit{ JHEP} \textbf{ 07} (2002) 034,
  \href{http://dx.doi.org/10.1088/1126-6708/2002/07/034}{\doi{10.1088/1126-6708/2002/07/034}},
\href{http://www.arXiv.org/abs/hep-ph/0206021}{\texttt{arXiv:hep-ph/0206021}}.

\bibitem{delAguila:1982fs}
\hrefCMSnoop {}{F.~del Aguila and M.~J. Bowick, ``The possibility of new
  fermions with {$\Delta I = 0$} mass'',} \textit{ Nucl. Phys. B} \textbf{ 224}
  (1983) 107,
\href{http://dx.doi.org/10.1016/0550-3213(83)90316-4}{\doi{10.1016/0550-3213(83)90316-4}}.

\bibitem{Aad:2019mbh}
\hrefCMSnoop {}{{ATLAS Collaboration}, ``Combined measurements of {Higgs} boson
  production and decay using up to $80$ fb$^{-1}$ of proton-proton collision
  data at $\sqrt{s}=$ 13 {TeV} collected with the {ATLAS} experiment'',}
  \textit{ Phys. Rev. D} \textbf{ 101} (2020) 012002,
  \href{http://dx.doi.org/10.1103/PhysRevD.101.012002}{\doi{10.1103/PhysRevD.101.012002}},
\href{http://www.arXiv.org/abs/1909.02845}{\texttt{arXiv:1909.02845}}.

\bibitem{Sirunyan:2018sgc}
\hrefCMSnoop {}{{CMS Collaboration}, ``Measurement and interpretation of
  differential cross sections for {Higgs} boson production at $\sqrt{s} =$ 13
  {TeV}'',} \textit{ Phys. Lett. B} \textbf{ 792} (2019) 369,
  \href{http://dx.doi.org/10.1016/j.physletb.2019.03.059}{\doi{10.1016/j.physletb.2019.03.059}},
  \href{http://www.arXiv.org/abs/1812.06504}{\texttt{arXiv:1812.06504}}.

\bibitem{Aguilar-Saavedra:2013qpa}
\hrefCMSnoop {}{J.~A. Aguilar-Saavedra, R.~Benbrik, S.~Heinemeyer, and
  M.~P{\'e}rez-Victoria, ``Handbook of vectorlike quarks: {Mixing} and single
  production'',} \textit{ Phys. Rev. D} \textbf{ 88} (2013) 094010,
  \href{http://dx.doi.org/10.1103/PhysRevD.88.094010}{\doi{10.1103/PhysRevD.88.094010}},
\href{http://www.arXiv.org/abs/1306.0572}{\texttt{arXiv:1306.0572}}.

\bibitem{delAguila:2000rc}
\hrefCMSnoop {}{F.~del Aguila, M.~P{\'e}rez-Victoria, and J.~Santiago,
  ``Observable contributions of new exotic quarks to quark mixing'',} \textit{
  JHEP} \textbf{ 09} (2000) 011,
  \href{http://dx.doi.org/10.1088/1126-6708/2000/09/011}{\doi{10.1088/1126-6708/2000/09/011}},
  \href{http://www.arXiv.org/abs/hep-ph/0007316}{\texttt{arXiv:hep-ph/0007316}}.

\bibitem{Aguilar-Saavedra:2013wba}
\hrefCMSnoop {}{J.~A. Aguilar-Saavedra, ``Mixing with vector-like quarks:
  constraints and expectations'',} \textit{ EPJ Web Conf.} \textbf{ 60} (2013)
  16012,
  \href{http://dx.doi.org/10.1051/epjconf/20136016012}{\doi{10.1051/epjconf/20136016012}},
  \href{http://www.arXiv.org/abs/1306.4432}{\texttt{arXiv:1306.4432}}.

\bibitem{Atre:2008iu}
\hrefCMSnoop {}{A.~Atre, M.~Carena, T.~Han, and J.~Santiago, ``Heavy quarks
  above the top at the {Tevatron}'',} \textit{ Phys. Rev. D} \textbf{ 79}
  (2009) 054018,
  \href{http://dx.doi.org/10.1103/PhysRevD.79.054018}{\doi{10.1103/PhysRevD.79.054018}},
  \href{http://www.arXiv.org/abs/0806.3966}{\texttt{arXiv:0806.3966}}.

\bibitem{Atre:2011ae}
A.~Atre\hrefCMSnoop {}{ {et~al.}, ``Model-independent searches for new quarks
  at the {LHC}'',} \textit{ JHEP} \textbf{ 08} (2011) 080,
  \href{http://dx.doi.org/10.1007/JHEP08(2011)080}{\doi{10.1007/JHEP08(2011)080}},
  \href{http://www.arXiv.org/abs/1102.1987}{\texttt{arXiv:1102.1987}}.

\bibitem{Aaboud:2018wxv}
\hrefCMSnoop {}{{ATLAS Collaboration}, ``Search for pair production of heavy
  vector-like quarks decaying into hadronic final states in $pp$ collisions at
  $\sqrt{s} = 13$ {TeV} with the {ATLAS} detector'',} \textit{ Phys. Rev. D}
  \textbf{ 98} (2018) 092005,
  \href{http://dx.doi.org/10.1103/PhysRevD.98.092005}{\doi{10.1103/PhysRevD.98.092005}},
\href{http://www.arXiv.org/abs/1808.01771}{\texttt{arXiv:1808.01771}}.

\bibitem{Sirunyan:2019sza}
\hrefCMSnoop {}{{CMS Collaboration}, ``Search for pair production of vectorlike
  quarks in the fully hadronic final state'',} \textit{ Phys. Rev. D} \textbf{
  100} (2019) 072001,
  \href{http://dx.doi.org/10.1103/PhysRevD.100.072001}{\doi{10.1103/PhysRevD.100.072001}},
  \href{http://www.arXiv.org/abs/1906.11903}{\texttt{arXiv:1906.11903}}.

\bibitem{Aaboud:2018pii}
\hrefCMSnoop {}{{ATLAS Collaboration}, ``Combination of the searches for
  pair-produced vector-like partners of the third-generation quarks at
  $\sqrt{s} =$ 13 {TeV} with the {ATLAS} detector'',} \textit{ Phys. Rev.
  Lett.} \textbf{ 121} (2018) 211801,
  \href{http://dx.doi.org/10.1103/PhysRevLett.121.211801}{\doi{10.1103/PhysRevLett.121.211801}},
  \href{http://www.arXiv.org/abs/1808.02343}{\texttt{arXiv:1808.02343}}.

\bibitem{CMS-NOTE-2011-005}
\href {https://cds.cern.ch/record/1379837}{{The ATLAS Collaboration, The CMS
  Collaboration, The LHC Higgs Combination Group}, ``Procedure for the {LHC}
  {Higgs} boson search combination in {Summer} 2011'',} Technical Report
  CMS-NOTE-2011-005, ATL-PHYS-PUB-2011-11, 2011.

\bibitem{Khachatryan:2016kdb}
\hrefCMSnoop {}{{CMS Collaboration}, ``Jet energy scale and resolution in the
  {CMS} experiment in pp collisions at 8 {TeV}'',} \textit{ JINST} \textbf{ 12}
  (2017) P02014,
  \href{http://dx.doi.org/10.1088/1748-0221/12/02/P02014}{\doi{10.1088/1748-0221/12/02/P02014}},
\href{http://www.arXiv.org/abs/1607.03663}{\texttt{arXiv:1607.03663}}.

\bibitem{Khachatryan:2016bia}
\hrefCMSnoop {}{{CMS Collaboration}, ``{The CMS trigger system}'',} \textit{
  JINST} \textbf{ 12} (2017) P01020,
  \href{http://dx.doi.org/10.1088/1748-0221/12/01/P01020}{\doi{10.1088/1748-0221/12/01/P01020}},
\href{http://www.arXiv.org/abs/1609.02366}{\texttt{arXiv:1609.02366}}.

\bibitem{Chatrchyan:2008zzk}
\hrefCMSnoop {}{{CMS Collaboration}, ``The {CMS} experiment at the {CERN}
  {LHC}'',} \textit{ JINST} \textbf{ 3} (2008) S08004,
  \href{http://dx.doi.org/10.1088/1748-0221/3/08/S08004}{\doi{10.1088/1748-0221/3/08/S08004}}.

\bibitem{CMS-PAS-LUM-17-001}
\href {https://cds.cern.ch/record/2257069}{{CMS Collaboration}, ``{CMS}
  luminosity measurement for the 2016 data taking period'',} CMS Physics
  Analysis Summary CMS-PAS-LUM-17-001, 2017.

\bibitem{CMS-PAS-LUM-17-004}
\href {https://cds.cern.ch/record/2621960}{{CMS Collaboration}, ``{CMS}
  luminosity measurement for the 2017 data-taking period at $\sqrt{s}$ = 13
  {TeV}'',} CMS Physics Analysis Summary CMS-PAS-LUM-17-004, 2018.

\bibitem{CMS-PAS-LUM-18-002}
\href {https://cds.cern.ch/record/2676164}{{CMS Collaboration}, ``{CMS}
  luminosity measurement for the 2018 data-taking period at $\sqrt{s} =
  13~\mathrm{TeV}$'',} CMS Physics Analysis Summary CMS-PAS-LUM-18-002, 2019.

\bibitem{Alwall:2014hca}
J.~Alwall\hrefCMSnoop {}{ {et~al.}, ``The automated computation of tree-level
  and next-to-leading order differential cross sections, and their matching to
  parton shower simulations'',} \textit{ JHEP} \textbf{ 07} (2014) 079,
  \href{http://dx.doi.org/10.1007/JHEP07(2014)079}{\doi{10.1007/JHEP07(2014)079}},
\href{http://www.arXiv.org/abs/1405.0301}{\texttt{arXiv:1405.0301}}.

\bibitem{Ball:2014uwa}
\hrefCMSnoop {}{{NNPDF} Collaboration, ``Parton distributions for the {LHC Run
  II}'',} \textit{ JHEP} \textbf{ 04} (2015) 040,
  \href{http://dx.doi.org/10.1007/JHEP04(2015)040}{\doi{10.1007/JHEP04(2015)040}},
\href{http://www.arXiv.org/abs/1410.8849}{\texttt{arXiv:1410.8849}}.

\bibitem{Sjostrand:2014zea}
T.~Sj{\"o}strand\hrefCMSnoop {}{ {et~al.}, ``An introduction to {PYTHIA}
  8.2'',} \textit{ Comput. Phys. Commun.} \textbf{ 191} (2015) 159,
  \href{http://dx.doi.org/10.1016/j.cpc.2015.01.024}{\doi{10.1016/j.cpc.2015.01.024}},
\href{http://www.arXiv.org/abs/1410.3012}{\texttt{arXiv:1410.3012}}.

\bibitem{Khachatryan:2015pea}
\hrefCMSnoop {}{{CMS Collaboration}, ``Event generator tunes obtained from
  underlying event and multiparton scattering measurements'',} \textit{ Eur.
  Phys. J. C} \textbf{ 76} (2016) 155,
  \href{http://dx.doi.org/10.1140/epjc/s10052-016-3988-x}{\doi{10.1140/epjc/s10052-016-3988-x}},
\href{http://www.arXiv.org/abs/1512.00815}{\texttt{arXiv:1512.00815}}.

\bibitem{Sirunyan:2019dfx}
\hrefCMSnoop {}{{CMS Collaboration}, ``Extraction and validation of a new set
  of {CMS} {PYTHIA8} tunes from underlying-event measurements'',} \textit{ Eur.
  Phys. J. C} \textbf{ 80} (2020) 4,
  \href{http://dx.doi.org/10.1140/epjc/s10052-019-7499-4}{\doi{10.1140/epjc/s10052-019-7499-4}},
\href{http://www.arXiv.org/abs/1903.12179}{\texttt{arXiv:1903.12179}}.

\bibitem{Czakon:2013goa}
\hrefCMSnoop {}{M.~Czakon, P.~Fiedler, and A.~Mitov, ``Total top-quark
  pair-production cross section at hadron colliders through
  {$O(\alpha^4_S)$}'',} \textit{ Phys. Rev. Lett.} \textbf{ 110} (2013) 252004,
  \href{http://dx.doi.org/10.1103/PhysRevLett.110.252004}{\doi{10.1103/PhysRevLett.110.252004}},
\href{http://www.arXiv.org/abs/1303.6254}{\texttt{arXiv:1303.6254}}.

\bibitem{Whalley:2005nh}
\hrefCMSnoop {}{M.~R. Whalley, D.~Bourilkov, and R.~C. Group, ``The {Les
  Houches} accord {PDFs} ({LHAPDF}) and {LHAGLUE}'',} in \textit{ {HERA and the
  LHC: A workshop on the implications of HERA for LHC physics. Proceedings,
  Part B}}, p.~575.
\newblock 2005.
\newblock
\href{http://www.arXiv.org/abs/hep-ph/0508110}{\texttt{arXiv:hep-ph/0508110}}.
\newblock

\bibitem{Sirunyan:2018omb}
\hrefCMSnoop {}{{CMS Collaboration}, ``Search for vector-like {T} and {B} quark
  pairs in final states with leptons at $\sqrt{s} =$ 13 {TeV}'',} \textit{
  JHEP} \textbf{ 08} (2018) 177,
  \href{http://dx.doi.org/10.1007/JHEP08(2018)177}{\doi{10.1007/JHEP08(2018)177}},
  \href{http://www.arXiv.org/abs/1805.04758}{\texttt{arXiv:1805.04758}}.

\bibitem{Aaij:2018okq}
\hrefCMSnoop {}{{LHCb Collaboration}, ``Measurement of the inelastic $pp$
  cross-section at a centre-of-mass energy of 13 {TeV}'',} \textit{ JHEP}
  \textbf{ 06} (2018) 100,
  \href{http://dx.doi.org/10.1007/JHEP06(2018)100}{\doi{10.1007/JHEP06(2018)100}},
  \href{http://www.arXiv.org/abs/1803.10974}{\texttt{arXiv:1803.10974}}.

\bibitem{Agostinelli:2002hh}
\hrefCMSnoop {}{{GEANT4} Collaboration, ``{\GEANTfour}---a simulation
  toolkit'',} \textit{ Nucl. Instrum. Meth. A} \textbf{ 506} (2003) 250,
\href{http://dx.doi.org/10.1016/S0168-9002(03)01368-8}{\doi{10.1016/S0168-9002(03)01368-8}}.

\bibitem{Allison:2006ve}
\hrefCMSnoop {}{J.~Allison {et~al.}, ``{Geant4} developments and
  applications'',} \textit{ IEEE Trans. Nucl. Sci.} \textbf{ 53} (2006) 270,
\href{http://dx.doi.org/10.1109/TNS.2006.869826}{\doi{10.1109/TNS.2006.869826}}.

\bibitem{CMS-DP-2018-058}
\href {http://cds.cern.ch/record/2646773}{{CMS Collaboration}, ``Performance of
  the {DeepJet} b tagging algorithm using 41.9/fb of data from proton-proton
  collisions at 13 {TeV} with phase 1 {CMS} detector'',} CMS Detector
  Performance Summary CMS-DP-2018-058, 2018.

\bibitem{Sirunyan:2017ezt}
\hrefCMSnoop {}{{CMS Collaboration}, ``Identification of heavy-flavour jets
  with the {CMS} detector in pp collisions at 13 {TeV}'',} \textit{ JINST}
  \textbf{ 13} (2018) P05011,
  \href{http://dx.doi.org/10.1088/1748-0221/13/05/P05011}{\doi{10.1088/1748-0221/13/05/P05011}},
\href{http://www.arXiv.org/abs/1712.07158}{\texttt{arXiv:1712.07158}}.

\bibitem{Sirunyan:2017ulk}
\hrefCMSnoop {}{{CMS Collaboration}, ``Particle-flow reconstruction and global
  event description with the {CMS} detector'',} \textit{ JINST} \textbf{ 12}
  (2017) P10003,
  \href{http://dx.doi.org/10.1088/1748-0221/12/10/P10003}{\doi{10.1088/1748-0221/12/10/P10003}},
\href{http://www.arXiv.org/abs/1706.04965}{\texttt{arXiv:1706.04965}}.

\bibitem{Cacciari:2008gp}
\hrefCMSnoop {}{M.~Cacciari, G.~P. Salam, and G.~Soyez, ``The anti-\kt jet
  clustering algorithm'',} \textit{ JHEP} \textbf{ 04} (2008) 063,
  \href{http://dx.doi.org/10.1088/1126-6708/2008/04/063}{\doi{10.1088/1126-6708/2008/04/063}},
\href{http://www.arXiv.org/abs/0802.1189}{\texttt{arXiv:0802.1189}}.

\bibitem{Cacciari:2011ma}
\hrefCMSnoop {}{M.~Cacciari, G.~P. Salam, and G.~Soyez, ``{FastJet} user
  manual'',} \textit{ Eur. Phys. J. C} \textbf{ 72} (2012) 1896,
  \href{http://dx.doi.org/10.1140/epjc/s10052-012-1896-2}{\doi{10.1140/epjc/s10052-012-1896-2}},
\href{http://www.arXiv.org/abs/1111.6097}{\texttt{arXiv:1111.6097}}.

\bibitem{Bertolini:2014bba}
\hrefCMSnoop {}{D.~Bertolini, P.~Harris, M.~Low, and N.~Tran, ``Pileup per
  particle identification'',} \textit{ JHEP} \textbf{ 10} (2014) 059,
  \href{http://dx.doi.org/10.1007/JHEP10(2014)059}{\doi{10.1007/JHEP10(2014)059}},
\href{http://www.arXiv.org/abs/1407.6013}{\texttt{arXiv:1407.6013}}.

\bibitem{CMS-PAS-JME-16-003}
\href {https://cds.cern.ch/record/2256875}{{CMS Collaboration}, ``Jet
  algorithms performance in 13 {TeV} data'',} CMS Physics Analysis Summary
  CMS-PAS-JME-16-003, 2017.

\bibitem{Dokshitzer:1997in}
\hrefCMSnoop {}{Y.~L. Dokshitzer, G.~D. Leder, S.~Moretti, and B.~R. Webber,
  ``Better jet clustering algorithms'',} \textit{ JHEP} \textbf{ 08} (1997)
  001,
  \href{http://dx.doi.org/10.1088/1126-6708/1997/08/001}{\doi{10.1088/1126-6708/1997/08/001}},
\href{http://www.arXiv.org/abs/hep-ph/9707323}{\texttt{arXiv:hep-ph/9707323}}.

\bibitem{Wobisch:1998wt}
\href {https://inspirehep.net/record/484872}{M.~Wobisch and T.~Wengler,
  ``Hadronization corrections to jet cross-sections in deep inelastic
  scattering'',} in \textit{ {Proceedings of the Workshop on Monte Carlo
  Generators for HERA Physics, Hamburg, Germany}}, p.~270.
\newblock 1998.
\newblock
\href{http://www.arXiv.org/abs/hep-ph/9907280}{\texttt{arXiv:hep-ph/9907280}}.
\newblock

\bibitem{Dasgupta:2013ihk}
\hrefCMSnoop {}{M.~Dasgupta, A.~Fregoso, S.~Marzani, and G.~P. Salam, ``Towards
  an understanding of jet substructure'',} \textit{ JHEP} \textbf{ 09} (2013)
  029,
  \href{http://dx.doi.org/10.1007/JHEP09(2013)029}{\doi{10.1007/JHEP09(2013)029}},
\href{http://www.arXiv.org/abs/1307.0007}{\texttt{arXiv:1307.0007}}.

\bibitem{Butterworth:2008iy}
\hrefCMSnoop {}{J.~M. Butterworth, A.~R. Davison, M.~Rubin, and G.~P. Salam,
  ``Jet substructure as a new {Higgs} search channel at the {LHC}'',} \textit{
  Phys. Rev. Lett.} \textbf{ 100} (2008) 242001,
  \href{http://dx.doi.org/10.1103/PhysRevLett.100.242001}{\doi{10.1103/PhysRevLett.100.242001}},
\href{http://www.arXiv.org/abs/0802.2470}{\texttt{arXiv:0802.2470}}.

\bibitem{Larkoski:2014wba}
\hrefCMSnoop {}{A.~J. Larkoski, S.~Marzani, G.~Soyez, and J.~Thaler, ``Soft
  drop'',} \textit{ JHEP} \textbf{ 05} (2014) 146,
  \href{http://dx.doi.org/10.1007/JHEP05(2014)146}{\doi{10.1007/JHEP05(2014)146}},
\href{http://www.arXiv.org/abs/1402.2657}{\texttt{arXiv:1402.2657}}.

\bibitem{10.2307/1913018}
\hrefCMSnoop {}{F.~M. Fisher, ``Tests of equality between sets of coefficients
  in two linear regressions: An expository note'',} \textit{ Econometrica}
  \textbf{ 38} (1970) 361,
  \href{http://dx.doi.org/10.2307/1913018}{\doi{10.2307/1913018}}.

\bibitem{Butterworth:2015oua}
\hrefCMSnoop {}{J.~Butterworth {et~al.}, ``{PDF4LHC} recommendations for {LHC
  Run II}'',} \textit{ J. Phys. G} \textbf{ 43} (2016) 023001,
  \href{http://dx.doi.org/10.1088/0954-3899/43/2/023001}{\doi{10.1088/0954-3899/43/2/023001}},
\href{http://www.arXiv.org/abs/1510.03865}{\texttt{arXiv:1510.03865}}.

\bibitem{CLS2}
\hrefCMSnoop {}{T.~Junk, ``Confidence level computation for combining searches
  with small statistics'',} \textit{ Nucl. Instrum. Meth. A} \textbf{ 434}
  (1999) 435,
  \href{http://dx.doi.org/10.1016/S0168-9002(99)00498-2}{\doi{10.1016/S0168-9002(99)00498-2}},
\href{http://www.arXiv.org/abs/hep-ex/9902006}{\texttt{arXiv:hep-ex/9902006}}.

\bibitem{CLS1}
\hrefCMSnoop {}{A.~L. Read, ``Presentation of search results: The
  {CL$_{\text{s}}$} technique'',} \textit{ J. Phys. G} \textbf{ 28} (2002)
  2693,
\href{http://dx.doi.org/10.1088/0954-3899/28/10/313}{\doi{10.1088/0954-3899/28/10/313}}.

\bibitem{Cowan:2010js}
\hrefCMSnoop {}{G.~Cowan, K.~Cranmer, E.~Gross, and O.~Vitells, ``Asymptotic
  formulae for likelihood-based tests of new physics'',} \textit{ Eur. Phys. J.
  C} \textbf{ 71} (2011) 1554,
  \href{http://dx.doi.org/10.1140/epjc/s10052-011-1554-0}{\doi{10.1140/epjc/s10052-011-1554-0}},
  \href{http://www.arXiv.org/abs/1007.1727}{\texttt{arXiv:1007.1727}}.
[Erratum: \DOI{10.1140/epjc/s10052-013-2501-z}].

\end{thebibliography}\endgroup
